\documentclass[12pt]{article}
\pdfoutput=1
\usepackage{multicol}
\usepackage{hyperref}
\usepackage{mathrsfs}
\usepackage{amsmath}
\usepackage{amsfonts}
\usepackage{mathrsfs}
\usepackage{mathtools}
\usepackage{rotating}
\usepackage{slashed}
\usepackage{epigraph}
\usepackage{fancyhdr}
\usepackage{amssymb}
\usepackage{graphicx}
\usepackage{makecell}
\usepackage{amscd}
\usepackage{bm}
\usepackage{color}
\usepackage{array}
\usepackage{verbatim}
\usepackage{cite}
\usepackage{subfig}
\usepackage{setspace}
\usepackage{breakurl}
\usepackage{authblk}
\usepackage{multirow}
\usepackage{xspace}
\usepackage{fullpage}
\usepackage{url}
\usepackage{style_files/tabu}
\usepackage{cancel}
\usepackage[utf8]{inputenc}
\usepackage[normalem]{ulem}
\usepackage{pdflscape}
\usepackage{soul}
\def\beq {\begin{equation}}
\def\eeq {\end{equation}}
\def\bea {\begin{eqnarray}}
\def\eea {\end{eqnarray}}
\def\bc {\begin{center}}
\def\ec {\end{center}}
\def\beaa {\begin{align}}
\def\eeaa {\end{align}}

\def\nn {\nonumber}

\def\hedv{\bar c \, \gamma^\mu \, {\rm P}_L \, b}
\def\heda{\bar c \, \gamma^\mu \, {\rm P}_R \, b}
\def\heds{\bar c  \, {\rm P}_L \, b}
\def\hedp{\bar c  \, {\rm P}_R \, b}
\def\hedt{\bar c \, \sigma^{\mu\nu} \,  b}
\def\Dst{{D^*}}
\def\bra{\langle}
\def\ket{\rangle}

\def\Eqn#1{Eq.~(\ref{#1})}
\def\PDSQ{|p_D|^2}
\def\PDd{|p_D|}

\def\FzeroSQ{{\bf F_0^2}}
\def\FoneSQ{{\bf  F_+^2}}
\def\FtwoSQ{{\bf  F_T^2}}

\def\CSLSq{|{\bf C_{SL}^{\ell}|^2}}
\def\CSRSq{|{\bf C_{SR}^{\ell}|^2}}
\def\CVLSq{|{\bf C_{VL}^{\ell}|^2}}
\def\CVRSq{|{\bf C_{VR}^{\ell}|^2}}
\def\CTLSq{|{\bf C_{TL}^{\ell}|^2}}
\def\CTRSq{|{\bf C_{TR}^{\ell}|^2}}

\def\pds{\left| p_{D^\ast}\right|^2}
\def\mds{M_{D^\ast}^2}
\def\cvl{C_{VL}^\ell}
\def\cvr{C_{VR}^\ell}
\def\cvrconj{C_{VR}^{\ell\ast}}
\def\call{C_{AL}^\ell}

\def\calconj{C_{AL}^{\ell\ast}}
\def\cpl{C_{PL}^\ell}
\def\cpr{C_{PR}^\ell}
\def\cplconj{C_{PL}^{\ell\ast}}

\def\csr{C_{SR}^\ell}
\def\cvlsq{\left|C_{VL}^\ell\right|^2}

\def\calsq{\left|C_{AL}^\ell\right|^2}

\def\cplsq{\left|C_{PL}^\ell\right|^2}

\def\azsq{A_0^2}
\def\aosq{A_1^2}
\def\atsq{A_2^2}
\def\vsq{V^2}
\def\mbmd{\left(M_B + M_{D^\ast}\right)}
\def\mbmc{\left(m_b + m_c\right)}
\def\mbmdq2{\left(M_B^2 - M_{D^\ast}^2 - q^2\right)}
\def\re{\mathcal{R}}
\def\nn{\nonumber}

\def\ctl{C_{TL}^\ell}
\def\ctlconj{C_{TL}^{\ell\ast}}

\def\ctrconj{C_{TR}^{\ell\ast}}
\def\ctlsq{\left|C_{TL}^\ell\right|^2}
\def\ctrsq{\left|C_{TR}^\ell\right|^2}
\def\mbmdsq{\left(M_B^2 - M_{D^\ast}^2 \right)}
\def\mbmdqsq{\left(M_B^2 - M_{D^\ast}^2 - q^2\right)}
\def\mbmdqthsq{\left(M_B^2 + 3 M_{D^\ast}^2 - q^2\right)}
\def\bdtaunu{\overline{B} \to D \tau \bar{\nu}_\tau}

\def\bdstaunu{\overline{B} \to D^* \tau \bar{\nu}_\tau}
\def\bdslnu{\overline{B} \to D^* \ell \bar{\nu}_\ell}

\def\cpr{C_{PR}^\ell}
\def\cvr{C_{VR}^\ell}
\def\car{C_{AR}^\ell}

\def\ctrconj{C_{TR}^{\ell\ast}}
\def\mB{M_B}
\def\mDs{M_{D^\ast}}
\def\pDmag{|p_{D^\ast}|}
\def\ctrsq{\left|C_{TR}^\ell\right|^2}
\allowdisplaybreaks
\begin{document}
\thispagestyle{empty}
\begin{center}
\vspace*{1cm}
{\LARGE\bf A closer look at the $R_D$ and $R_{D^*}$ anomalies\\}
\bigskip
{\large Debjyoti Bardhan}\,$^{a,1}$, \, \, 
{\large Pritibhajan Byakti}\,$^{b,2}$, \, \, 
{\large Diptimoy Ghosh}\,$^{c,3}$
\\
\bigskip 
\bigskip
{\small
$^a$ Department of Theoretical Physics, Tata Institute of Fundamental 
Research, \\ 1 Homi Bhabha Road, Mumbai 400005, India. \\[2mm]
$^b$ Department of Theoretical Physics, Indian Association for the Cultivation of Science,
\\ 2A \& 2B, Raja S.C. Mullick Road, Jadavpur, Kolkata 700 032, India. \\[2mm]
$^c$ Department of Particle Physics and Astrophysics, Weizmann Institute of Science, 
\\ Rehovot 76100, Israel. 
}
\end{center}
\vspace*{1.6cm}
{\bf Abstract} 
The measurement of $R_D$ ($R_{D^*}$), the ratio of the branching fraction of $\bdtaunu (\bdstaunu)$ to that of 
$\overline{B} \to D l \bar{\nu}_l (\overline{B} \to D^* l \bar{\nu}_l)$, shows $1.9 \sigma$ $(3.3 \sigma)$ deviation from 
its Standard Model (SM) prediction. The 
combined deviation is at the level of $4 \sigma$ according to the Heavy Flavour Averaging Group (HFAG). 
In this paper, we perform an effective field theory analysis (at the dimension 6 level) of these potential New Physics 
(NP) signals assuming $\rm SU(3)_{C} \times SU(2)_{L} \times U(1)_{Y}$ gauge invariance. We first show that, in general, 
$R_D$  and $R_{D^*}$ are theoretically independent observables and hence, their theoretical predictions 
are not correlated. We identify the operators that can explain the  experimental measurements of $R_D$  and 
$R_{D^*}$ individually and also together. Motivated by the recent measurement of the 
$\tau$ polarisation in $\bdstaunu$ decay, $P_\tau (D^*)$ by the Belle collaboration, we study the 
impact of a more precise measurement of  $P_\tau (D^*)$ (and a measurement of $P_\tau (D)$) on the various possible NP 
explanations. Furthermore, we show that the measurement of $R_{D^*}$ in bins of $q^2$, the square of the invariant mass of 
the lepton-neutrino system, along with the information on $\tau$ polarisation and the forward-backward asymmetry of the 
$\tau$ lepton,  can completely distinguish the various operator structures. 
We also provide the full expressions of the double differential decay widths for the individual 
$\tau$ helicities in the presence of all the 10 dimension-6 operators that can contribute to these decays.
\vspace*{-0.35in}
\begin{quotation}
\noindent    
\end{quotation}
\bigskip
%
%
\vfill
%
%
\bigskip
\hrule
\vspace*{0.1in}
\hspace*{-4mm}
$^1$ debjyoti@theory.tifr.res.in ~~
$^2$ tppb@iacs.res.in ~~
$^3$ diptimoy.ghosh@weizmann.ac.il 
\newpage 
\tableofcontents
\section{Introduction}

In recent years, a number of experimental measurements involving $B$ meson decays have shown interesting deviations 
from their Standard Model (SM) expectations. Deviations have been seen both in the neutral current $b \to s$ decays 
\cite{Aaij:2014ora,Aaij:2015oid}\footnote{For theoretical implications, see for example 
\cite{Altmannshofer:2008dz,Alok:2009tz,Alok:2010zd,Alok:2011gv,DescotesGenon:2011yn,Descotes-Genon:2013wba,Altmannshofer:2013foa,
Datta:2013kja,Ghosh:2014awa,Mandal:2014kma,Altmannshofer:2014rta,Jager:2014rwa, Descotes-Genon:2015uva,Ciuchini:2015qxb} and 
the references therein.} as well as the charged current $b \to c$ processes. 
The most statistically significant deviation, at the $4 \sigma$ level \cite{Amhis:2014hma}, is seen in the combination of $R_D$ and $R_{D^*}$ 
which are defined as,
\bea
R_{D^{(*)}} = \frac{\mathcal{B} \left(\overline{B} \to D^{(\ast)} \tau \bar{\nu}_{\tau} \right)}{\mathcal{B} \left(\overline{B} \to 
D^{(\ast)} l \bar{\nu}_l \right)} \, ,
\label{def-rd}
\eea
where $l = e$ or $\mu$.  
In Table \ref{exp-values}, we collect all the relevant experimental results related to the $\overline{B} \to D^{(\ast)} \ell \nu_\ell$ decay 
processes. 

\begin{table}[ht!]
\tabulinesep=1.05mm
\hspace*{-0.3cm}\begin{tabu}{ |l|l|cl|c| }
\hline
\multicolumn{5}{ |c| }{List of Observables} \\
\hline
\multirow{2}{*}{Observable} & \multicolumn{3}{c|}{ Experimental Results} & \multirow{2}{*}{SM Prediction} \\
\cline{2-4}
 & Experiment & \multicolumn{2}{ c| }{Measured value} &  \\ \hline
\multirow{4}{*}{$R_D$} & Belle & 0.375 $\pm$ 0.064 $\pm$ 0.026 & \cite{Huschle:2015rga} & 
\multirow{1}{*}{0.299 $\pm$ 0.011 \cite{Lattice:2015rga}} \\
& BaBar & 0.440 $\pm$ 0.058 $\pm$ 0.042 & \cite{Lees:2012xj,Lees:2013uzd} &  0.300 $\pm$ 0.008 \cite{Na:2015kha} \\
\cline{2-4}
& \multirow{2}{*}{HFAG average} &  \multirow{2}{*}{0.397 $\pm$ 0.040 $\pm$ 0.028} & \multirow{2}{*}{\cite{Amhis:2014hma}} & $ 0.299 \pm 0.003$ \cite{Bigi:2016mdz} \\ 
  &&& & ${\bf 0.300 \pm 0.011}$ \\ \hline
\multirow{5}{*}{$R_{D^\ast}$} & Belle & 0.293 $\pm$ 0.038 $\pm$ 0.015 & \cite{Huschle:2015rga} & \multirow{7}{*}{0.252 $\pm$ 0.003 \cite{Fajfer:2012vx}} \\
& Belle & 0.302 $\pm$ 0.030 $\pm$ 0.011 & \cite{Abdesselam:2016cgx} & \\
& BaBar & 0.332 $\pm$ 0.024 $\pm$ 0.018 & \cite{ Lees:2012xj,Lees:2013uzd}  & \\
& LHCb & 0.336 $\pm$ 0.027 $\pm$ 0.030 & \cite{Aaij:2015yra} &  \\
\cline{2-4}
& HFAG average & 0.316 $\pm$ 0.016 $\pm$ 0.010 & \cite{Amhis:2014hma} & ${\bf 0.254 \pm 0.004}$ \\
\cline{2-4}
& Belle & 0.276 $\pm$ 0.034 ${}^{+0.029}_{-0.026}$ & \cite{Abdesselam:2016xqt}  & \\
\cline{2-4}
& Our average & $0.310 \pm 0.017$ &     & \\
\hline
$\mathcal{B} \left(\overline{B} \to D \tau \bar{\nu}_\tau \right) $& BaBar & 1.02 $\pm$ 0.13 $\pm$ 0.11 \% & \cite{Lees:2012xj} 
& ${ \bf 0.633 \pm 0.014 \, \%}$  \\ \hline
$\mathcal{B} \left(\overline{B} \to D^\ast \tau \bar{\nu}_\tau \right)$ & BaBar & 1.76 $\pm$ 0.13 $\pm$ 0.12 \% & \cite{Lees:2012xj} 
&  ${\bf 1.28 \pm 0.09}$ \%\\ \hline 
\multirow{1}{*}{$\mathcal{B} \left(\overline{B} \to D l \bar{\nu}_l \right)$} & HFAG average & 2.13 $\pm$ 0.03 $\pm$ 
0.09 \%& \cite{Amhis:2014hma} &  ${\bf 2.11^{+ 0.12}_{-0.10} \, \%}$ \\ \hline
\multirow{1}{*}{$\mathcal{B} \left(\overline{B} \to D^\ast l \bar{\nu}_l \right)$} &  HFAG average & 4.93 $\pm$ 0.01 $\pm$ 0.11 \%
& \cite{Amhis:2014hma} & ${\bf 5.04^{+0.44}_{-0.42} } \% $\\ \hline
\multirow{2}{*}{$P_\tau \left(\overline{B} \to D \tau \bar{\nu}_\tau \right)$}  &   &   &  & $0.325 \pm 0.009$ \cite{Tanaka:2010se} \\ 
    &   &   &  &  ${ \bf 0.325 \pm {0.012}}$\\ \hline
\multirow{2}{*}{$P_\tau \left(\overline{B} \to D^\ast \tau \bar{\nu}_\tau \right)$}  & \multirow{2}{*}{Belle} & 
\multirow{2}{*}{$-0.44$ $\pm$ 0.47 ${}^{+0.20}_{-0.17}$ }
& \multirow{2}{*}{\cite{Abdesselam:2016xqt}} &  $-0.497 \pm 0.013$ \cite{Tanaka:2012nw,Abdesselam:2016xqt} \\
 & &  & &  ${\bf -0.497 \pm 0.008}$   \\ \hline
$\mathcal{A}_{FB}^D$  &   &   &  &  ${\bf -0.360 ^{+0.002}_{-0.001} }$ \\
 \hline
$\mathcal{A}_{FB}^{D^*}$  &   &   &  & ${ \bf 0.064 \pm 0.014 }$ \\ \hline
\end{tabu}
\caption{The relevant observables, their experimental measurements and the SM predictions are shown. While computing the 
branching ratios, we have used $V_{cb} = 0.04$. As HFAG has not yet included the latest Belle measurement of $R_{D^*}$ in their 
global average, we have taken a naive weighted average of the latest Belle result and the average given by HFAG. However, since the 
recent Belle result has a large uncertainty, it does not affect the previous world average in any significant way. The values given in 
boldface are our results for the SM predictions. Note that, for the $\bdslnu$ SM predictions, the uncertainties correspond to $2\sigma$ 
uncertainties in the form factor parameters, see section \ref{ff-2} for more details. \label{exp-values}}
\end{table}

Note that, we have used the notation $\ell$ to denote any lepton (e, $\mu$ or $\tau$) and $l$ to denote only the 
light leptons, e and $\mu$.  

The large statistical significance 
of the anomaly in $R_D$ and $R_{D^\ast}$ has spurred a lot of interest in this decay modes in the last few years 
\cite{Nierste:2008qe,Fajfer:2012vx,Datta:2012qk,Sakaki:2012ft,Crivellin:2012ye,Choudhury:2012hn,Tanaka:2012nw,Celis:2012dk,
Sakaki:2013bfa,Dorsner:2013tla,Duraisamy:2013kcw,Biancofiore:2013ki,Duraisamy:2014sna,
Freytsis:2015qca,Greljo:2015mma,Calibbi:2015kma,Bhattacharya:2015ida,Hati:2015awg,
Bauer:2015knc,Barbieri:2015yvd,Ivanov:2015tru,Cline:2015lqp,Das:2016vkr,Bordone:2016tex,Alonso:2016gym,
Nandi:2016wlp,Feruglio:2016gvd,Alok:2016qyh,Boucenna:2016wpr,Boucenna:2016qad,Sahoo:2016pet,
Faroughy:2016osc,Ligeti:2016npd,Ivanov:2016qtw,Dorsner:2016wpm,Becirevic:2016yqi} and  various possible theoretical explanations have been proposed.  

The main purpose of this work is to identify observables which can help distinguish the different NP Lorentz structures that can potentially 
solve  the $R_D$ and $R_{D^*}$ anomalies. 
We first perform an operator analysis of these potential NP signals by considering  all the dimension-6 operators 
that are consistent with SM gauge invariance. We compute the values of the relevant Wilson coefficients (WCs) that explain the 
experimental measurements within their $1 \sigma$ ranges. It is important to note that we consider the presence of NP only in the tau-channel and not for the electron or the muon channels. Thus, in our calculations of $R_D$ and $R_{D^\ast}$, we use the SM values of the WCs in the denominator. For these values of the WCs, we compute the predictions for a few observables 
that have the potential to distinguish between the various NP operators. Although we provide numerical results only for the operators that 
are consistent with SM gauge invariance, we provide the analytical expressions for the double differential 
decay rates for the individual $\tau$ helicities for all the 10 independent dimension-6 operators contributing to these decays. 
To our knowledge, we are the first in the literature to provide the full expressions.

As we show later, $R_D$ and $R_{D^\ast}$ are in general theoretically independent observables and the anomalies
can exist independently. A future measurement might reveal a greater anomaly in one of them without affecting
the other. Hence, in this paper, we attempt to explain each without worrying about the other initially, but then also point out how 
both can be explained together.

Very recently, the Belle collaboration reported the first measurement of the $\tau$-polarisation  in the decay 
 $\overline{B} \to D^\ast \tau \bar{\nu}_\tau$ \cite{Abdesselam:2016xqt}. While the uncertainty in this measurement is rather large now, 
motivated by the possibility of more precise measurements in the future, we  investigate how such a measurement  can distinguish the various NP 
explanations of $R_D$ and $R_{D^*}$. Furthermore, we show that measurements of $R_{D^*}$ in bins of $q^2$ can provide 
important information about the nature of short distance physics. In fact, a combination of binwise $R_{D^*}$ and more precise 
measurements (that can be done in Belle II, for example) of $\tau$ polarisation in both the $\bdtaunu$ and $\bdstaunu$ decays can 
completely distinguish all the different NP operators. 
Moreover, we show that the forward-backward asymmetry of the $\tau$ lepton (in the $\tau$ - $\nu_\tau$ rest frame) also has the 
potential to differentiate the various NP Lorentz structures.

The paper is organised as follows: In section \ref{op-basis} we write down all the operators relevant for this study and define 
 the notations for the corresponding WCs. The various observables of our interest are defined in section \ref{observables}. 
 The sections \ref{ff-1} and \ref{ff-2} discuss the form factors required for the calculation of the decay amplitudes. 
The analytic expressions for the double differential decay widths for the individual lepton helicities are shown in sections 
\ref{b2d-formulae} and \ref{b2ds-formulae}. In the following section (section \ref{results}), we present all our numerical results. 
Finally, we summarise our findings in section \ref{conclusion}.

The full expressions for the double differential decay widths are shown in the appendices \ref{b2d-full} and \ref{b2ds-full}, and 
the contribution of the tensor operator ${\cal O}_{\rm TL}$ is discussed in appendix \ref{tensorL}. In appendix \ref{warsaw}, 
we show how our operators are related to the dimension-6 operators of \cite{Grzadkowski:2010es}. The renormalisation 
group equations for the WCs are computed in appendix \ref{RG}.

\section{Operator basis}
\label{op-basis}

The effective Lagrangian for the $b \to c \, \ell \, \bar{\nu}$ process at the dimension 6 level is given by, 
\begin{eqnarray}
{\cal L}^{b \to c \, \ell \, \nu}_{\rm eff} =  \frac{2 G_F V_{cb}}{\sqrt{2}}&\Big(& C^{cb\ell}_9 \, {\mathcal O}^{cb\ell}_9 + C^{cb\ell \, '}_9 \, {\mathcal O}^{cb\ell \, '}_9 + 
C^{cb\ell}_{10} \, {\mathcal O}^{cb\ell}_{10}  + C^{cb\ell \, '}_{10} \, {\mathcal O}^{cb\ell \, '}_{10} + C^{cb\ell}_s \, {\mathcal O}^{cb\ell}_s  + C^{cb\ell \, '}_s \, {\mathcal O}^{cb\ell \, '}_s   \nonumber \\
&+&  C^{cb\ell}_p \, {\cal O}^{cb\ell}_p   + C^{cb\ell \, '}_p \, {\cal O}^{cb\ell \, '}_p + C^{cb\ell}_T \, {\cal O}^{cb\ell}_T + C^{cb\ell}_{T5} \, {\cal O}^{cb\ell}_{T5} \, \, \Big) 
\label{eff-lag}
\end{eqnarray}
where ${\mathcal O}^{cb\ell}_i$ constitute a complete basis of 6-dimensional operators and $C^{cb\ell}_i$ are the corresponding Wilson coefficients defined at the renormalization scale $\mu = m_b$. In the SM, $C^{cb\ell}_9 = - C^{cb\ell}_{10} = 1$ and all the other WCs vanish. The full set of operators is given by:
\begin{multicols}{2}
\noindent
\begin{align}
\label{b2c-basis}
{\cal O}^{cb \ell}_9       &=    [\hedv ] [\bar \ell \, \gamma_\mu \, \nu]  \nonumber \\ 
{\cal O}^{cb\ell}_{10}   &=    [\hedv ] [\bar \ell \, \gamma_\mu \gamma_5 \, \nu] \nonumber \\  
{\cal O}^{cb\ell}_s       &=    [\heds] [\bar \ell  \, \nu] \nonumber \\ 
{\cal O}^{cb\ell}_p       &=    [\heds] [[\bar \ell \, \gamma_5 \, \nu] \nonumber \\ 
{\cal O}^{cb\ell}_T       &=   [\hedt] [\bar \ell \, \sigma_{\mu\nu} \, \nu] \nonumber 
\end{align}
\begin{align}
{\cal O}^{cb\ell \, '}_9       &=    [\heda ] [\bar \ell \, \gamma_\mu \, \nu]  \nonumber \\ 
{\cal O}^{cb\ell \, '}_{10}   &=    [\heda ] [\bar \ell \, \gamma_\mu \gamma_5 \, \nu] \nonumber \\  
{\cal O}^{cb\ell \, '}_s       &=    [\hedp] [\bar \ell  \, \nu]  \\ 
{\cal O}^{cb\ell \, '}_p       &=    [\hedp] [[\bar \ell \, \gamma_5 \, \nu] \nonumber \\ 
 {\cal O}^{cb\ell}_{T5}  &=   [\hedt] [\bar \ell \, \sigma_{\mu\nu}\gamma_5 \, \nu] \nonumber
\end{align}
\end{multicols}

The other possible tensor structures are related to  ${\cal O}^{cb\ell}_{T}$ and ${\cal O}^{cb\ell}_{T5}$  in the following way,
\begin{eqnarray}
\epsilon_{\mu \nu \alpha \beta} [\bar c \, \sigma^{\mu\nu} \,  b] [\bar \ell \, \sigma^{\alpha \beta} \, \nu] &=& -2 i {\cal O}^{cb\ell}_{T5} \\ \relax
[\bar c \, \sigma^{\mu\nu} \gamma_5 \,  b] [\bar \ell \, \sigma_{\mu\nu} \gamma_5 \, \nu] &=&  {\cal O}^{cb\ell}_T \\ \relax
[\bar c \, \sigma^{\mu\nu} \gamma_5 \,  b] [\bar \ell \, \sigma_{\mu\nu} \, \nu] &=& {\cal O}^{cb\ell}_{T5} \, .
\end{eqnarray}

Note that the above basis of operators is different from the one used in some earlier literature \cite{Datta:2012qk,Sakaki:2013bfa}. 
For example, the reference \cite{Datta:2012qk}  uses the following set of operators, 
\begin{multicols}{2}
\noindent
\begin{align}
\label{b2c-basis}
{\cal O}^{cb \ell}_{\rm VL}      &=    [\bar{c} \, \gamma^\mu \, b] [\bar \ell \, \gamma_\mu \, P_L \, \nu]  \nonumber \\ 
{\cal O}^{cb\ell}_{\rm AL}       &=    [\bar{c} \, \gamma^\mu \, \gamma_5 \, b] [\bar \ell \, \gamma_\mu \, P_L \, \nu] \nonumber \\  
{\cal O}^{cb\ell}_{\rm SL}       &=    [\bar{c} \, b] [\bar \ell  \, P_L \, \nu] \nonumber \\ 
{\cal O}^{cb\ell}_{\rm PL}       &=    [\bar{c} \, \gamma_5 \, b] [[\bar \ell \, P_L \, \nu] \nonumber \\ 
{\cal O}^{cb\ell}_{\rm TL}       &=   [\hedt] [\bar \ell \, \sigma_{\mu\nu} \, P_L \, \nu] \nonumber 
\end{align}
\begin{align}
{\cal O}^{cb \ell}_{\rm VR}      &=    [\bar{c} \, \gamma^\mu \, b] [\bar \ell \, \gamma_\mu \, P_R \, \nu]  \nonumber \\ 
{\cal O}^{cb\ell}_{\rm AR}       &=    [\bar{c} \, \gamma^\mu \, \gamma_5 \, b] [\bar \ell \, \gamma_\mu \, P_R \, \nu] \nonumber \\  
{\cal O}^{cb\ell}_{\rm SR}       &=    [\bar{c} \, b] [\bar \ell  \, P_R \, \nu] \\ 
{\cal O}^{cb\ell}_{\rm PR}       &=    [\bar{c} \, \gamma_5 \, b] [[\bar \ell \, P_R \, \nu] \nonumber \\ 
{\cal O}^{cb\ell}_{\rm TR}       &=   [\hedt] [\bar \ell \, \sigma_{\mu\nu} \, P_R \, \nu] \nonumber 
\end{align}
\end{multicols}

The Wilson coefficients of these two basis of operators are related through the following equations, 
\begin{multicols}{2}
\noindent
\begin{align}
C_{\rm VL}^{cb\ell}  &=  \frac12 \left(C_9^{cb\ell} - C_{10}^{cb\ell} + 
C_{9}^{cb\ell \, '} - C_{10}^{cb\ell \, '}\right)\nonumber \\ 
C_{\rm AL}^{cb \ell}  &=  \frac12 \left(-C_9^{cb\ell} + 
C_{10}^{cb\ell} + C_{9}^{cb\ell \, '} - C_{10}^{cb\ell \, '}\right)  \nonumber 
\end{align}
\begin{align}
C_{\rm SR}^{cb\ell}  &=  \frac12 \left( C_s^{cb\ell} + C_p^{cb\ell} + 
C_s^{cb\ell \, '} + C_p^{cb\ell \, '} \right) \\ 
C_{\rm PR}^{cb\ell}  &=  \frac12 \left(- C_s^{cb\ell} -
C_p^{cb\ell} + C_s^{cb\ell \, '} + C_p^{cb\ell \, '} \right) \nonumber 
\end{align}
\end{multicols}

\begin{multicols}{2}
\noindent
\begin{align}
C_{\rm SL}^{cb \ell}   &=  \frac12 \left( C_s^{cb\ell} - C_p^{cb\ell} + 
C_s^{cb\ell \, '} - C_p^{cb\ell \, '} \right)\nonumber \\ 
C_{\rm PL}^{cb \ell}   &=  \frac12 \left(- C_s^{cb\ell} + 
C_p^{cb\ell} + 
C_s^{cb\ell \, '} - C_p^{cb\ell \, '} \right)\nonumber \\ 
C_{\rm TL}^{cb \ell} &=  \left(C_T^{cb\ell}-C_{T5}^{cb\ell}\right) \nonumber
\end{align}
\begin{align}
C_{\rm VR}^{cb\ell} &=  \frac12 \left(C_9^{cb\ell} +C_{10}^{cb\ell} + 
C_{9}^{cb\ell \, '} + C_{10}^{cb\ell \, '}\right)  \label{wc-LR} \\ 
C_{\rm AR}^{cb\ell} &=  \frac12 \left(-C_9^{cb\ell} 
-C_{10}^{cb\ell} + 
C_{9}^{cb\ell \, '} + C_{10}^{cb\ell \, '}\right) \nonumber \\ 
C_{\rm TR}^{cb\ell} &=  \left(C_T^{cb\ell} + C_{T5}^{cb\ell}\right) \nonumber
\end{align}
\end{multicols}

We now assume the neutrino in the final state to be left handed. This implies that the WCs 
in eq.~\eqref{eff-lag} satisfy the following relations, 
\bea
C^{cb\ell}_9 &=& - C^{cb\ell}_{10}  \\
C^{cb\ell \, '}_9 &=& - C^{cb\ell \, '}_{10} \\
C^{cb\ell}_s &=& - C^{cb\ell}_p \\
C^{cb\ell \, '}_s &=& - C^{cb\ell \, '}_p \\
C^{cb\ell}_T &=& -C^{cb\ell}_{T5} \, .
\label{nu-left}
\eea
Consequently, all the WCs in the right hand column of eq.~\ref{wc-LR} vanish. Note that, the operators on the left hand column 
of eq.~\ref{wc-LR} are the only ones that are consistent with the full gauge invariance of the SM. In appendix \ref{warsaw}, we 
show how these WCs are related to the 6-dimensional operators listed in \cite{Grzadkowski:2010es}. 
Moreover, since many microscopic models do not generate the tensor operator, we neglect them in the main text and 
study its effect  only in the appendix (see appendix \ref{tensorL}). 

Although, we do not study the effects of the operators with a right handed neutrino (the ones in the right hand column of eq.~\ref{wc-LR}), 
we compute the full analytic expressions considering all the 10 operators for the first time in the literature. The results are presented in 
appendices \ref{b2d-full} and \ref{b2ds-full}.  
\section{Observables}
\label{observables}

The double differential branching fractions for the decays $\overline{B} \to D \ell \bar{\nu}_\ell$ and 
$\overline{B} \to D^\ast \ell \bar{\nu}_\ell$ can be written as
\bea
\frac{d^2 {\mathcal B}^{D^{(*)}}_\ell}{d q^2 \, d(\cos\theta)} &=&  {\mathcal N} \, |p_{D^{(*)}}| \, \left( a_\ell^{D^{(*)}} +
 b_\ell^{D^{(*)}} \cos\theta +  c_\ell^{D^{(*)}} \cos^2\theta \right) \, .
\eea
The normalisation factor, $\cal{N}$ and the absolute value of the $D^{(*)}$-meson momentum, $|p_{D^{(*)}}|$ are  given by,
\bea
{\cal N}  &=& \frac{\tau_B \, G_F^2 |V_{cb}|^2q^2}{256 \pi^3 M_B^2}  \, \left( 1 - \frac{m_\ell^2}{q^2} \right)^2 \\
|p_{D^{(*)}}| &=& \frac{\sqrt{\lambda(M_B^2, M_{D^{(*)}}^2,q^2)}}{2 M_B},
\eea
where $\lambda(a,b,c)= a^2 + b^2 +c^2 -2 (ab + bc + ca)$.
The angle $\theta$ is defined as the angle between the lepton and $D^{(*)}$-meson in the lepton-neutrino centre-of-mass frame,
and $q^2$ is the invariant mass squared of the lepton-neutrino system. 

The total branching fraction is given by,
\bea
 {\cal B}^{D^{(*)}}_\ell &=& \int {\cal N} \, |p_{D^{(*)}}| \, \left( 2 a_\ell^{D^{(*)}} +  \frac{2}{3}  c_\ell^{D^{(*)}} \right) d q^2
\eea

The observables $R_D$ and $R_{D^*}$ have already been defined in eq.~\eqref{def-rd}. We now define binned $R_{D^{(*)}}$ in the 
following way,
\bea
R_{D^{(*)}}[q^2 \, {\rm bin}] = \frac{{\cal B}^{D^{(*)}}_\tau [q^2 \, {\rm bin}]}{{\cal B}^{D^{(*)}}_{l} [q^2 \, {\rm bin}]} \, 
\eea

For the decays with $\tau$ lepton in the final state, the polarisation of the $\tau$ also constitutes an useful observable and can potentially 
be used to distinguish the NP Lorentz structures. The $\tau$ polarisation fraction is defined in the following way, 
\bea
P_{\tau} (D^{(*)}) = \frac{\Gamma^{D^{(*)}}_\tau (+) ~ - ~ \Gamma^{D^{(*)}}_\tau (-)}{\Gamma^{D^{(*)}}_\tau (+) ~ + ~ \Gamma^{D^{(*)}}_\tau (-)}
\eea
where, $\Gamma^{D^{(*)}}_\tau (+)$ and $\Gamma^{D^{(*)}}_\tau (-)$ are the decay widths for positive and negative helicity $\tau$ 
leptons respectively.\\
The $\tau$ forward-backward asymmetry, $\mathcal{A}_{FB}^{D^(\ast)}$ is defined as
\bea
\mathcal{A}_{FB}^{D^{(\ast)}} &=& \frac{\int_0^{\pi/2} \frac{d\Gamma^{D^{(\ast)}}}{d\theta} d\theta- \int_{\pi/2}^{\pi} 
\frac{d\Gamma^{D^{(\ast)}}}{d\theta} d\theta}{\int_0^{\pi/2} \frac{d\Gamma^{D^{(\ast)}}}{d\theta} d\theta + \int_{\pi/2}^{\pi} \frac{d\Gamma^{D^{(\ast)}}}{d\theta}d \theta} \nn \\
&=& \frac{\int b_\tau^{D^{(\ast)}}(q^2) dq^2}{\Gamma^{D^{(\ast)}}}
\label{def-fba}
\eea
where $\Gamma^{D^{(\ast)}}$ is  the total decay width of $D^{(\ast)}$ and the angle $\theta$ has already been defined above.  
Note that, while the branching fractions depend on the functions $a_\ell^{D^{(*)}}$ and $c_\ell^{D^{(*)}}$, the forward-backward asymmetry 
depends only on $b_\ell^{D^{(*)}}$. Hence, they provide complementary information on the nature of the short distance physics. 
 
\section{$\bar{B} \to D$ form factors}
\label{ff-1}

The hadronic matrix elements for $\bar{B} \to D$ transition are parametrised 
by\footnote{We use the convention $\epsilon^{0123} =1$. This implies $\epsilon_{0123} =-1$. }

\bea\label{a:e:vf}
\bra D(p_D, M_D)| \bar c \gamma^\mu b |\bar B(p_B, M_B)\ket &=& F_+(q^2) \Big[(p_B+p_D)^\mu -\frac{M_B^2 - M_D^2}{q^2} q^\mu \Big] \nonumber \\*
&&  + F_0(q^2) \frac{M_B^2- M_D^2}{q^2} q^\mu  \\*
\bra D(p_D, M_D)| \bar c \gamma^\mu \gamma_5 b |\bar B(p_B, M_B)\ket &=& 0  \label{b2d-ff-av}\\*
\label{a:e:sf}
\bra D(p_D, M_D)| \bar c b |\bar B(p_B, M_B)\ket &=& F_0(q^2) \frac{M_B^2 -  
M_D^2}{m_b - m_c}  \\*
\bra D(p_D, M_D)| \bar c \gamma_5 b |\bar B(p_B, M_B)\ket &=& 0 \label{b2d-ff-ps} \\*
\label{a:e:tf1}
{\bra D(p_D, M_D)| \bar c \sigma^{\mu\nu}  b |\bar B(p_B, M_B)\ket} &=&  
-i (p_B^\mu p_D^\nu - p_B^\nu p_D^\mu) \frac{2F_T(q^2)}{ M_B + M_D}  \\*
\label{a:e:tf2}
{\bra D(p_D, M_D)| \bar c \sigma^{\mu\nu} \gamma_5 b |\bar B(p_B, M_B)\ket} 
&=& \varepsilon^{\mu\nu\rho\sigma}  
p_{B\rho} p_{D\sigma} \frac{2F_T(q^2)}{ M_B + M_D} 
\eea

Note that \Eqn{a:e:sf} and \Eqn{a:e:tf2} are not  independent equations and follow from \Eqn{a:e:vf} and  \Eqn{a:e:tf1} respectively. 
Multiplying the left hand side of \Eqn{a:e:vf} by $q_\mu$ one gets
\bea
q_\mu \bra D(p_D, m_D)| \bar c \gamma^\mu b |\bar B(p_B, M_B)\ket
&=& \mbox{Inverse Fourier transform of } \bra D| i \partial_\mu (\bar c
\gamma^\mu b) |B \ket \nonumber\\
&=& \mbox{Inverse Fourier transform of } 
\bra D|  (i \partial_\mu\bar c \gamma^\mu b + \bar c \gamma^\mu i\partial_\mu
b) |B \ket\nonumber\\
&=&  (m_b-m_c) \bra D(p_D, M_D)| \bar c b
|\bar B(p_B, M_B)\ket 
\label{a:e:vsf}
\eea
Similarly, the term proportional to $F_+$ in the right hand side of \Eqn{a:e:vf} vanishes 
upon multiplication by $q_\mu$ and gives
\bea
{\rm RHS} = F_0(q^2) (M_B^2- M_D^2).
\label{a:e:vfrhs}
\eea 
Thus, \Eqn{a:e:vsf} and \Eqn{a:e:vfrhs} taken together give us \Eqn{a:e:sf}.

In order to get \Eqn{a:e:tf2} from \Eqn{a:e:tf1} one has to use the identity,
\bea
\sigma^{\mu\nu} \gamma_5=\frac{i}{2}\varepsilon^{\mu\nu\alpha\beta}\sigma_{\alpha\beta} \,. 
\eea 
Substituting the above identity into the  left hand side of \Eqn{a:e:tf2} one gets,
\bea
{\bra D(p_D, M_D)| \bar c \sigma^{\mu\nu} \gamma_5 b |\bar B(p_B, M_B)\ket} 
&=& \frac{i}{2}\varepsilon^{\mu\nu\alpha\beta} {\bra D(p_D, M_D)| \bar c \sigma_{\alpha\beta} b |\bar B(p_B, M_B)\ket} \\ 
&=& \frac{i}{2}\varepsilon^{\mu\nu\alpha\beta} \left( -i (p_{B\alpha} p_{D\beta} - p_{B\beta} p_{D\alpha}) \frac{2F_T(q^2)}{ M_B + M_D}   \right) \\
&=& \varepsilon^{\mu\nu\alpha\beta}  p_{B\alpha} p_{D\beta} \frac{2F_T(q^2)}{ M_B + M_D} 
\eea

The form factors $F_0(q^2)$ and $F_+(q^2)$ have been calculated using lattice QCD techniques in 
\cite{Lattice:2015rga}\footnote{There has been another Lattice calculation of these form factors with similar results \cite{Na:2015kha}.}. They are 
given by the following expressions, 
\bea
F_+(z) &=& \frac{1}{\phi_+(z)} \sum_{k=0}^3 a_k^+ \, z^k \, ,  \\
F_0(z) &=& \frac{1}{\phi_0(z)} \sum_{k=0}^3 a_k^0 \, z^k \, ,
\eea
where 
\begin{align*}
z \equiv  z(q^2)  = \frac{\sqrt{(M_B + M_D)^2 - q^2} - \sqrt{4 M_B M_D}}{\sqrt{(M_B + M_D)^2 - q^2} + \sqrt{4 M_B M_D}} \, . 
\end{align*}

The functions $\phi_+(z)$ and $\phi_0(z)$ are given by, 
\bea
\phi_+(z) &=& 1.1213  \frac{(1+z)^2 (1-z)^{1/2}}{\left[(1+r)(1-z) + 2 \sqrt{r}(1+z)\right]^5} \, , \\
\phi_0(z) &=& 0.5299 \frac{(1+z)(1 - z)^{3/2}}{\left[(1+r)(1-z) + 2 \sqrt{r} (1+z)\right]^4} \, , 
\eea
where, $r = M_D/M_B$.

The central values, uncertainties, and correlation matrix for the parameters $a_k^0$ and $a_k^+$ are shown in 
tables~\ref{B2D-FF-numbers} and \ref{B2D-FF-corr}. 

\begin{figure}[t]
\begin{center}
\begin{tabular}{c}
\includegraphics[scale=0.55]{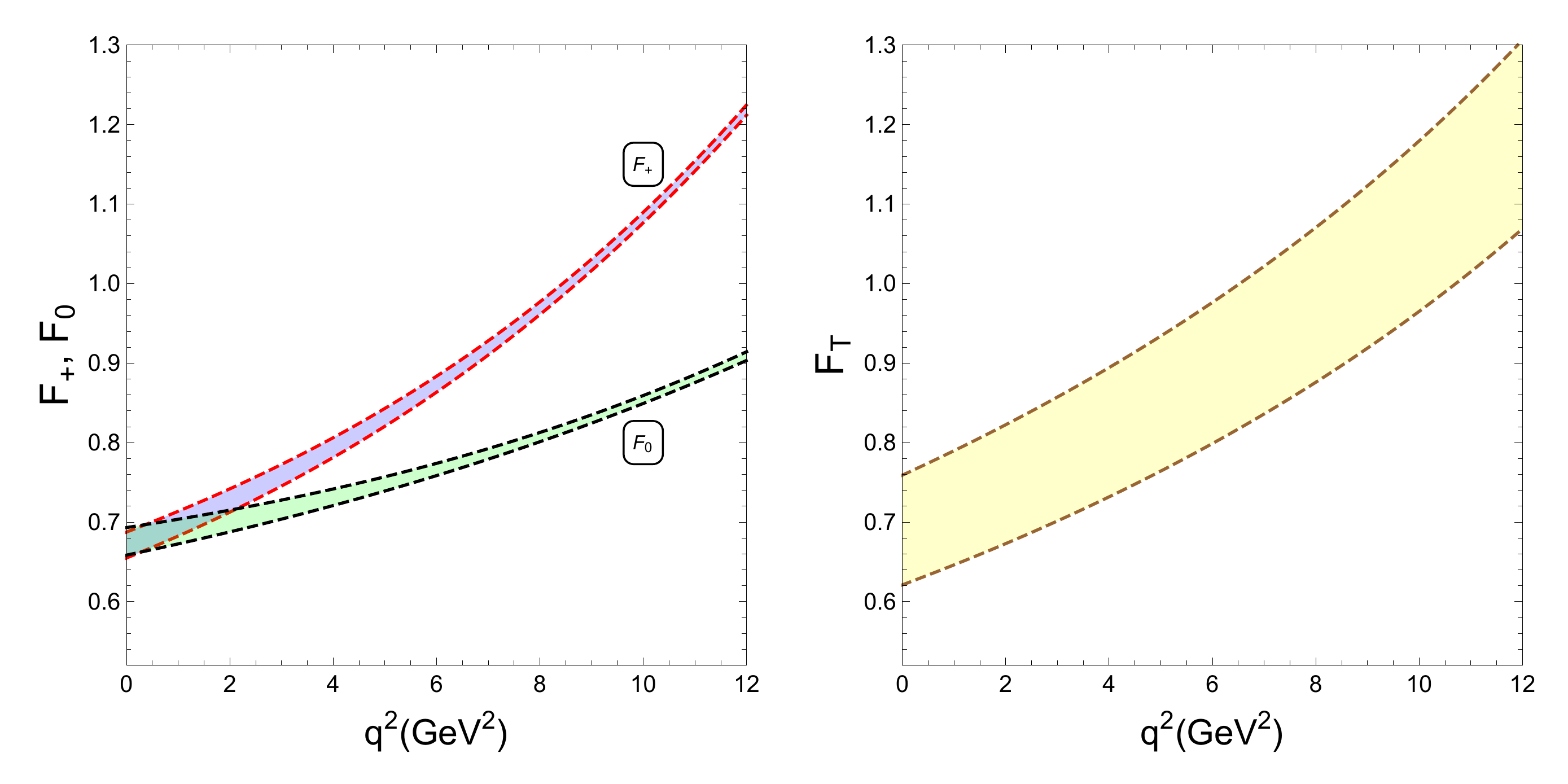}
\end{tabular}
\caption{The $q^2$ dependence of the form factors $F_0$, $F_+$ and $F_T$. The uncertainty bands for $F_0$ and $F_+$ correspond to 
a $\chi^2 \leq 1.646$ where the $\chi^2$ is computed using the expression  
$\chi^2({\bf x}) = \left({\bf x - x_0}\right)^T {\bf V}^{-1} \left({\bf x - x_0}\right)$
where ${\bf x} = (a_0^+, a_1^+, a_2^+, a_3^+,a_0^0,a_1^0,a_2^0,a_3^0 )$ and $x_0$ consists of the central values given in 
table~\ref{B2D-FF-numbers}.  The covariance matrix $V$ is computed from the correlation matrix $\rho_{ij}$ given in 
table~\ref{B2D-FF-corr} using the formula $V_{ij} = \sigma_i({\bf x}) \rho_{ij} \sigma_j({\bf x})$
where $\sigma({\bf x})$ is the vector of uncertainties given  in tables~\ref{B2D-FF-numbers}.
The uncertainty band for $F_T$ is obtained by simply taking a 
$\pm 10\%$ uncertainty on the central value.}
\label{fig:B2D-FF-fig}
\end{center}
\end{figure}

\begin{table}[h!]
\centering
\begin{tabular}{|l|cccc|cccc|}
\hline
&$a_0^+$& $a_1^+$& $a_2^+$& $a_3^+$& $a_0^0$& $a_1^0$& $a_2^0$& $a_3^0$ \\
\hline
Values & 0.01261 & -0.0963 & 0.37& -0.05 & 0.01140 & -0.0590 & 0.19 & -0.03 \\
Uncertainties & 0.00010 & 0.0033 & 0.11 & 0.90 & 0.00009 & 0.0028 & 0.10 & 0.87 \\
\hline
\end{tabular}
\caption{The central values and uncertainties for the parameters $a_k^0$ and $a_k^+$ from ref.~\cite{Lattice:2015rga} 
(table XI of their arXiv version 1).\label{B2D-FF-numbers}}
\label{as}
\end{table}

\begin{table}[h!]
\centering
\begin{tabular}{r|rrrrrrrr}
\hline\hline
& $a^+_0$    & $a^+_1$    & $a^+_2 $ & $a^+_3$ &  $a^0_0$  & $a^0_1$    & $a^0_2$    & $a^0_3$ \\
 \hline
 $a^+_0$ & $  1.00000 $ & $ 0.24419 $ & $  -0.08658 $ & $  0.01207 $ & 0.00000 & $  0.23370 $ & $  0.03838 $ & $ -0.05639 $ \\
 $a^+_1$ & $                $ & $ 1.00000 $ & $ -0.57339 $ & $  0.25749 $ & 0.00000 & $  0.80558 $ & $ -0.25493 $ & $ -0.15014 $ \\
 $a^+_2$ & $                $ & $               $ & $  1.00000 $ & $ -0.64492 $ & 0.00000 & $ -0.44966 $ & $  0.66213 $ & $  0.05120 $ \\
 $a^+_3$ & $                $ & $               $ & $                $ & $  1.00000 $ & 0.00000 & $ 0.11311 $ & $ -0.20100 $ & $  0.23714 $ \\
 $a^0_0$ & $                $ & $               $ & $                $ & $                $ & 1.00000  & $ 0.00000 $ & $  0.00000 $ & $  0.00000 $ \\
 $a^0_1$ & $                $ & $               $ & $                $ & $                $ &               & $  1.00000 $ & $ -0.44352 $ & $ 0.02485 $ \\
 $a^0_2$ & $                $ & $               $ & $                $ & $                $ &               & $                $ & $  1.00000 $ & $ -0.46248 $ \\
 $a^0_3$ & $                $ & $               $ & $                $ & $                $ &               & $                $ & $                $ & $  1.00000 $ \\
\hline
\hline
\end{tabular}
\caption{The correlation matrix for the parameters $a_k^0$ and $a_k^+$ from ref.~\cite{Lattice:2015rga} 
(table XI of their arXiv version 1).
\label{B2D-FF-corr}}
\end{table}

As the tensor form factor $F_T$ has not been computed from lattice QCD,  we have taken them from \cite{Melikhov:2000yu}. 
Following \cite{Melikhov:2000yu}, we write $F_T(q^2)$ as,  
\bea
F_T(q^2) &=& \frac{0.69}{\left(1- \frac{q^2}{(6.4 {\rm GeV})^2} \right) \left(1 - 0.56 \frac{q^2}{(6.4 {\rm GeV})^2}\right)} \, .
\eea
In fig.~\ref{fig:B2D-FF-fig}, we show the $q^2$ dependences of $F_0$, $F_+$ and $F_T$ following the above expressions.

\section{$\bar{B} \to D^*$ form factors}
\label{ff-2}

The hadronic matrix elements for $\bar{B} \to D^*$ transition are parametrised by
\bea
\bra D^\ast(p_{D^\ast}, M_{D^\ast})| \bar c \gamma_\mu b |\bar B(p_B, M_B)\ket &=&  
i \varepsilon_{\mu\nu\rho\sigma} \epsilon^{ \nu \ast} p_B^\rho p_{D^\ast}^\sigma  \, \frac{2 V(q^2)}{M_B+M_{D^\ast}} \\[3mm]
\bra D^\ast(p_{D^\ast}, M_{D^\ast})| \bar c \gamma_\mu \gamma_5 b |\bar B(p_B,
M_B)\ket &=&  2 M_{D^\ast}  \frac{\epsilon^\ast.q }{q^2 } q_\mu A_0(q^2) +
(M_B+M_{D^\ast})  \Big[\epsilon_{\mu}^\ast - \frac{\epsilon^\ast.q}{q^2 } q_\mu   \Big] A_1(q^2) \nonumber\\*
&& \hspace*{-2mm} - \frac{\epsilon^\ast.q}{M_B+M_{D^\ast}} \Big[(p_B + p_{D^\ast})_\mu  -\frac{M_B^2-M_{D^\ast}^2}{q^2} q_\mu  \Big] A_2(q^2) \\[3mm]
\bra D^\ast(p_{D^\ast}, M_{D^\ast})| \bar c b |\bar B(p_B, M_B)\ket &=&  0    \label{rds-ff-s} \\[3mm]
\bra D^\ast(p_{D^\ast}, M_{D^\ast})| \bar c \gamma_5 b |\bar B(p_B, M_B)\ket &=&  
- \epsilon^\ast.q \, \frac{2 M_{D^\ast}}{m_b + m_c}  \, A_0(q^2)   \\[3mm]
\bra D^\ast(p_{D^\ast}, M_{D^\ast})| \bar c\sigma_{\mu\nu}  b |\bar B(p_B,
M_B)\ket &=&    -\varepsilon_{\mu \nu \alpha\beta } 
\Big[-\epsilon^{\alpha\ast}(p_{D^\ast} + p_B)^\beta T_1(q^2) \nonumber\\*
&& + \frac{M_B^2-M_{D^\ast}^2}{q^2}\epsilon^{\ast\alpha} q^\beta \left(T_1(q^2)
- T_2(q^2)\right) \\*
 && +  2 \frac{\epsilon^\ast.q }{q^2 }   p_B^\alpha p_{D^\ast}^\beta 
\left(T_1(q^2) -T_2(q^2) - \frac{q^2}{M_B^2-M_{D^\ast}^2} T_3(q^2)\right) \Big] \nonumber \\[3mm]
\bra D^\ast(p_{D^\ast}, M_{D^\ast})| \bar c\sigma_{\mu\nu} q^\nu b |\bar B(p_B,M_B)\ket &=& 
- 2 \varepsilon_{\mu \nu \rho \sigma} \epsilon^{\ast \nu} p_{B}^\rho p_{D^\ast}^\sigma T_1(q^2)
\eea

None of the form factors $V, A_0, A_1, A_2, T_1, T_2, T_3$ has been calculated in Lattice QCD. We used the heavy quark effective 
theory (HQET) form factors based on \cite{Caprini:1997mu}.
These form factors can be written in terms of the HQET form factors in the following way 
\cite{Caprini:1997mu,Sakaki:2013bfa},

\begin{equation}
   \begin{split}
      V(q^2) =& { M_B + M_\Dst \over 2\sqrt{M_B M_\Dst} } \, h_V(w(q^2)) \,, \\
      A_1(q^2) =& { ( M_B + M_\Dst )^2 - q^2 \over 2\sqrt{M_B M_\Dst} ( M_B + M_\Dst ) } \, h_{A_1}(w(q^2)) \\
      A_2(q^2) =& { M_B+M_\Dst \over 2\sqrt{M_B M_\Dst} } \left[ h_{A_3}(w(q^2)) + { M_\Dst \over M_B } h_{A_2}(w(q^2)) \right] \\
      A_0(q^2) =& { 1 \over 2\sqrt{M_B M_\Dst} } \left[ { ( M_B + M_\Dst )^2 - q^2 \over 2M_\Dst } \, h_{A_1}(w(q^2)) \right. \\
                & -\left. { M_B^2 - M_\Dst^2 + q^2 \over 2M_B } \, h_{A_2}(w(q^2)) - { M_B^2 - M_\Dst^2 - q^2 \over 2M_\Dst } \, h_{A_3}(w(q^2)) \right] \\
                      T_1(q^2) =& { 1 \over 2\sqrt{M_B M_\Dst} } \left[ ( M_B + M_\Dst ) h_{T_1}(w(q^2)) - ( M_B - M_\Dst ) h_{T_2}(w(q^2)) \right] 
   \end{split}
\end{equation}

\begin{equation}
   \begin{split}
      T_2(q^2) =& { 1 \over 2\sqrt{M_B M_\Dst} } \left[ { ( M_B + M_\Dst )^2 - q^2 \over M_B + M_\Dst } \, h_{T_1}(w(q^2)) \right. \\
                & \quad\quad\quad\quad\quad\quad \left. - { ( M_B - M_\Dst )^2 - q^2 \over M_B - M_\Dst } \, h_{T_2}(w(q^2)) \right]  \\
      T_3(q^2) =&  { 1 \over 2\sqrt{M_B M_\Dst} } \left[ ( M_B - M_\Dst ) h_{T_1}(w(q^2)) - ( M_B + M_\Dst ) h_{T_2}(w(q^2)) \right. \\
                & \quad\quad\quad\quad\quad\quad \left.- 2 { M_B^2 -M_\Dst^2 \over M_B } h_{T_3}(w(q^2)) \right] \,,
   \end{split} \nn 
\end{equation}
where, 
\begin{align}
      \begin{split}
         h_V(w) =& R_1(w) h_{A_1}(w) \\
         h_{A_2}(w) =& { R_2(w)-R_3(w) \over 2\,r_\Dst } h_{A_1}(w) \\
         h_{A_3}(w) =& { R_2(w)+R_3(w) \over 2 } h_{A_1}(w) \\
         h_{T_1}(w) =& { 1 \over 2 ( 1 + r_\Dst^2 - 2r_\Dst w ) } \left[ { m_b - m_c \over M_B - M_\Dst } ( 1 - r_\Dst )^2 ( w + 1 ) \, h_{A_1}(w) \right. \\
                     & \quad\quad\quad\quad\quad\quad\quad\quad\quad \left. - { m_b + m_c \over M_B + M_\Dst } ( 1 + r_\Dst )^2 ( w - 1 ) \, h_V(w) \right]  \\
         h_{T_2}(w) =& { ( 1 - r_\Dst^2 ) ( w + 1 ) \over 2 ( 1 + r_\Dst^2 - 2r_\Dst w ) } \left[ { m_b - m_c \over M_B - M_\Dst } \, h_{A_1}(w) - 
         { m_b + m_c \over M_B + M_\Dst } \, h_V(w) \right] 
      \end{split}
   \end{align}
   
   \begin{align}
      \begin{split}
         h_{T_3}(w) =& -{ 1 \over 2 ( 1 + r_\Dst ) ( 1 + r_\Dst^2 - 2r_\Dst w ) } \left[2 { m_b - m_c \over M_B - M_\Dst } r_\Dst ( w + 1 ) \, h_{A_1}(w) \right. \\
                     & - { m_b - m_c \over M_B - M_\Dst } ( 1 + r_\Dst^2 - 2r_\Dst w ) ( h_{A_3}(w) - r_\Dst h_{A_2}(w) ) \\
                     & \left. - { m_b + m_c \over M_B + M_\Dst } ( 1 + r_\Dst )^2 \, h_V(w) \right] 
      \end{split}
   \end{align}

\begin{equation}
   \begin{split}
      h_{A_1}(w) =& h_{A_1}(1) [ 1 - 8\rho_\Dst^2 z + (53\rho_\Dst^2-15) z^2 - (231\rho_\Dst^2-91) z^3 ]  \\
      R_1(w) =& R_1(1) - 0.12(w-1) + 0.05(w-1)^2  \\
      R_2(w) =& R_2(1) + 0.11(w-1) - 0.06(w-1)^2  \\
      R_3(w) =& 1.22 - 0.052(w-1) + 0.026(w-1)^2 
      \label{eq:HQET_parametrization}
   \end{split}
\end{equation}
Here, $r_{D^\ast} = M_{D^\ast}/M_B$, $w(q^2)=(M_B^2+M_{D^{*}}^2-q^2)/2M_B M_{D^{*}}$ and 
$z(w) = ( \sqrt{w+1} - \sqrt2 ) / ( \sqrt{w+1} + \sqrt2 )$.

\begin{figure}[t!]
\begin{center}
\begin{tabular}{c}
\hspace*{-10mm}\includegraphics[scale=0.6]{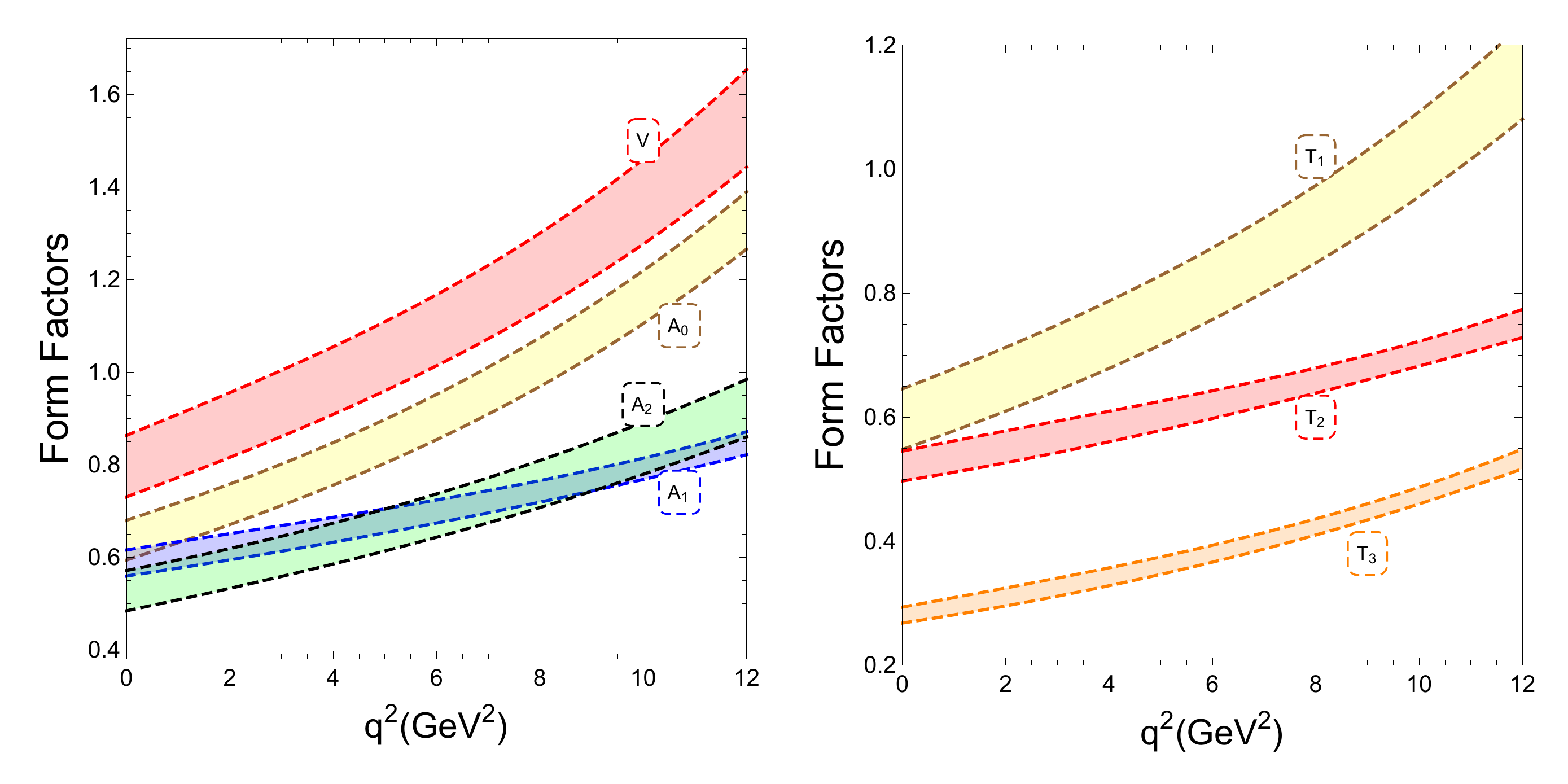}
\end{tabular}
\caption{The $q^2$ dependence of the $B \to D^*$ form factors. The bands correspond to two times the uncertainties given in 
Eq.~\ref{bds-ff-param} \label{fig:B2Dstar-FF-plot}.}
\end{center}
\end{figure}

The numerical values of the relevant parameters of the form factors along with their respective $1 \sigma$ errors
are given by
\bea
\label{bds-ff-param}
&&R_1(1) = 1.406 \pm 0.033,  ~ R_2(1) = 0.853 \pm 0.020,  ~ \rho_{D^*}^2 =1.207 \pm  0.026  \text{\cite{Amhis:2014hma}} \nonumber \\
&& \hspace{4cm}   h_{A_1}(1) = 0.906 \pm 0.013 \text{ \cite{Bailey:2014tva}} \, .
\eea
In Fig.~\ref{fig:B2Dstar-FF-plot} we show the $q^2$ dependence of the form factors using these numerical values. 
As there have been no lattice calculations of these form factors, in order to be conservative, we use two times larger uncertainties 
than those quoted above.


\section{Expressions for $a_\ell^D$, $b_\ell^D$ and $c_\ell^D$ for $\overline{B} \to D \ell \bar{\nu}_\ell$}
\label{b2d-formulae}

The quantities $a_\ell^D$, $b_\ell^D$ and $c_\ell^D$ for positive helicity lepton are given by:


\bea
a^D_\ell(+) &=&\frac{ 2\left(M_B^2-M_D^2\right)^2}{\left(m_b-m_c\right){}^2}  \, {\bf {\CSLSq} {F_0^2}}  \nonumber\\
&& + m_\ell \left[\frac{4(M_B^2-M_D^2)^2}{q^2 \left(m_b -m_c \right)}  \re \left( {\rm \bf C_{\bf VL}^\ell C_{SL}^{\ell *}} \right)  {\bf F_0^2 } \right] \nonumber\\
&& + m_\ell^2 \left[\frac{2\left(M_B^2-M_D^2\right)^2}{q^4} \, {\bf \CVLSq F_0^2} \right] \label{alD+}\\
b^D_\ell(+) &=& - m_\ell \left[ \frac{8{\PDd} M_B \left(M_B^2-M_D^2\right)  }{q^2 \left(m_b-m_c\right)} 
\re \left({\bf C_{\rm \bf SL}^\ell C_{\rm\bf VL}^{\ell *}}\right) {\bf F_0} {\bf F_+} \right] \nonumber \\
&&- m_\ell^2 \left[ \frac{ 8 {\PDd} M_B
\left(M_B^2-M_D^2\right)}{q^4}  { \CVLSq} {\bf F_0} {\bf 
F_+} \right]  \label{blD+} \\
c^D_\ell (+) &=&
m_\ell^2 \left[\frac{8{\PDd}^2 M_B^2}{q^4}{ \CVLSq} {\bf F_+^2}   \right]  \label{clD+}
\eea

Their expressions for the negative helicity lepton are,
\bea
a^D_\ell (-) &=& \frac{ 8 M_B^2 {\PDSQ}  }{q^2}  \,  {\bf \CVLSq} {\bf F_+^2} \label{alD-} \\ 
b^D_\ell (-) &=& 0 \\
c^D_\ell (-) &=&  - \frac{8 M_B^2 {\PDd}^2}{q^2} {\CVLSq} {\bf F_+^2}  \label{clD-}
\eea

Note that, the WCs $C_{AL}^\ell$ and $C_{PL}^\ell$ do not contribute to this decay. This is because the corresponding 
QCD matrix elements vanish, as can be seen from eqs.~\eqref{b2d-ff-av} and \eqref{b2d-ff-ps}.

The lepton mass dependence of the various terms can also be understood easily. 
As the vector operators do not change the chirality of the fermion line, because of the left chiral nature of the neutrino, 
the outgoing (negatively charged) lepton also has negative chirality (and hence negative helicity in the massless limit).
Thus the production of a left-handed lepton through the vector operator does not need a mass insertion. 
By a similar argument, one can see that the production of a right-handed lepton through the scalar operator does not 
need any mass insertion. The amplitude for the production of a right-handed lepton through a vector operator, 
on the other hand, clearly requires a mass insertion in order to flip the lepton helicity. This explains why the terms 
proportional to $|C_{VL}^\ell |^2$ in Eqs.~\ref{alD+}-\ref{clD+} have $m_\ell^2$ and the interference terms 
proportional to $\re \left({ C_{\rm SL}^\ell C_{\rm VL}^{\ell *}}\right)$ have $m_\ell$ in front, while there is no such 
dependence in Eqs.~\ref{alD-}-\ref{clD-}.

The full expressions for $a_\ell^D$, $b_\ell^D$ and $c_\ell^D$ including all the operators in Eq.~\eqref{b2c-basis} are 
shown in appendix \ref{b2d-full}.

\section{Expressions for $a_\ell^{D^*}$, $b_\ell^{D^*}$ and $c_\ell^{D^*}$ for $\overline{B} \to D^\ast \ell \bar{\nu}_\ell$}
\label{b2ds-formulae}
The quantities $a_\ell^{D^*}$, $b_\ell^{D^*}$ and $c_\ell^{D^*}$ for positive and negative helicitiy leptons are given by,
\begin{eqnarray}
a_\ell^{D^\ast} (-) &=& \frac{8 M_B^2 \pds}{\mbmd^2} {\bf \cvlsq \vsq} + 
\frac{\mbmd^2 ( 8 M_{D^*}^2 q^2 + \lambda)}{2 M_{D^*}^2 q^2} {\bf \calsq \aosq} \nonumber \\
&&+\frac{8 M_B^4 |p_{D^\ast}|^4 }{\mds \mbmd ^2 q^2} {\bf \calsq \atsq} \nn\\
&&-  \frac{4\pds M_B^2 \mbmdq2 }{\mds q^2} {\bf \calsq  \left(A_1 A_2\right)}\\[4mm]
b_\ell^{D^\ast} (-) &=& 
-16 |p_{D^\ast}| M_B {\bf \re \left(\cvl \calconj \right)  \left(V A_1\right)}\\[4mm]
c_\ell^{D^\ast} (-) &=& \frac{8 \pds M_B^2}{\mbmd^2} {\bf{\cvlsq \vsq}} - 
\frac{\mbmd^2 \lambda}{2 \mds q^2} {\bf \calsq \aosq} \nn\\
&& - \frac{8 |p_{D^\ast}|^4 M_B^4}{\mbmd^2 \mds q^2} {\bf \calsq \atsq} \nn\\
&&+ \frac{4 \pds M_B^2 \mbmdq2}{\mds q^2} {\bf \calsq \left(A_1 A_2 \right)} 
\end{eqnarray}

\begin{eqnarray}
a_\ell^{D^\ast} (+) &=& \frac{8 \pds M_B^2}{\mbmc^2} {\bf \cplsq \azsq} \nn\\
&&- m_\ell \left[\frac{16 \pds M_B^2}{\mbmc q^2}{\bf \re \left(\call \cplconj \right)\azsq}\right] \nn\\ && + m_\ell^2 \left[\frac{8 \pds M_B^2}{q^4} {\bf \calsq \azsq} 
+\frac{8 \pds M_B^2}{\mbmd^2 q^2} {\bf \cvlsq \vsq} \right. \nn\\
&&\left. \quad + \frac{2 \mbmd^2}{q^2} {\bf \calsq \aosq} \right]
\end{eqnarray}
\begin{eqnarray}
b_\ell^{D^\ast} (+) &=& m_\ell \left[\frac{ 4 |p_{D^\ast}| M_B \mbmd\mbmdq2}{M_{D^\ast} \mbmc q^2} {\bf \re \left( \call \cplconj\right) A_0 A_1} \right.\nn\\
&&\left. - \frac{16}{\mbmc}\frac{|p_{D^\ast}|^3 M_B^3}{\mbmd M_{D^\ast} q^2} {\bf \re \left(\call \cplconj\right) A_0 A_2}\right] \nn\\
&& + m_\ell^2 \left[- \frac{4|p_{D^\ast}| M_B \mbmd}{M_{D^\ast} q^4} \mbmdq2 {\bf \calsq A_0 A_1} \right.\nn\\
&&\left. + \frac{16 |p_{D^\ast}|^3 M_B^3}{\mbmd M_{D^\ast} q^4}{\bf \calsq A_0 A_2} \right] \\
c_\ell^{D^\ast} (+) &=& m_\ell^2 \left[ - \frac{8 \pds M_B^2}{\mbmd^2 q^2} {\bf \cvlsq \vsq} +
\frac{\mbmd^2 \lambda}{2 \mds q^4} {\bf \calsq \aosq} \right. \nn\\
&& \left. + \frac{8 |p_{D^\ast}|^4 M_B^4}{\mds \mbmd^2 q^4} {\bf \calsq \atsq} \right.\nn\\ 
&& \left.- \frac{4 \pds M_B^2}{\mds q^4} \mbmdq2 {\bf \calsq \left(A_1 A_2 \right)}\right]
\end{eqnarray}


The WC $C_{SL}^\ell$ does not contribute to this decay because the corresponding QCD matrix element vanishes as can be seen 
from eq.~\eqref{rds-ff-s}. The lepton mass dependence of the various terms can be understood in the same way as the 
$\overline{B} \to D \ell \bar{\nu}_\ell$ decay. Note also the absence of interference terms proportional to 
$\re \left(C_{VL}^{\ell} \cplconj \right)$ in the above expressions. 

We provide the completely general result taking into account all the operators in Eq.~\eqref{b2c-basis} in appendix \ref{b2ds-full}.
\section{Results}
\label{results}

\subsection{Explaining $R_D$ alone}
\label{rd-alone}

\begin{figure}[h!]
\centering
\includegraphics[scale=0.6]{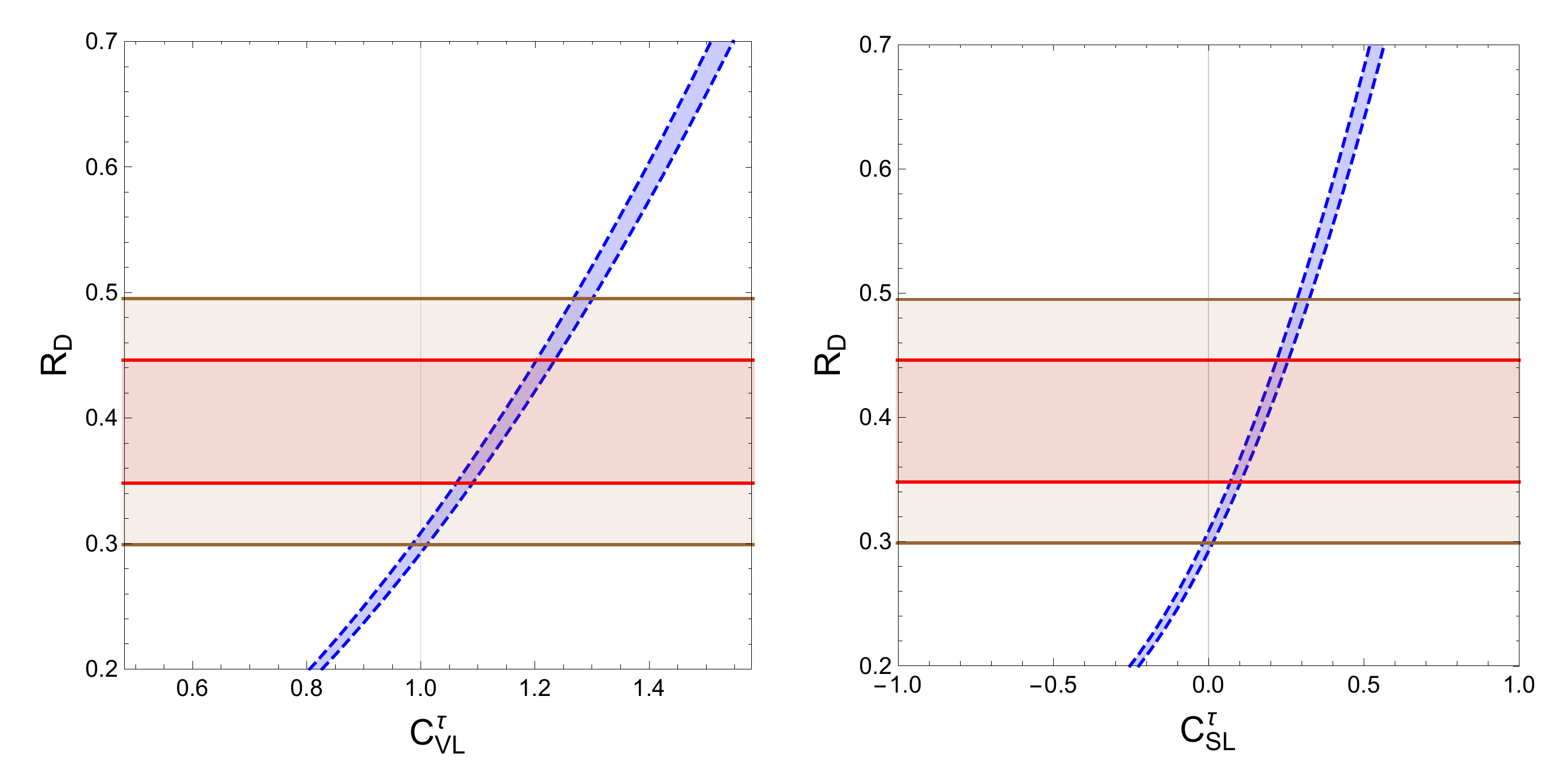}
\caption{The dependence of $R_D$ with respect to the variation of the WCs $C_{VL}^\tau$ (left) and 
$C_{SL}^\tau$ (right). \label{rd-vs-wc}}
\end{figure}

As mentioned in sec.~\ref{b2d-formulae}, the $\bdtaunu$  amplitude depends only on the WCs  $C_{VL}^\tau$ and $C_{SL}^\tau$. 
In Fig.~\ref{rd-vs-wc}, we show $R_D$ as function of $C_{VL}^\tau$ and $C_{SL}^\tau$. In the right plot, we set $C_{VL}^\tau$ to its 
SM value $C_{VL}^\tau |_{\rm SM} = 1$ and vary $C_{SL}^\tau$, while in the left plot, we hold $C_{SL}^\tau$ fixed at its SM value 
$C_{SL}^\tau |_{\rm SM} = 0$ and change $C_{VL}^\tau$. The red and brown shades correspond to the experimentally allowed 
$1 \sigma$ and $2 \sigma$ ranges (see Table~\ref{exp-values}), for which we have added the statistical and systematic 
uncertainties in quadrature.  

\begin{figure}[h!]
\begin{center}
\includegraphics[scale=0.6]{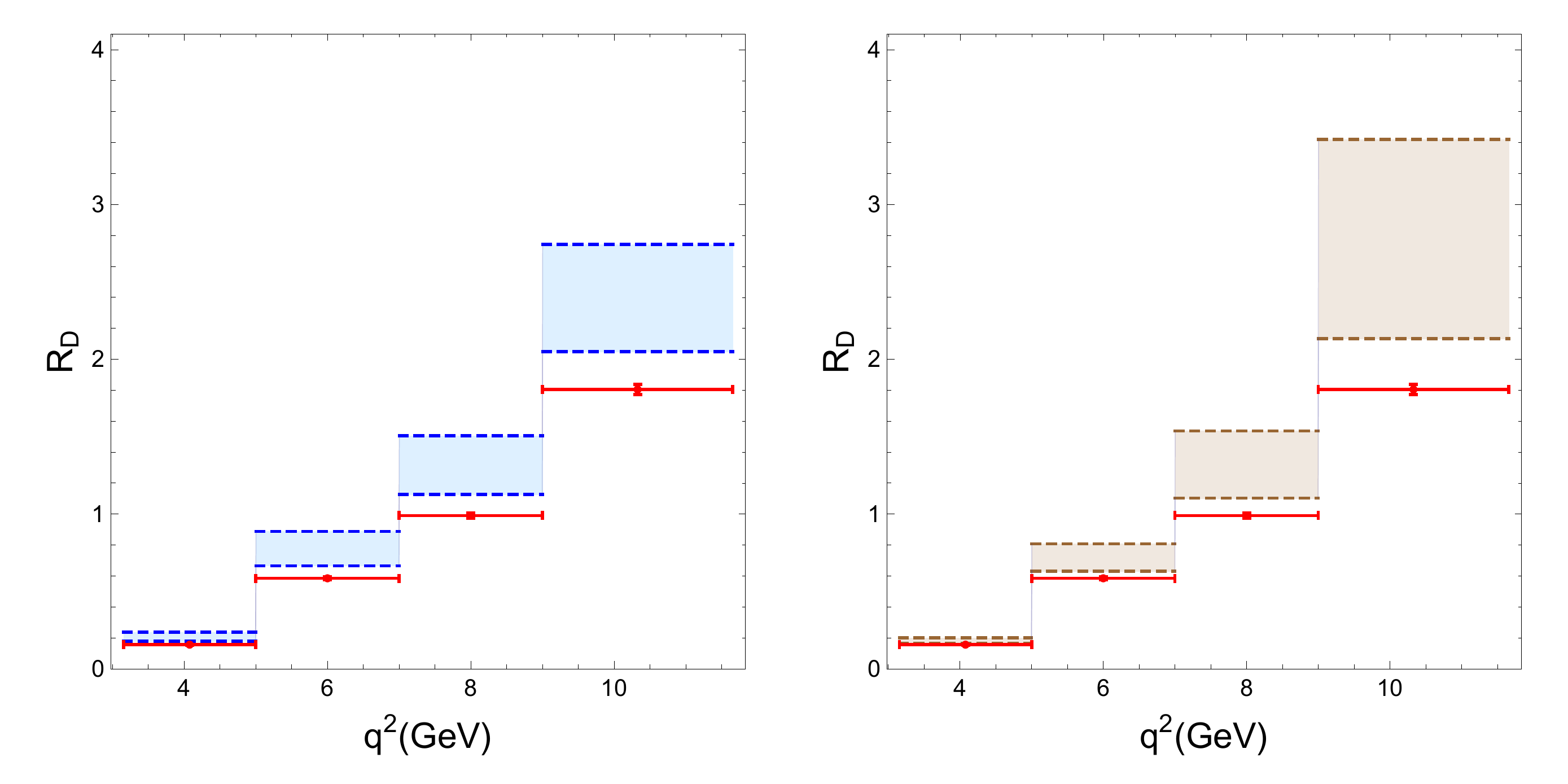}
\caption{The binwise $R_D$ for four $q^2$ bins. On the left, $C_{VL}^\tau$ is varied, while on the right, 
$C_{SL}^\tau$ is varied within their $1 \sigma$ allowed ranges.\label{fig:binned-rd}}
\end{center}
\end{figure}

The ranges of $C_{VL}^\tau$ and $C_{SL}^\tau$ that are consistent with $R_D$ at $1\sigma$ are shown in the second 
row of Table \ref{rdbin-table}. 
In the rows 3, 4 and 5-8, we also show the predictions for $P_\tau (D)$, ${\cal A}_{FB}^D$ and $R_D$ in four 
different bins for the allowed ranges of $C_{VL}^\tau$ and $C_{SL}^\tau$. 
Note that, ${\cal A}_{FB}^D$ and  the polarisation fraction $P_\tau(D)$ are independent of $C_{VL}^\tau$ if $C_{SL}^\tau$ is set to zero. This is because, 
in this case the differential decay rate is proportional to $|C_{VL}^\tau|^2$ and hence, the dependence on $C_{VL}^\tau$ drops out in 
$P_\tau(D)$ and ${\cal A}_{FB}^D$. This is why the ranges for $P_\tau(D)$ and ${\cal A}_{FB}^D$ in the third and fourth columns are identical.
The binwise $R_D$ values are also graphically represented in Fig.~\ref{fig:binned-rd}. The left and the right panels correspond 
to the WCs 
$C_{VL}^\tau$ and $C_{SL}^\tau$ respectively. The SM predictions are shown in red.  One can conclude from 
Fig.~\ref{fig:binned-rd} that the binwise $R_D$ does not help distinguish the two WCs $C_{VL}^\tau$ and $C_{SL}^\tau$.

\begin{table}[h!]
\begin{center}
\tabulinesep=1.2mm
\begin{tabu}{|c|l|c|c|c|}
\hline
              &                                                  &        \multirow{2}{*}{SM}          &          ${\bf C_{VL}}$             &       ${\bf C_{SL}}$             \\
              &                                                  &                                                &        (${\bf C_{SL}} = 0$)       &       (${\bf C_{VL}} =1$)     \\
\cline{2-5}              
              & \multirow{2}{*}{$1 \sigma $ range of the WC}                   &           &     \multirow{2}{*}{[1.073, 1.222]}     & \multirow{2}{*}{[0.067, 0.253]}            \\                     
              &                                                                           &           &                                                         &                \\ 

\cline{2-5}
              & $P_\tau (D)$      &                      [0.313, 0.336] &    [0.313, 0.336]    &   [0.388, 0.563]                                \\
\hline
              & \multirow{2}{*}{$\mathcal{A}_{FB}^D$}  &   \multirow{2}{*}{[$-0.361$, $-0.358$]}        &     \multirow{2}{*}{[$-0.361$, $-0.358$]}     &  \multirow{2}{*}{[-0.351, -0.318]}         \\                     
              &                                                                           &           &                                                         &            \\        \hline
\multirow{4}{*}{$R_D$ [bin] }
              & $[m_\tau^2-5]$ GeV$^2$         &                      [0.154, 0.158] &  [0.178, 0.236]       &  [0.164, 0.199] \\
              & $[5-7]$ GeV$^2$    &   [0.578, 0.593] & [0.665, 0.888] & [0.630, 0.808]     \\
              & $[7-9]$ GeV$^2$    &   [0.980, 1.003]  & [1.127, 1.505] & [1.102, 1.536]           \\
              & $[9-(M_B-M_D)^2]$ GeV$^2$  &  [1.776, 1.823]      & [2.049, 2.741]    &  [2.133, 3.420]        \\
\hline
\end{tabu}
\end{center}
\caption{The values of the WCs consistent with the $1 \sigma$ experimental range for $R_D$ are shown in the second row. The subsequent rows 
show the predictions for $P_\tau(D)$, $\mathcal{A}_{FB}^D$ and  $R_D$ in four $q^2$ bins for the WC ranges shown in the second row. \label{rdbin-table}}
\end{table}

\begin{figure}[h!]
\begin{center}
\begin{tabular}{cc}
\hspace*{-0.9cm}\includegraphics[scale=0.44]{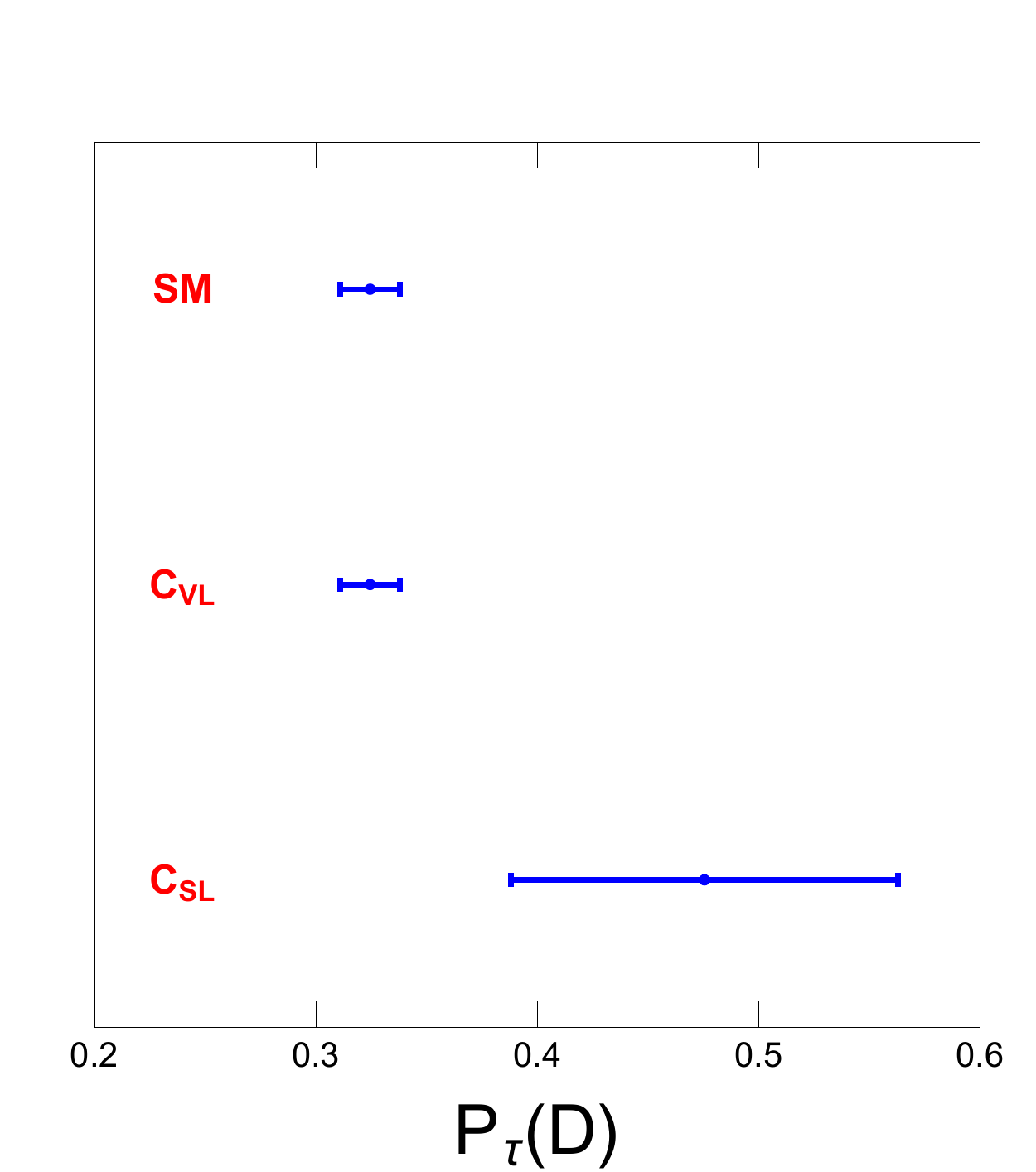}
\includegraphics[scale=0.45]{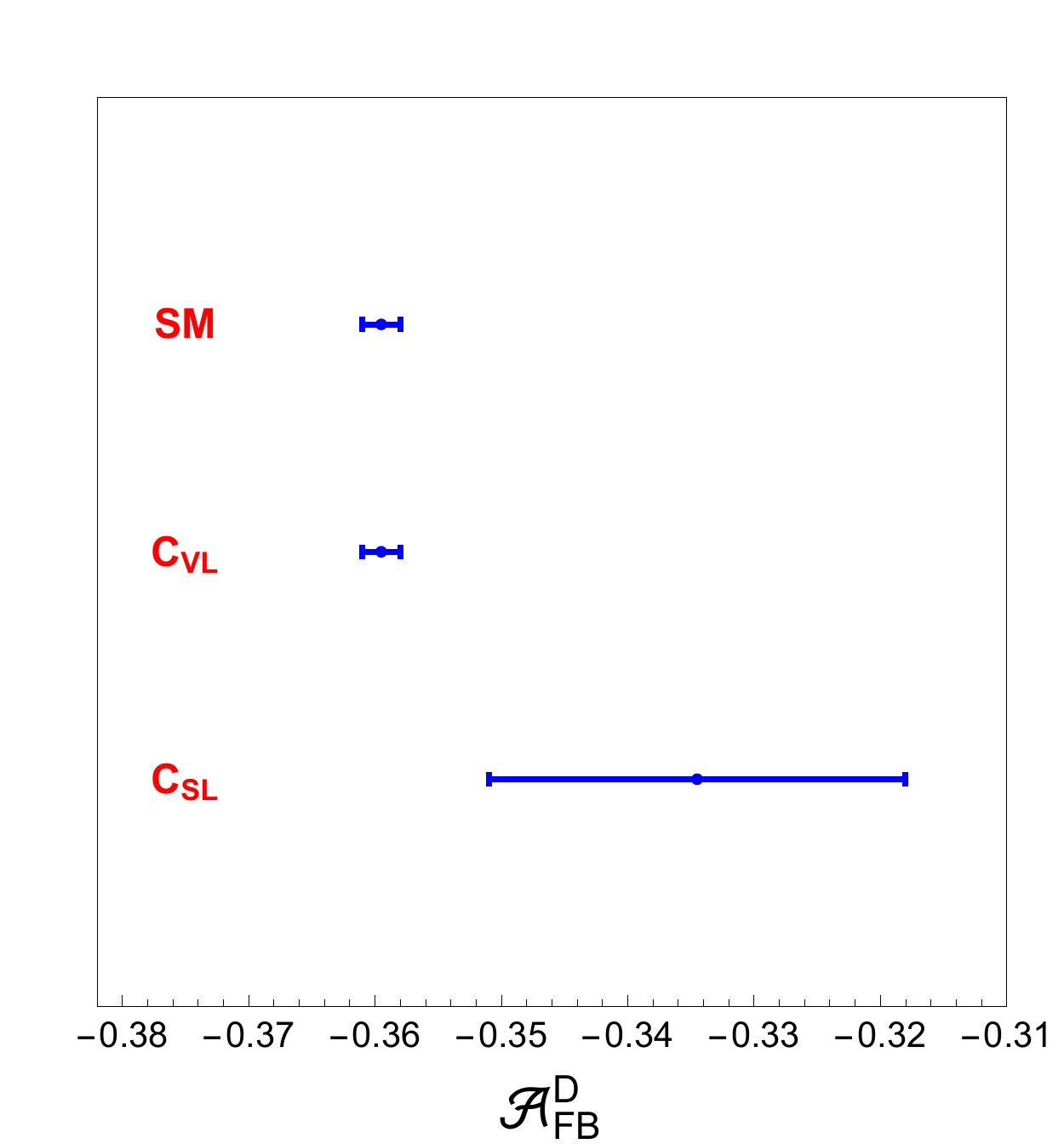}
\includegraphics[scale=0.45]{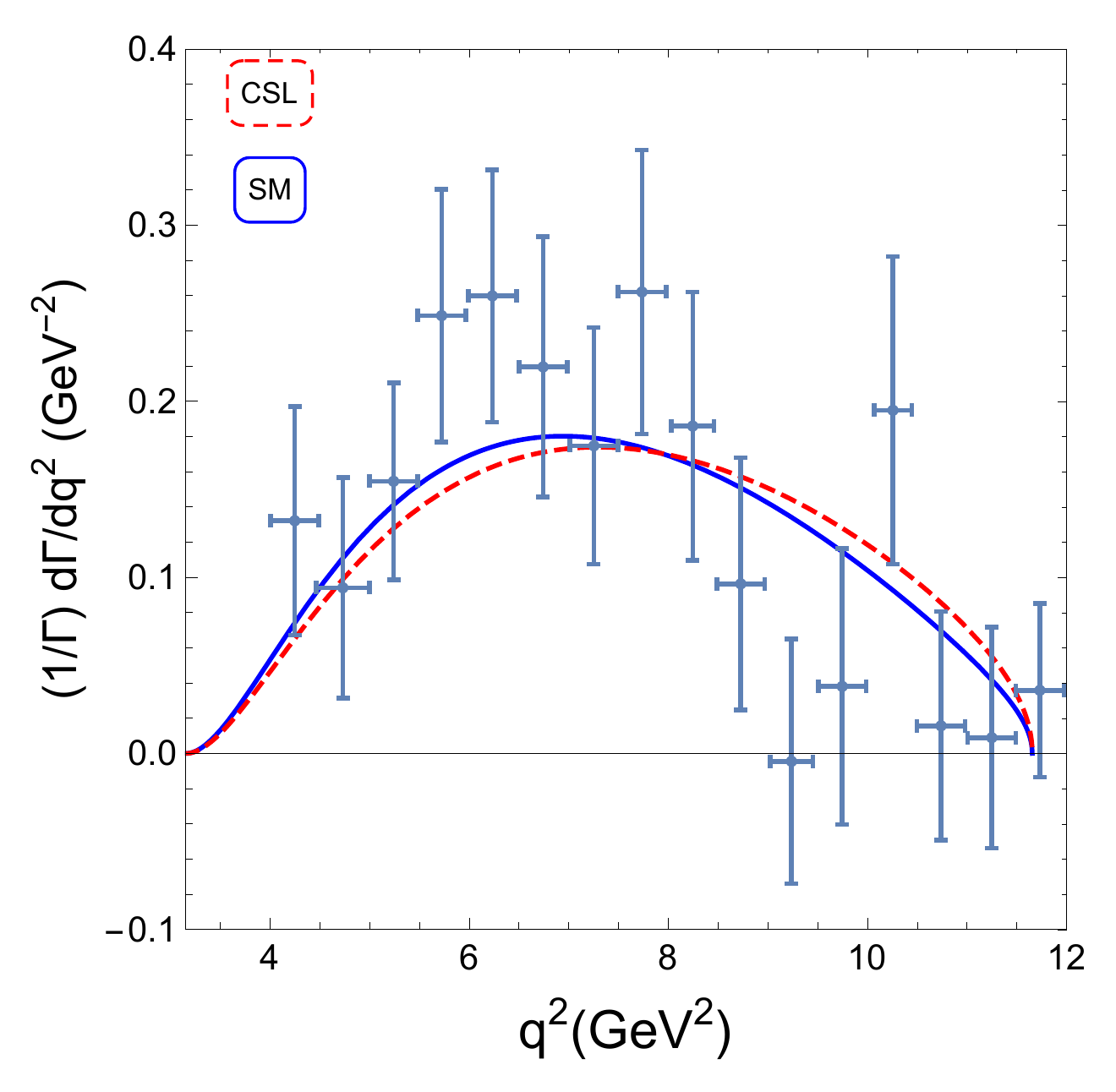}
\end{tabular}
\caption{Predictions for the polarisation fraction $P_\tau(D)$ (left), $\mathcal{A}_{FB}^D$ (middle) and the differential decay width (right). 
In the right graph showing the normalised differential decay width, the solid blue line is the SM prediction. The dashed red
line corresponds to  $C_{SL}^\tau = 0.16$. The data points shown on the right plot are due to the BaBar
collaboration and are taken from \cite{Lees:2013uzd}.
\label{fig:ptau-dbr}}
\end{center}
\end{figure}

The predictions for $P_\tau(D)$, $\mathcal{A}_{FB}^D$ are pictorially presented in the left and middle panel of Fig.~\ref{fig:ptau-dbr}. As mentioned earlier, 
in the absence of $C_{SL}^\tau$, $P_\tau(D)$ and $\mathcal{A}_{FB}^D$ are completely independent of $C_{VL}^\tau$. Hence, neither measurement
can distinguish between $C_{VL}^\tau = 1$ and other values of $C_{VL}^\tau$. However, the predictions 
are very different for $C_{SL}^\tau$. Therefore, a measurement of $P_\tau(D)$ will tell us whether NP in the form of scalar operator  ${\cal O}_{SL}^{cb\tau}$
exists or not. 

In the right panel of Fig.~\ref{fig:ptau-dbr}, we also show the normalised differential decay width as a function of $q^2$.  
As for the case of $P_\tau (D)$ and $\mathcal{A}_{FB}^D$ , the normalised differential decay width is independent of $C_{VL}^\tau$ for $C_{SL}^\tau =0$. 
The blue solid line is the SM prediction, and the red dashed line is the prediction for $C_{SL}^\tau = 0.16$. While producing these 
plots, we have used the central values of the form factors. The blue data points are from the BaBar measurement reported 
in \cite{Lees:2013uzd}. It is clear that the differential decay width is not a good discriminant 
of the various NP operators.

\subsection{Explaining $R_{D^\ast}$ alone}
\label{rds-alone}

\begin{figure}[!h]
\begin{center}
\begin{tabular}{l}
\hspace*{-1cm}\includegraphics[scale=0.55]{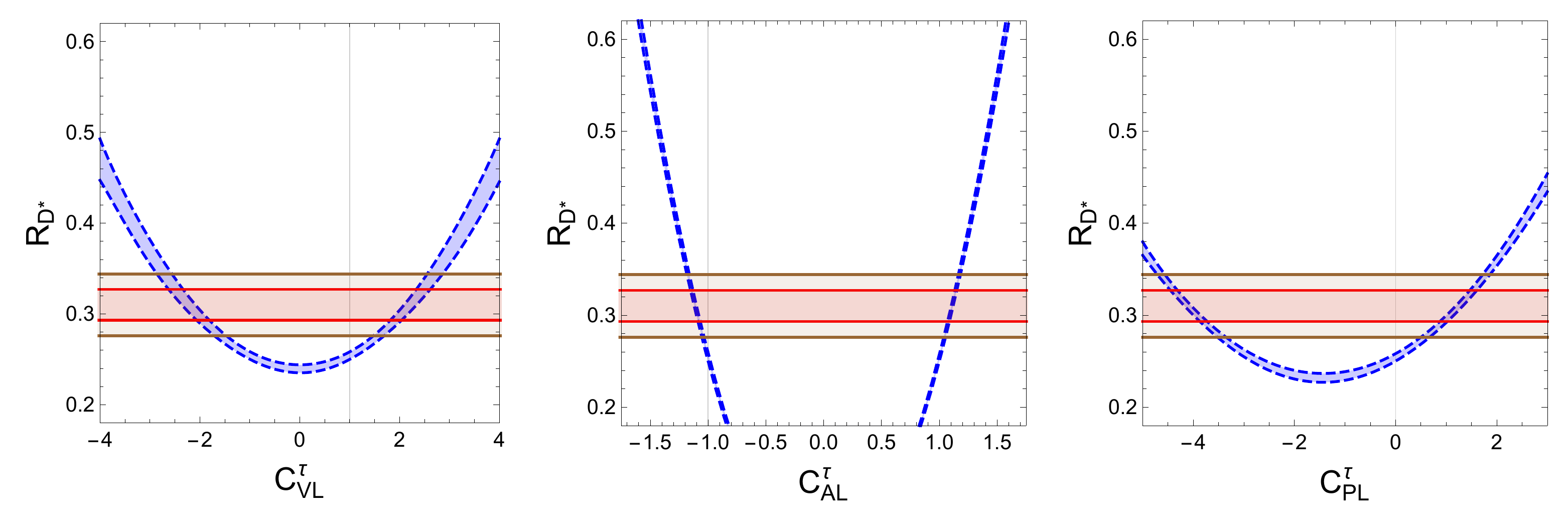} 
\end{tabular}
\caption[]{The dependence of $R_{D^*}$ with respect to the variation of the WCs  
$C_{VL}^\tau$ (left), $C_{AL}^\tau$ (middle) and $C_{PL}^\tau$ (right). A thin vertical line shows the SM values of the WCs.
\label{rds-vs-wc}}
\end{center} 
\end{figure}

The $\bdstaunu$ decay amplitude depends on three WCs, $C_{VL}^\tau, C_{AL}^\tau$ and  $C_{PL}^\tau$. 
In Fig.~\ref{rds-vs-wc}, we show $R_{D^*}$ as function of these WCs. 
In each of the plots, the WCs that are not varied are all set to their SM values. The red and brown shades 
correspond to the experimentally allowed $1 \sigma$ and $2 \sigma$ ranges respectively (see table~\ref{exp-values}). 

The ranges of $C_{VL}^\tau, C_{AL}^\tau$ and $C_{PL}^\tau$ that are consistent with the experimental value of $R_{D^*}$ at $1\sigma$ are shown 
in the second row of Table \ref{rdsbin-table}. We only show the ranges that are closest to the SM values of the WCs. 
In the rows 3, 4 and 5-8, we also show the predictions for $P_\tau (D^*)$, ${\cal A}_{FB}^{D^*}$ and $R_{D^*}$ in four different bins for these 
allowed ranges of $C_{VL}^\tau, C_{AL}^\tau$ and $C_{PL}^\tau$.

\begin{figure}[h!]
\begin{center}
\hspace*{-1cm}\includegraphics[scale=0.58]{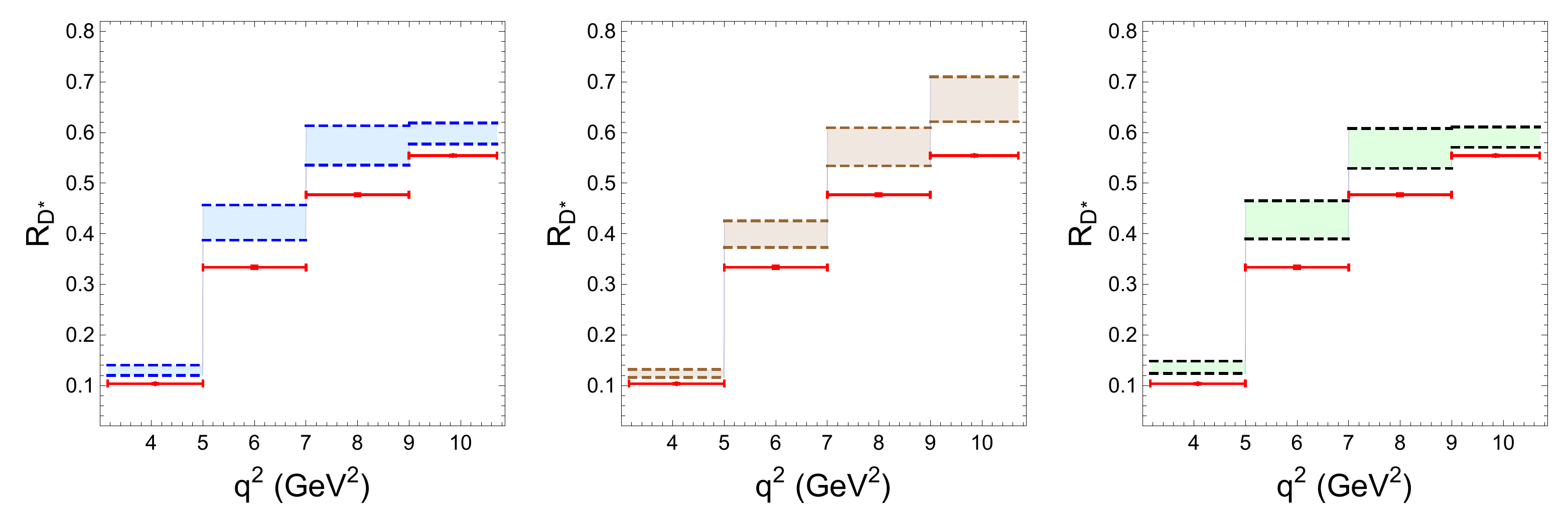}
\caption{The binwise $R_D^*$ for four $q^2$ bins. On the left, $C_{VL}^\tau$ is varied, in the middle $C_{AL}^\tau$ 
is varied, annd on the right, $C_{PL}^\tau$ is varied within their $1 \sigma$ allowed ranges. The SM predictions are shown in red. 
\label{fig:binned-rds}}
\end{center}
\end{figure}

The binwise $R_{D^*}$ values are also plotted in Fig.~\ref{fig:binned-rds}. The left, middle and the right panels correspond to the variation of WCs 
$C_{VL}^\tau, C_{AL}^\tau$ and $C_{SL}^\tau$ respectively. The $1 \sigma$ and $2 \sigma$ experimental values are shown in red and 
brown respectively.  It can be seen that $R_{D^*}$ in the 
last bin can be used to distinguish between $C_{VL}^\tau$(or $C_{PL}^\tau$) and  $C_{AL}^\tau$.


\begin{table}[h!]
\begin{center}
\tabulinesep=1.5mm
\hspace*{-1.4cm}\begin{tabu}{|c|l|c|c|c|c|}
\hline
              &                                                  &        \multirow{2}{*}{SM}                                     &          ${\bf C_{VL}}$                      &       ${\bf C_{AL}}$                &    ${\bf C_{PL}}$ \\
              &                                                  &            &          ${\bf C_{AL,PL}} = -1,0$      &       ${\bf C_{VL,PL}}=1,0$    &    ${\bf C_{VL,AL}}=1,-1$ \\
\cline{2-6}              
             & Range  in WC                 &      &    [1.856, 2.569]    & [$-1.149$, $-1.073$]    & [0.890, 1.583] \\ 
\cline{2-6}
               & $P_\tau (D^*)$                        &   [$-0.505$, $-0.490$] &    [$-0.530$, $-0.509$]    & [$-0.505$, $-0.488$]    & [$-0.322$, $-0.144$]  \\
\hline
               & $\mathcal{A}_{FB}^{D^\ast}$               &   [$0.050$, $0.078$]  &    [$0.191$, $0.297$]    & [$0.028$, $0.062$]    & [$-0.078$, $-0.007$]  \\
\hline
\multirow{4}{*}{$R_{D^{\ast}}$}
             & $[m_\tau^2-5]$ GeV$^2$         &   [0.103, 0.105] &    [0.120, 0.140]    & [0.116, 0.132]    & [0.124, 0.148] \\
             & $[5-7]$ GeV$^2$              &   [0.331, 0.336] &    [0.387, 0.457]    & [0.373,  0.425]    & [0.390, 0.465] \\
   $[\rm bin]$           & $[7-9]$ GeV$^2$     &   [0.475, 0.479] &    [0.535, 0.613]    & [0.535,  0.613]    & [0.534, 0.610] \\
              & $[9-(M_B-M_{D^\ast})^2]$ GeV$^2$  &   [0.554, 0.556] &    [0.577, 0.619]    & [0.621, 0.710]    & [0.571, 0.611] \\
\hline
\end{tabu}
\end{center}
\caption{The values of the WCs consistent with the $1 \sigma$ experimental range for $R_D^*$ are shown in the second row. 
We only show the ranges that are closest to the SM values of the WCs.  
The subsequent rows show the predictions for $P_\tau(D^\ast)$, $\mathcal{A}_{FB}^{D^\ast}$ and  $R_{D^\ast}$ in four $q^2$ bins for the WC ranges shown in the 
second row. \label{rdsbin-table}}
\end{table}

\begin{figure}[h!]
\begin{center}
\begin{tabular}{cc}
\hspace*{-8mm}\includegraphics[scale=0.44]{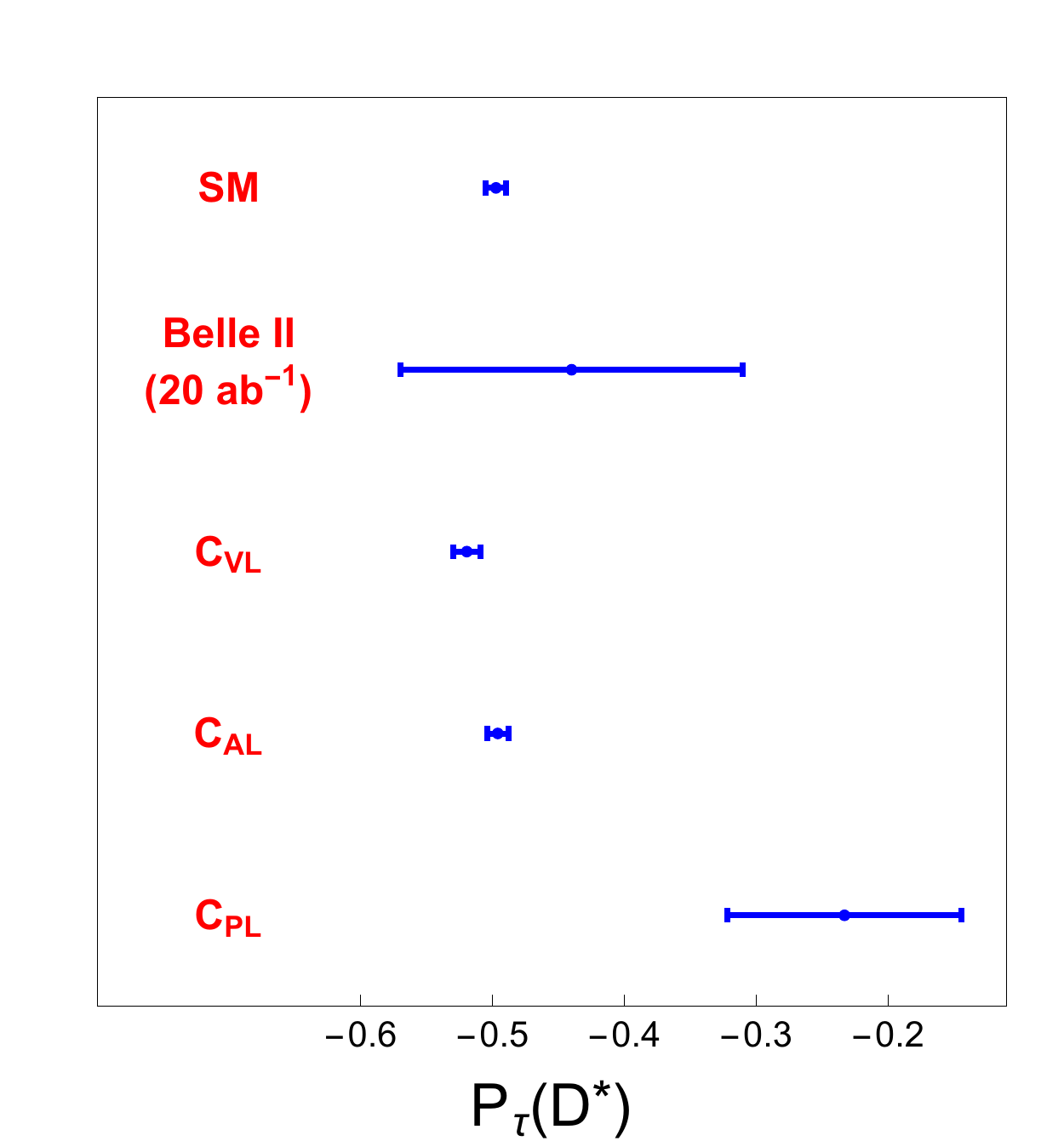}
\includegraphics[scale=0.44]{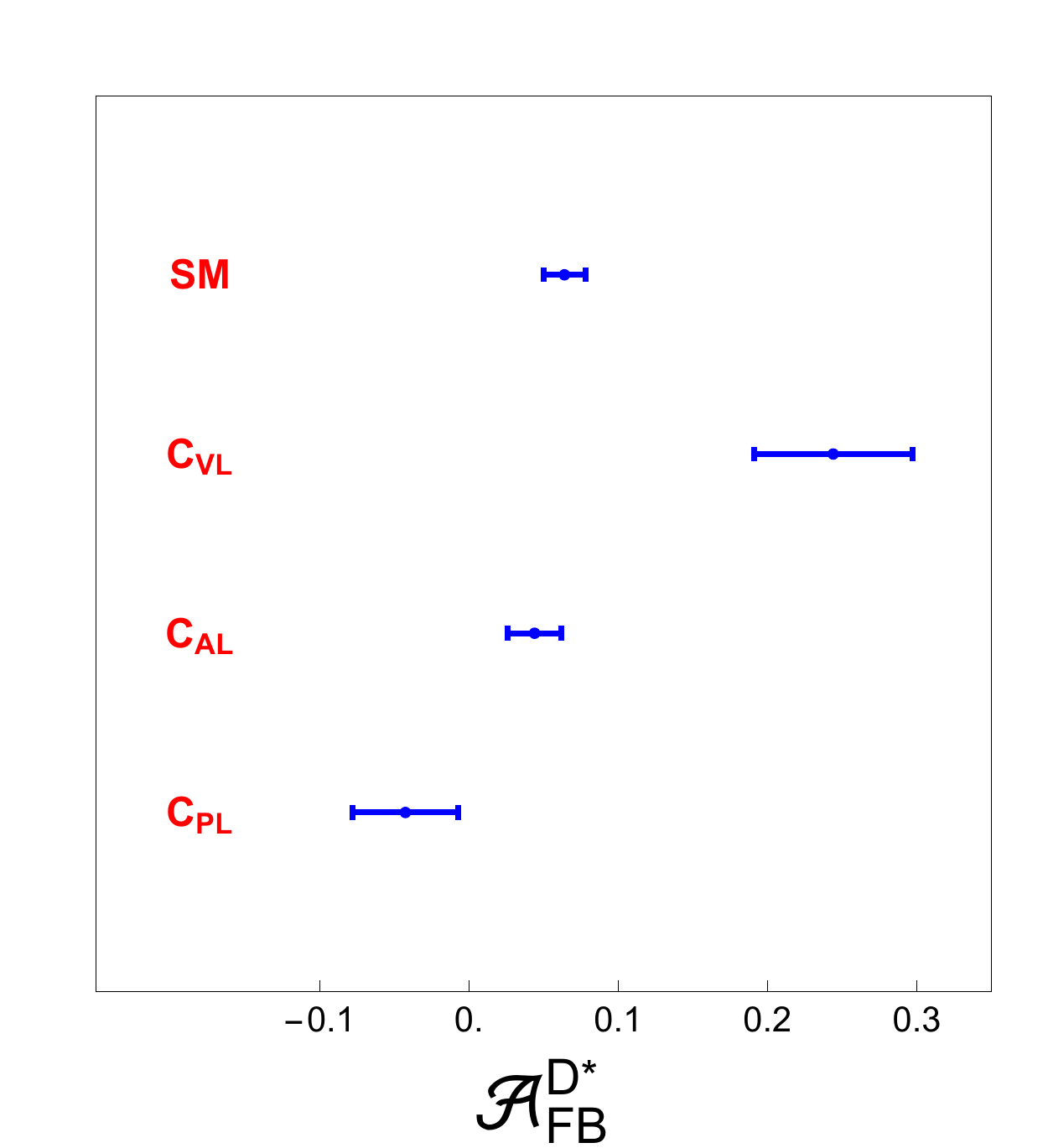}
\includegraphics[scale=0.45]{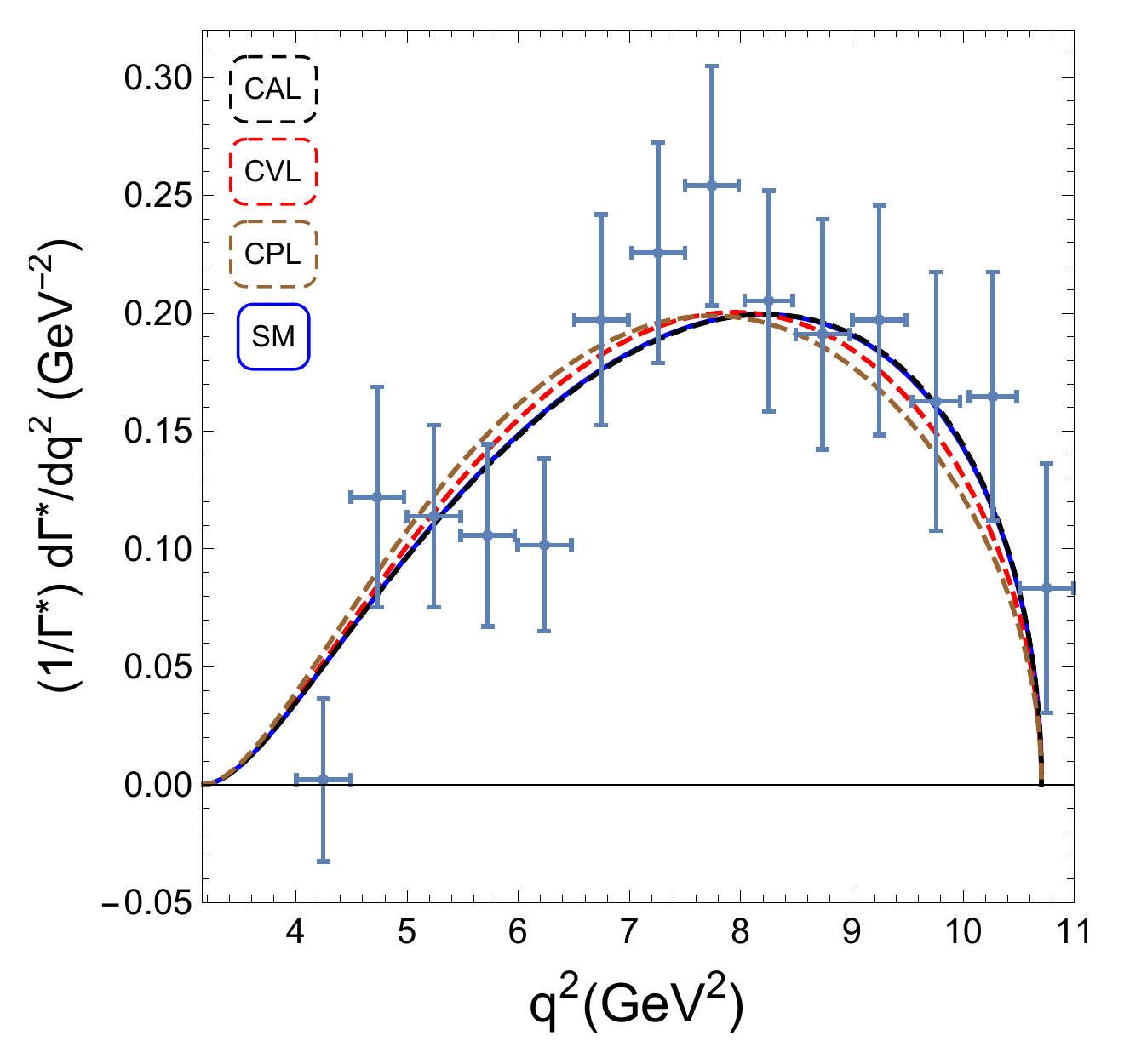}
\end{tabular}
\caption{Predictions for the polarisation fraction $P_\tau(D^*)$ (left), $\mathcal{A}_{FB}^{D^\ast}$ (middle)  and the differential decay width (right). In the left plot, the Belle II 20 ab$^{-1}$ 
projection is obtained by i) scaling down the statistical uncertainty of the recent Belle measurement by the ratio of the luminosities i.e., 
$\sqrt{20/0.71}$ ii) assuming the systematic uncertainty to go down by a factor of two, and adding them in quadrature. 
The central value is assumed to remain unchanged. On the right plot, The solid blue line is the SM prediction. 
The dashed black, red and  brown lines correspond to $C_{AL}^\tau = -1.12$, $C_{VL}^\tau = 1.9$ and $C_{PL}^\tau = 1.5$ respectively, where in each case every other WC is set to their SM values. Note that the black dashed curve is indistinguishable from the SM curve. The data is due to a BaBar measurement reported in \cite{Lees:2013uzd}.} 
\label{fig:ptau-dbr-ds}
\end{center}
\end{figure}

The predictions for $P_\tau(D^*)$ are pictorially presented in the left panel of Fig.~\ref{fig:ptau-dbr-ds}. We do not show the 
recent Belle measurements in this figure because the uncertainties are rather large. Instead, we show a projection 
for Belle II 20 ab$^{-1}$ (which is expected to be collected by the end of 2021 \cite{belle2}) assuming that the systematic uncertainty will 
go down by a factor of two compared to that in the recent Belle measurement. It is then possible to distinguish 
$C_{PL}^\tau$ from the other WCs. 
The middle panel of Fig.~\ref{fig:ptau-dbr-ds} shows the predictions of $\mathcal{A}_{FB}^{D^\ast}$ pictorially. It can be seen that 
a measurement of $\mathcal{A}_{FB}^{D^\ast}$ can also potentially differentiate the various operators.
In the right panel of Fig.~\ref{fig:ptau-dbr-ds}, we show the normalised differential decay width as a function of $q^2$ for some 
representative values of the WCs from Table \ref{rdsbin-table}. It can be seen that the shape of the distribution does not change 
dramatically across the various NP  explanations of $R_{D^*}$.

\begin{figure}[h!]
\begin{center}
\hspace*{-1.5cm}
\begin{tabular}{c}
\includegraphics[scale=0.45]{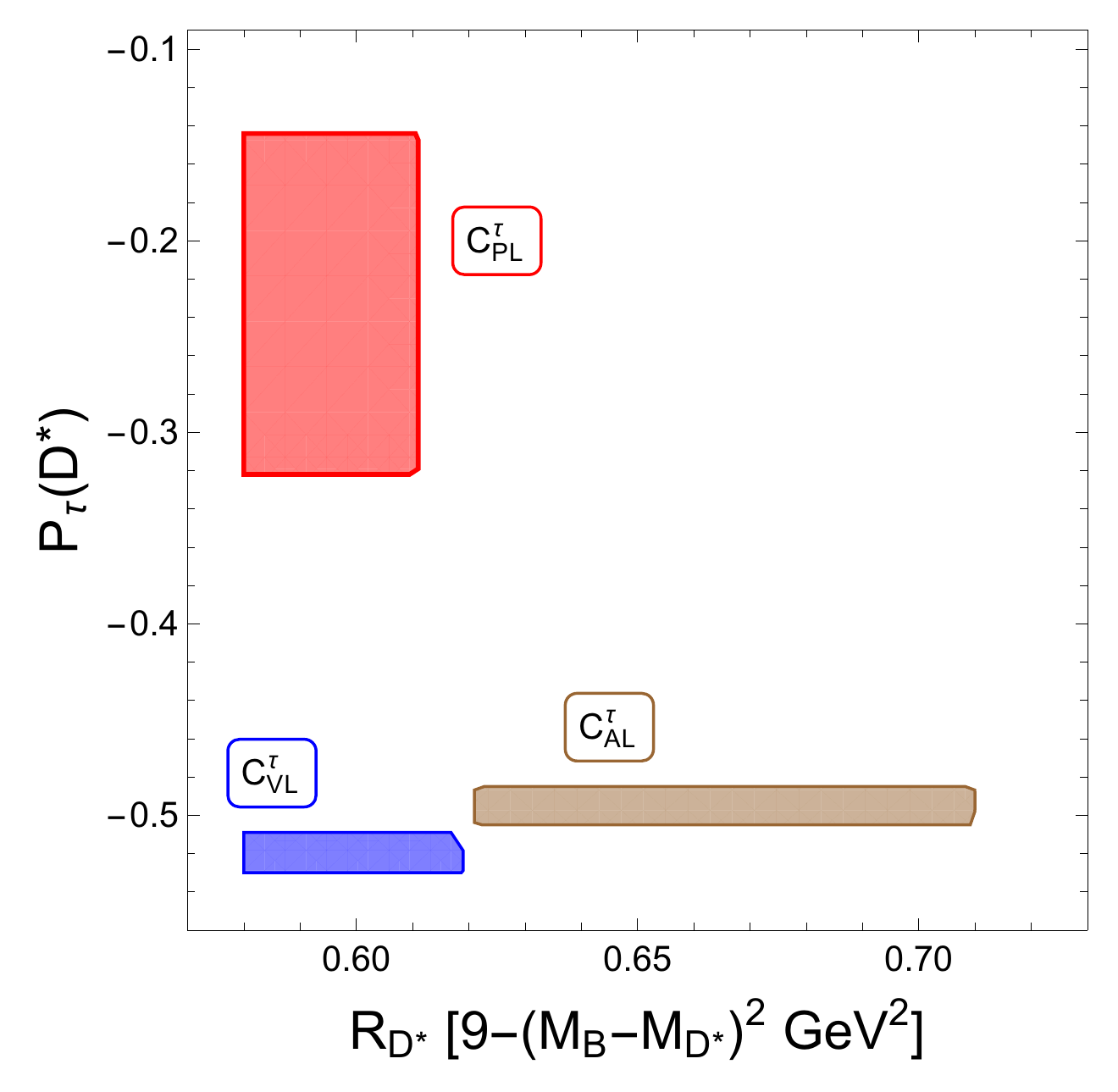}
\includegraphics[scale=0.45]{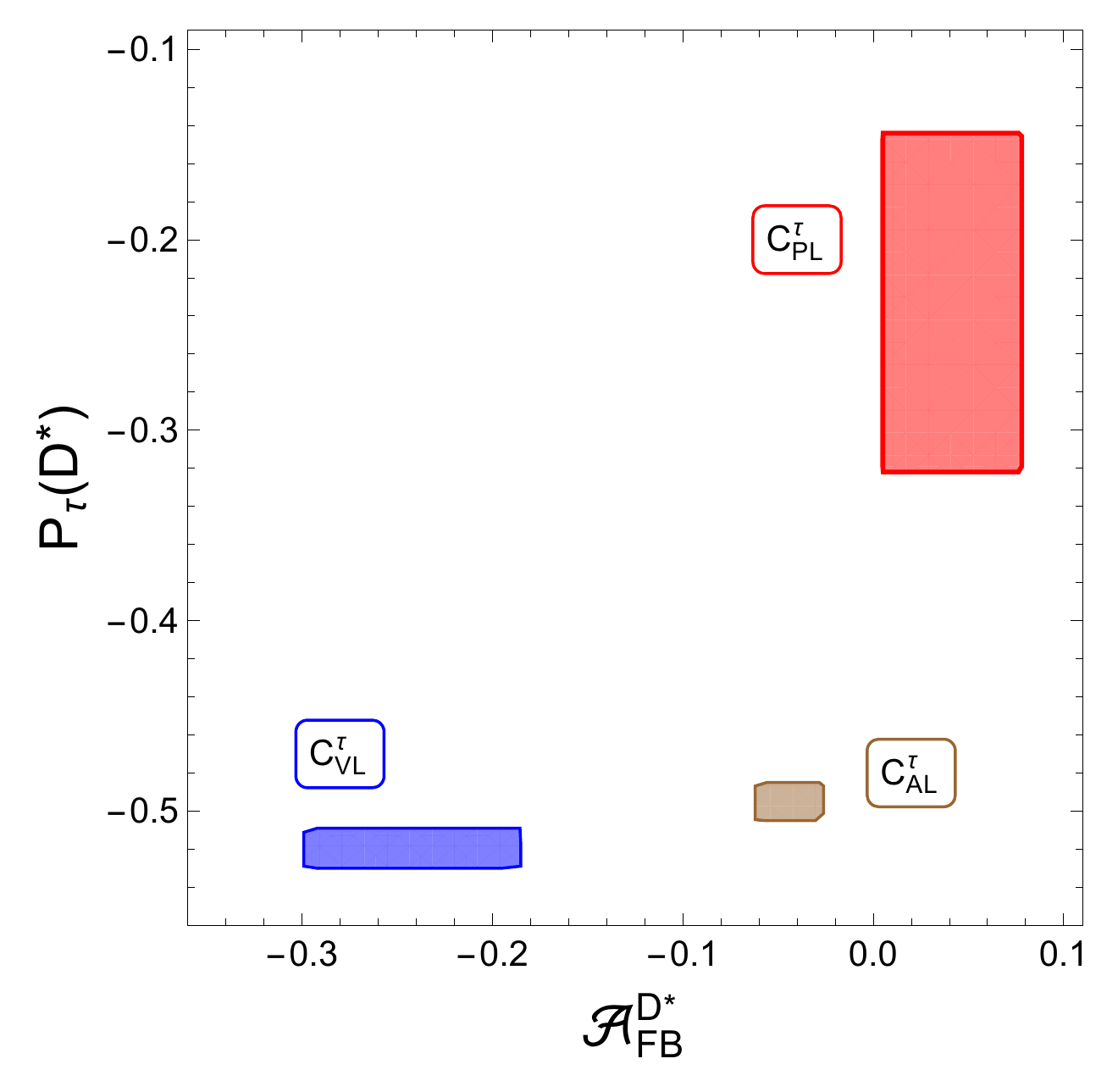}
\includegraphics[scale=0.45]{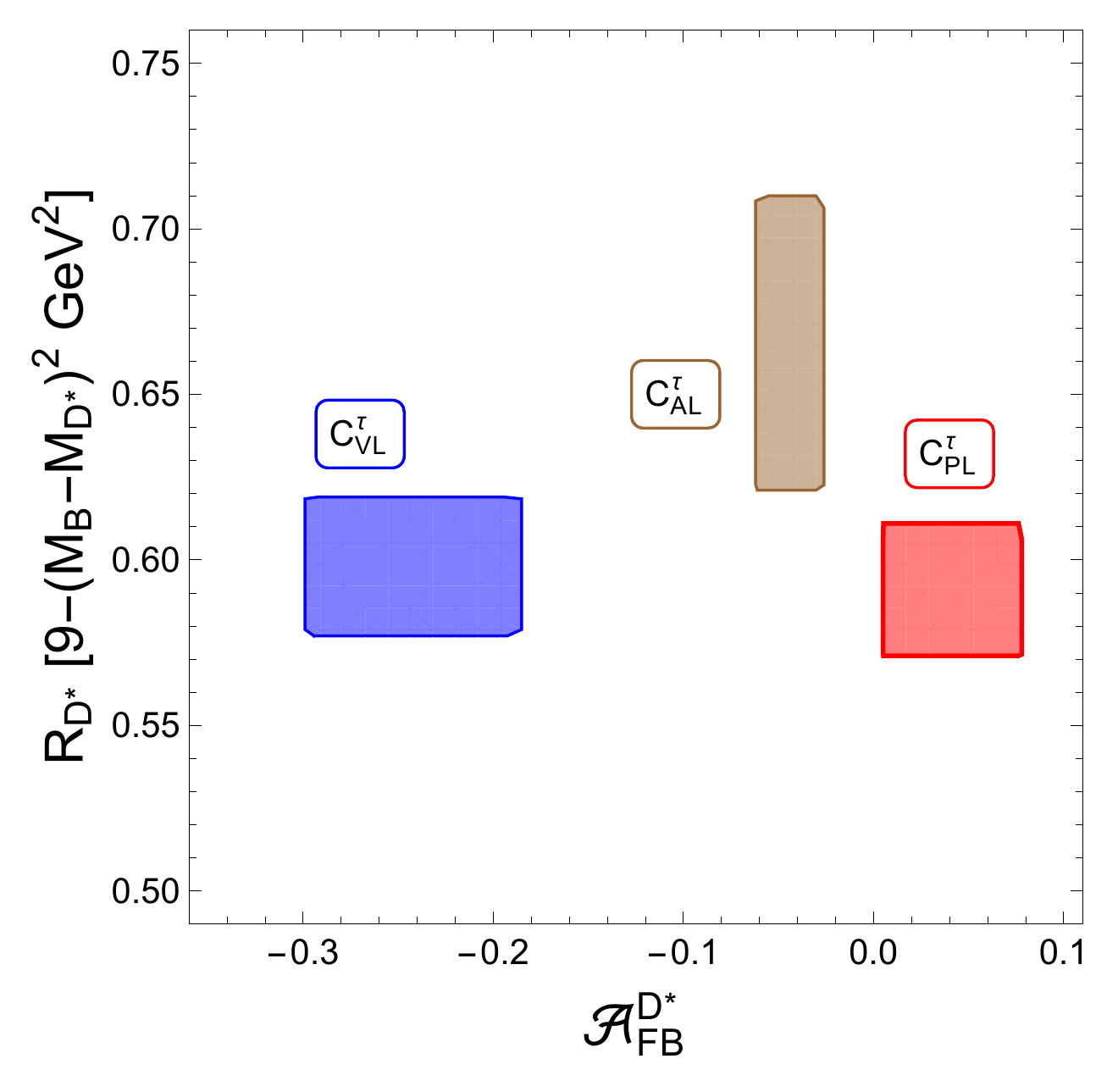}
\end{tabular}
\caption{ The predictions for $P_\tau(D^*)$, $R_{D^*}$ in the last bin and $\mathcal{A}_{FB}^{D^\ast}$ are shown in three different planes
 for the ranges of the three WCs $C_{VL}^\tau$, $C_{AL}^\tau$ and $C_{PL}^\tau$ given in Table \ref{rdsbin-table}.
 We remind the readers that, we have inflated the uncertainties in the form factor parameters in Eq.~\eqref{bds-ff-param} by a factor of two. 
 Hence, the ranges of $P_\tau(D^*)$ and $R_{D^*}$ shown here are rather conservative. \label{fig:RDs-Ptau}}
\end{center}
\end{figure}

In Fig.~\ref{fig:RDs-Ptau}, we show the predictions for $P_\tau(D^*)$, $R_{D^*}$ in the last bin and $\mathcal{A}_{FB}^{D^{\ast}}$ 
in three different planes  for the three WCs $C_{VL}^\tau$, $C_{AL}^\tau$ and $C_{PL}^\tau$ when their 
values are restricted to the ranges shown in Table~\ref{rdsbin-table}. 
Interestingly, we find that each of the three pairs of observables can potentially distinguish between the  WCs unambiguously.
Hence, the measurements of these observables by the experimental collaborations ought to be very much on the cards in their future runs. 

\subsection{Explaining $R_D$ and $R_{D^\ast}$ together}

\begin{figure}[h!]
\begin{center}
\begin{tabular}{c}
\includegraphics[scale=0.7]{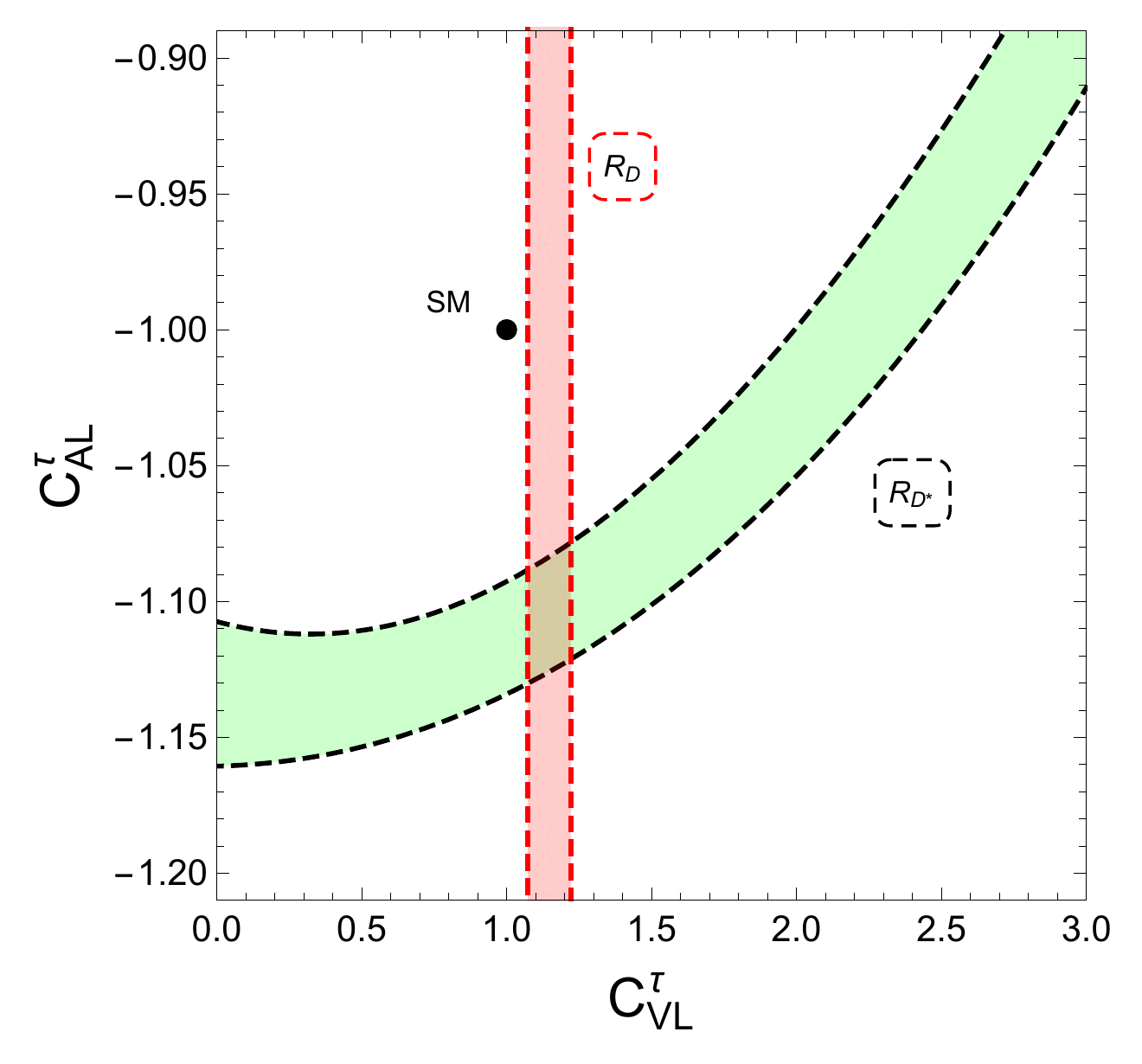}
\end{tabular}
\caption{Allowed region in the $C_{VL}^\tau - C_{AL}^\tau$ plane by $R_D$ and $R_{D^*}$ measurements \label{fig:cvl-al-combined}.}
\end{center}
\end{figure}

We have seen from section \ref{rd-alone} and \ref{rds-alone} that while $R_D$ gets contributions from $C_{VL}^\tau$ and $C_{SL}^\tau$, 
$R_{D^*}$ is affected by $C_{VL}^\tau$, $C_{AL}^\tau$ and $C_{PL}^\tau$. Therefore, in general, these two observables are theoretically 
independent. In the basis of WCs defined by \{$C_{VL}^\tau, C_{AL}^\tau, C_{SL}^\tau, C_{PL}^\tau$\},  the $C_{VL}^\tau$ direction 
is the only direction that affects both. However, as can be seen from tables \ref{rdbin-table} and \ref{rdsbin-table}, the range of $C_{VL}^\tau$ 
( i.e., [1.073, 1.222] ) that explains $R_D$ within $1\sigma$  is different from the range ( i.e., [1.849, 2.648] ) that explains $R_{D^*}$ 
successfully within $1\sigma$. Thus $R_{D}$ and $R_{D^*}$ can not be explained simultaneously by invoking NP only of type  $C_{VL}^\tau$. Fig.~\ref{fig:cvl-al-combined} shows the allowed region in the $C_{VL}^\tau - C_{AL}^\tau$ plane by the $R_D$ and $R_{D^*}$ 
measurements. As $C_{AL}^\tau$ does not contribute to the $\bdtaunu$ decay, the allowed region for $C_{VL}^\tau$ from $R_D$ (the 
red region) is independent of the value of $C_{AL}^\tau$. On the other hand, both the WCs $C_{VL}^\tau$ and 
$C_{AL}^\tau$ contribute to the $\bdstaunu$ decay and hence the values of these WCs allowed by $R_{D^*}$ measurement are correlated. 
The overlap of the red and the green regions correspond to $C_{VL}^\tau  \in [1.073, 1.222]$ and $C_{AL}^\tau  \in [-1.144,-1.062]$. 

Hence, a minimum value of $C_{VL}^\tau  \approx  - C_{AL}^\tau \approx 1.07$ which translates to $\Delta (C_9 - C_{10}) \approx 0.15$ 
(i.e, 15\% shift from the SM values) can explain both $R_D$ and $R_D^{*}$ successfully. This correspond to the operator 
$[\hedv ] [\bar \ell \, \gamma_\mu {\rm P}_L \, \nu]$ with a coefficient $g_{NP}^2/\Lambda^2$ where $\Lambda$ is given by 
$\Lambda \approx g_{_{\rm NP}} \, 2.25~\rm TeV$.

The predictions for $P_\tau (D^*)$, $\mathcal{A}_{FB}^{D^\ast}$ and binwise $R_{D^*}$ for the above ranges of $C_{VL}^\tau$ and 
$C_{AL}^\tau$ are given in table \ref{rdsbin-combined}.
\begin{table}[h!]
\footnotesize
\tabulinesep=1.2mm
\hspace*{-1cm}
\begin{tabu}{|c|c|c|c|c|c|}
\hline
$C_{VL}^\tau$ & $P_\tau (D^\ast)$ & \multicolumn{4}{c|}{$R_{D^\ast}$ [bin]} \\ 
\cline{3-6}
$\in$ [1.073, 1.222]  & $\in$    [-0.507, -0.489]   & $[m_\tau^2-5]$ GeV$^2$ & $[5-7]$ GeV$^2$ & $[7-9]$ GeV$^2$& $[9-(M_B-M_{D^*})^2]$ GeV$^2$ \\
\cline{2-6}
$C_{AL}^\tau$     &     $\mathcal{A}_{FB}^{D^\ast}$  &  \multirow{2}{*}{[0.116, 0.131]} &  
\multirow{2}{*}{[0.373, 0.426]} &  \multirow{2}{*}{[0.535, 0.609]} &  \multirow{2}{*}{[0.616, 0.706]} \\
 $\in$ [-1.144, -1.067]    &  $\in$    [0.055, 0.092]          &                            &                            &                             &                             \\
\hline
\end{tabu}
\caption{Predictions for $P_\tau (D^*)$, $\mathcal{A}_{FB}^{D^\ast}$ and binwise $R_{D^*}$ for the values of WCs satisfying both the observations 
simultaneously. The $1\sigma$ range of the WCs is given in the first column. } \label{rdsbin-combined}
\end{table}

\section{Summary}
\label{conclusion}

In this paper we have performed a model independent analysis of the $R_D$ and $R_{D^\ast}$ anomalies using 
dimension-6 operators that arise in a gauge invariant way.  Among the four WCs {$C_{VL}^\tau$, $C_{AL}^\tau$,
$C_{SL}^\tau$ and $C_{PL}^\tau$}, only $C_{VL}^\tau$ and $C_{SL}^\tau$ contribute to $R_D$. On the other hand, 
$R_{D^\ast}$ gets contributions from $C_{VL}^\tau$, $C_{AL}^\tau$ and $C_{PL}^\tau$. Thus, $C_{VL}^\tau$ is the only 
WC that affects both (barring tensor operator that is discussed in appendix \ref{tensorL}) and hence, these two observables 
are in general theoretically independent. In view of this, initially we studied the solutions of $R_D$ and 
$R_{D^\ast}$ anomalies independent of each other. We obtained the ranges of the WCs that are allowed by the $R_D$ and $R_D^*$ 
measurements at $1 \sigma$. We also discussed the possibility of simultaneous solutions of these two anomalies.  

For the allowed ranges of the WCs, we computed the predictions for both $R_D$ and $R_{D^\ast}$ in four different $q^2$ 
bins, the forward-backward asymmetry, $\mathcal{A}_{FB}^{D^{(\ast)}}$ and the polarisation fraction of the final state $\tau$ lepton. 
We show that measuring the $\tau$ polarisation in 
$\bdstaunu$ decays along with the value of $R_{D^\ast}$ in the last $q^2$ bin can distinguish between 
the three WCs which contribute to this process. This is graphically presented in Fig.~\ref{fig:RDs-Ptau}. Similarly, as seen 
in Fig.~\ref{fig:ptau-dbr}, the measurement of the $\tau$ polarisation in $\bdtaunu$ decay can in principle be used to 
distinguish the two WCs $C_{VL}^\tau$ and $C_{SL}^\tau$. Furthermore, we find that the forward-backward asymmetry 
of the $\tau$ lepton is also a powerful discriminant of the various WCs (see Figs.~\ref{fig:ptau-dbr} and \ref{fig:RDs-Ptau}). 
We hope that the experimental collaborations will take a note of this and make these measurements in near future.

Additionally, in the appendix we also provide the analytic expressions for the double differential decay widths for individual 
$\tau$ helicities taking into account all the 10 dimension-6 operators listed out in section \ref{op-basis}. 
To our knowledge, we are the first to provide the full expressions in the literature. 

Although we have not considered the tensor operator ${\cal O}_{\rm TL}$ in the main text, we have explored 
its effects on the $R_D$ and $R_D^*$ anomalies in appendix \ref{tensorL}. We have shown that there exists a small 
range of $C_{\rm TL}$ that is consistent with both the anomalies.


\section*{Acknowledgement}
We thank Abhishek M. Iyer for collaboration in the very first stage of this work. 

\appendix
\begin{center}
\Large{{\bf Appendix}}
\end{center}
\vspace{-8mm}

\section{Full expressions for $a_\ell^D$, $b_\ell^D$ and $c_\ell^D$ }
\label{b2d-full}
{\bf For the negative helicity of the lepton:}
\bea
\frac{1}{8} a^D_\ell (-) &=& \frac{ M_B^2 {\PDSQ}  }{q^2}  \, \boxed{ {\bf \CVLSq  
\FoneSQ}} + 
\frac{\left(M_B^2-M_D^2\right)^2}{4\left(m_b-m_c\right)^2} \, {\bf \CSRSq 
\FzeroSQ } \nonumber\\*
&+& m_\ell \left[\frac{ \left(M_B^2-M_D^2\right)^2}{2 q^2 (m_b-m_c)} \mathcal{R} 
\left({\bf C_{SR}^{\ell}} {\bf C_{VR}^{\ell*}}\right)  {\FzeroSQ}   +  \frac{4 M_B^2 \PDSQ}{q^2\left(M_B+M_D \right)}
 \mathcal{R} \left({\bf C_{ \bf TL}^\ell} {\bf C_{VL}^{\ell*}}\right) 
{\bf F_+} {\bf F_T}     \right]\nonumber\\*
&+& m_\ell^2 \left[ \frac{ \left(M_B^2-M_D^2\right)^2}{4 q^4} {{\bf \CVRSq}} 
{\FzeroSQ} + \frac{4 {\PDSQ}M_B^2}{q^2\left(M_B+M_D \right)^2} {\CTLSq}  
{\FtwoSQ}   \right] 
\eea
\bea
\frac{1}{8} b^D_\ell (-) &=& \left[-2 {\PDd} M_B \frac{M_B-M_D}{m_b-m_c}  
 \re \left({\bf C_{SR}^\ell} 
{\bf \ctrconj} \right)  {\bf F_0} {\bf F_T}
\right]\nonumber\\*
&-& m_\ell \left[\frac{2 {\PDd} M_B \left(M_B-M_D\right) }{q^2
} \re 
\left({\bf \cvr} {\bf \ctrconj}\right) {\bf F_0} {\bf 
F_T}      \right.\nonumber\\*
&+&  \left.\frac{{\PDd} M_B \left(M_B^2-M_D^2\right) }{q^2 
\left(m_b-m_c\right)}  \re 
\left({\bf \csr} {\bf \cvrconj} \right) {\bf F_0} 
{\bf F_+}       \right]\nonumber\\*
&-& m_\ell^2 \left[ \frac{  {\PDd} 
M_B \left(M_B^2-M_D^2\right)}{q^4} {\bf \CVRSq} {\bf F_0} {\bf F_+}      
\right] \\
%
\frac{1}{8} c^D_\ell (-) &=& \left[ \frac{4 M_B^2 {\PDd}^2}{(M_B+M_D)^2}
{\CTRSq} {\FtwoSQ} - \frac{ M_B^2 {\PDd}^2}{q^2}\boxed{ {\CVLSq} {\FoneSQ}  }
\right]\nonumber\\*
&-&  m_\ell \left[ \frac{4  {\PDd}^2 M_B^2
} {q^2\left(M_B + M_D \right)} \left(\re \left({\bf \cvl} {\bf \ctl}\right) {\bf F_+}{\bf F_T} - \re \left({\bf \cvr} {\bf 
\ctrconj}\right) {\bf F_+} {\bf 
F_T}\right) \right]\nonumber\\*
&+& m_\ell^2 \left[ \frac{ {\PDd}^2 M_B^2}{q^4} {\CVRSq} {\FoneSQ}
- \frac{4 {\PDd}^2 M_B^2}{\left(M_B + M_D \right)^2 q^2}    {\CTLSq}   {\FtwoSQ}
  \right].
\eea

\vspace{2cm}
{\bf For the positive helicity of the lepton:}
\bea
\frac{1}{8} a^D_\ell(+) &=& \frac{ {M_B^2 \PDd}^2 }{q^2} \, {\bf {\CVRSq} {\FoneSQ}} + \frac{ \left(M_B^2-M_D^2\right)^2}{4\left(m_b-m_c\right){}^2}  \, {\bf {\CSLSq} {\FzeroSQ}}
 \nonumber \\
&+& m_\ell \left[ \frac{\left(M_B^2-M_D^2\right)^2}{2 q^2 (m_b-m_c)} \, {\bf \re 
\left({\bf C_{SL}^\ell} {\bf C_{VL}^{\ell*}}\right){\FzeroSQ}} + 
\frac{ 4 M_B^2 \PDd^2 }{q^2(M_B+M_D)} \, {\bf \re \left({\bf C_{VR}^\ell} 
{\bf C_{TR}^{\ell *}}\right) {\bf F_+} F_T }     \right]\nonumber\\
&+& m_\ell^2 \left[ \frac{\left(M_B^2-M_D^2\right)^2}{4 q^4} \, \boxed{{\bf \CVLSq \FzeroSQ}} + 
\frac{4 M_B^2 \PDd^2}{q^2 (M_B+M_D)^2} \,  {\bf |C^\ell_{TR}|^2 \FtwoSQ} \right] \\
\frac{1}{8} b^D_\ell(+) &=& \left[ -2 M_B \PDd \frac{M_B -M_D }{m_b - m_c}  
 \re \left({\bf C_{\bf SL}^\ell} {\bf C_{\bf TL}^{\ell 
*}}\right) {\bf F_0} {\bf F_T} \right]\nonumber\\*
&-& m_\ell \left[ \frac{2{\PDd} \left(M_B-M_D\right) M_B }{q^2} 
\re \left({\bf C_{VL}^\ell} {\bf C_{ \bf TL}^{\ell *}}\right) {\bf F_0} 
{\bf F_T} \right. \nonumber\\*
&+&  \left. \frac{{\PDd} M_B \left(M_B^2-M_D^2\right)  }{q^2 
\left(m_b-m_c\right)} \re \left({\bf C_{SL}^\ell C_{VL}^{\ell 
*} }\right) {\bf 
F_0} {\bf F_+}   \right] \nonumber\\*
&-& m_\ell^2 \left[ \frac{ {\PDd} M_B
\left(M_B^2-M_D^2\right)}{q^4} \boxed{ { \CVLSq} {\bf F_0} {\bf 
F_+}} \right] 
\eea
\bea
\frac{1}{8} c^D_\ell (+) &=&
 \left[\frac{{4 \PDd}^2 M_B^2}{\left(M_B + M_D \right)^2 }{\CTLSq} {\FtwoSQ} 
- \frac{{\PDd}^2 M_B^2}{q^2}  {\CVRSq}
{\FoneSQ} \right]\nonumber\\*
&-& m_\ell \left[ \frac{4 {\PDd}^2 M_B^2 }{\left(M_B + M_D \right) q^2}   \re 
\left({\bf C_{VR}^\ell} {\bf C_{TR}^{\ell *}}\right) {\bf F_+} {\bf 
F_T}- \frac{4 M_B^2 \PDSQ}{\left(M_B + M_D \right)q^2} \re \left({\bf 
C_{VL}^\ell} {\bf C_{TL}^{\ell *}}\right) {\bf F_+} {\bf F_T}    
\right]\nonumber\\*
&+& m_\ell^2 \left[\frac{{\PDd}^2 M_B^2}{q^4}\boxed{{ \CVLSq} 
{\FoneSQ}} - \frac{4{\PDd}^2 M_B^2}{\left(M_B +M_D  \right)^2q^2}  {\CTRSq} 
{\FtwoSQ}   \right]
\eea

\section{Full expressions for $a_\ell^{D^*}$, $b_\ell^{D^*}$ and $c_\ell^{D^*}$ }
\label{b2ds-full}

\begin{eqnarray}
a_\ell^{D^\ast} (-) &=& \frac{8 M_B^2 \pds}{\mbmd^2} {\bf \cvlsq \vsq} + 
\frac{\mbmd^2 ( 8 M_{D^*}^2 q^2 + \lambda)}{2 M_{D^*}^2 q^2} {\bf \calsq \aosq}
\nonumber \\
&&+\frac{8 M_B^4 |p_{D^\ast}|^4 }{\mds \mbmd ^2 q^2} {\bf \calsq \atsq} \nn\\
&&-  \frac{4\pds M_B^2 \mbmdqsq }{\mds q^2} {\bf \calsq  A_1 A_2}
\nn\\
&&{+\frac{ 32 \mB^2 {\pDmag}^2 }{q^2} {\bf {\ctrsq} T_1^2 } + \frac{ 8
\left(\mB^2-\mDs^2\right)^2 }{q^2} {\bf {\ctrsq} T_2^2 }}\nn\\
&& + m_\ell \left[\frac{32 M_B^2 \pds}{q^2 \mbmd} {\bf \re\left(\cvl \ctlconj \right)
V T_1} \nn \right. \\
&& \left. + \frac{8 \mbmd \left(2 \mds \mbmdsq + M_B^2 \pds \right)}{q^2 \mds}
{\bf \re\left(\call \ctlconj \right) A_1 T_2} \nn \right. \\
&&\left. - \frac{8 M_B^2 \mbmdqsq \pds}{q^2 \left(M_B - M_{D^\ast}\right) \mds}
{\bf \re \left(\call \ctlconj \right) A_1 T_3} \nonumber \right. \nn \\
&& \left. \quad - \frac{8 M_B^2 \mbmdqthsq \pds}{q^2 \mbmd \mds} {\bf \re
\left(\call \ctlconj \right) A_2 T_2} \right. \nn \\
&& \left. + \frac{32 M_B^4 |p_{D^\ast}|^4}{q^2 \mds \mbmd \mbmdsq} {\bf \re
\left(\call \ctlconj \right) A_2 T_3} \right.\nn\\
&& \left. +{\frac{32  \mB^2 {\pDmag}^2 }{(\mB+\mDs)
q^2} {\bf \re\left(\cvr \ctrconj\right) V T_1}}\right.\nn\\
&& -\frac{8 (\mB-\mDs) (\mB+\mDs)^2 }{q^2} {\bf
\re\left(\car \ctrconj\right) A_1 T_2} \bigg]\nn\\
&&+ m_\ell^2 \left[\frac{32 M_B^2 \pds}{q^4} {\bf \ctlsq T_1^2}  \right. \nn 
\eea
\bea
&& \left. + \frac{2 \left(8 \mds\left(2 \left(M_B^2 + M_{D^\ast}^2\right) -
q^2\right) q^2 + \left(4 \mds + q^2\right)\lambda\right)}{q^4 \mds} {\bf \ctlsq
T_2^2} \nonumber \right. \\
&& \hspace*{-8mm} \left. \quad + \frac{32 M_B^4 |p_{D^\ast}|^4}{q^2 M_{D^\ast}^2 \mbmdsq^2}
{\bf \ctlsq T_3^2} - \frac{16 M_B^2 \pds \mbmdqthsq}{q^2 \mds \mbmdsq} {\bf
\ctlsq T_2 T_3} \right]
\end{eqnarray}
\begin{eqnarray}
b_\ell^{D^\ast} (-) &=& 
-16 |p_{D^\ast}| M_B {\bf \re \left(\cvl \calconj \right)  V A_1 }
\nn\\
&& + { \frac{32 M_B^3 {\pDmag}^3 }{{\mbmc} {\mbmdsq}
\mDs}{\bf \re\left(\cpr \ctrconj\right) A_0 T_3 }}\nn\\
&& - \frac{8 \mB {\pDmag} \left(\mB^2+3
M_{D^\ast}^2-q^2\right) }{{\mbmc} \mDs} {\bf \re\left(\cpr \ctrconj\right) A_0 T_2}
\nn \\ 
&&  - m_\ell \left[\frac{32 M_B \left(M_B -
M_{D^\ast}\right)|p_{D^\ast}|}{q^2}
{\bf \re \left(\cvl \ctlconj \right) V T_2} \right. \nn \\ 
&& \left. + \frac{32 M_B \left(M_B + M_{D^\ast}\right) |p_{D^\ast}| }{q^2} {\bf
\re \left(\call \ctlconj\right) A_1 T_1} \right.\nn\\
&&+\frac{8 \mB {\pDmag} \left(M_B^2+3
\mDs^2-q^2\right) }{\mDs q^2} {\bf \re\left(\car \ctrconj \right) A_0 T_2}\nn\\
&& -\frac{ 32 \mB^3 \pDmag^3 }{(\mB-\mDs) \mDs
(\mB+\mDs) q^2} {\bf \re\left(\car \ctrconj\right) A_0 T_3} \bigg]\nn\\
&& - m_\ell^2 \left[\frac{64 M_B \mbmdsq |p_{D^\ast}|}{q^4} {\bf \ctlsq T_1
T_2}\right] 
\end{eqnarray}
\begin{eqnarray}
c_\ell^{D^\ast} (-) &=& \frac{8 \pds M_B^2}{\mbmd^2} {\bf{\cvlsq \vsq}} - 
\frac{\mbmd^2 \lambda}{2 \mds q^2} {\bf \calsq \aosq} \nn\\
&& - \frac{8 |p_{D^\ast}|^4 M_B^4}{\mbmd^2 \mds q^2} {\bf \calsq \atsq} \nn\\
&&+ \frac{4 \pds M_B^2 \mbmdqsq}{\mds q^2} {\bf \calsq A_1 A_2} \nn \\
&&{-\frac{32\mB^2 \mDs^2 \left(\mB^2-\mDs^2\right)^2
{\pDmag}^2}{\left(-\mB^2 \mDs+ \mDs^3\right)^2 q^2} {\bf
{\ctrsq}  T_1^2}}\nn\\
&&{\frac{2  \left(\mB^2-\mDs^2\right)^2
}{\mDs^2} {\bf {\ctrsq} T_2^2 }}\nn\\ && {-\frac{4 
\left(-\mB^2+\mDs^2\right)
\left(-\mB^4+\mDs^4+4 \mB^2 \pDmag^2\right) }{\mDs^2 q^2}
{\bf {\ctrsq} T_2^2}}\nn\\
&&{+\frac{ 32 \mB^4 {\pDmag}^4 }{\left(-\mB^2
\mDs + \mDs^3\right)^2} {\bf {\ctrsq} T_3^2}}\nn
\eea
\bea
&&{+\frac{ 16 \mB^2 {\pDmag}^2 \left(\mB^2+3
\mDs^2-q^2\right) }{-\mB^2 \mDs^2+\mDs^4} {\bf {\ctrsq} T_2 T_3}}\nn \\
&& + m_\ell \left[\frac{32 M_B^2 \pds}{q^2 \mbmd} {\bf \re\left(\cvl \ctlconj\right)
V T_1}  \nn \right. \\ 
&& \left. -\frac{8 M_B^2 \left(M_B + M_{D^\ast}\right)\pds }{q^2 \mds}{\bf
\re\left(\call \ctlconj \right) A_1 T_2} \right. \nn \\
&& \left.+ \frac{8 M_B^2 \mbmdqsq \pds}{q^2 \mds \left(M_B - M_{D^\ast}\right) }
{\bf \re\left(\call \ctlconj \right) A_1 T_3 } \nn \right. \\
&& \left. + \frac{8 M_B^2 \mbmdqthsq \pds}{q^2 \mds \mbmd }{\bf \re\left(\call
\ctlconj \right) A_2 T_2} \nn \right. \\ 
&& \left. - \frac{32 M_B^4 |p_{D^\ast}|^4}{q^2 \mds \mbmd \mbmdsq} {\bf
\re\left(\call \ctlconj \right) A_2 T_3} \right. \nn \\
&& \left.-{\frac{ 32 \mB^2 {\pDmag}^2 }{(\mB+\mDs)
q^2} {\bf\re\left(\cvr\ctrconj \right) V T_1}}\right.\nn\\
&& {+\frac{ 8 \mB^2 (\mB+\mDs) {\pDmag}^2 }{\mDs^2
q^2}{\bf \re\left(\car\ctrconj \right) A_1 T_2}}\nn\\
&&{-\frac{ 8 \mB^2 {\pDmag}^2
\left(-\mB^2+\mDs^2+q^2\right) }{(\mB-\mDs) \mDs^2 q^2} {\bf
\re\left(\cvr\ctrconj \right) A_1 T_3}}  \nn\\
&&{+\frac{8 \mB^2 {\pDmag}^2 \left(\mB^2+3
\mDs^2-q^2\right) }{\mDs^2 (\mB+\mDs) q^2} {\bf \re\left(\cvr\ctrconj \right) A_2
T_2}}\nn\\
&&{-\frac{ 32 \mB^4 {\pDmag}^4 }{(\mB-\mDs)
\mDs^2 (\mB+\mDs)^2 q^2} {\bf \re\left(\cvr\ctrconj \right) A_2 T_3}}\bigg]\nn\\
&& + m_\ell^2 \left[\frac{32 M_B^2 \pds}{q^4} {\bf \ctlsq T_1^2} + \frac{2
\left(4\mds - q^2\right)\lambda}{\mds q^4} {\bf \ctlsq T_2^2} \nn \right. \\
&& \left. - \frac{32 M_B^4 |p_{D^\ast}|^4 }{q^2 \mds \mbmdsq^2} {\bf \ctlsq
T_3^2} \nonumber \right. \\
&& \left. \quad + \frac{16 M_B^2 \pds \mbmdqthsq}{q^2 \mds \mbmdsq} {\bf \ctlsq
T_2 T_3} \right]
\end{eqnarray}

\begin{eqnarray}
a_\ell^{D^\ast} (+) &=& \frac{8 \pds M_B^2}{\mbmc^2} {\bf \cplsq \azsq} +
\frac{32 M_B^2 \pds}{q^2} {\bf \ctlsq T_1^2} + \frac{8 \mbmdsq^2}{q^2} {\bf
\ctlsq T_2^2} \nn \\
&&- m_\ell \bigg[\frac{16 \pds M_B^2}{\mbmc q^2}{\bf \re \left(\call \cplconj
\right)\azsq}  \nn 
\eea
\bea
&&  - \frac{32 M_B^2 \pds}{q^2 \mbmd} {\bf \re\left(\cvl \ctlconj\right) V
T_1} \nonumber\\
&&  - \frac{8(M_B + M_{D^\ast})\mbmdsq}{q^2}{\bf \re \left(\call
\ctlconj\right) A_1 T_2} \nn\\ 
&& - \frac{ 32\mB^2 \pDmag^2}{(\mB+\mDs) q^2}
{\bf \re\left(\cvr\ctrconj\right) V T_1}\nn\\
&& + \frac{8 (\mB+\mDs) \left(-2 \mDs^4+\mB^2
\left(2 \mDs^2+\pDmag^2\right)\right) }{\mDs^2 q^2}{\bf \re\left(\car \ctrconj\right)
A_1 T_2} \nn\\
&&+\frac{ 8 \mB^2 {\pDmag}^2 \left(-\mB^2+\mDs^2+q^2\right)
}{(\mB-\mDs) \mDs^2 q^2} {\bf \re\left(\car \ctrconj\right) A_1 T_3}  \nn\\
&&-\frac{ 8 \mB^2 \pDmag^2 \left(\mB^2+3
\mDs^2-q^2\right) }{\mDs^2 (\mB+\mDs) q^2} {\bf \re\left(\car \ctrconj\right) A_2
T_2}\nn\\
&&{+\frac{32\mB^4 \pDmag^4 }{(\mB-\mDs)
\mDs^2 (\mB+\mDs)^2 q^2} {\bf \re\left(\car \ctrconj\right) A_2 T_3}}\bigg]\nn\\
&& + m_\ell^2 \left[\frac{8 \pds M_B^2}{q^4} {\bf \calsq \azsq} 
+\frac{8 \pds M_B^2}{\mbmd^2 q^2} {\bf \cvlsq \vsq} \right. \nn\\
&&\left. \quad + \frac{2 \mbmd^2}{q^2} {\bf \calsq \aosq} \right.\nn\\
&&{+\frac{32 \mB^2 {\pDmag}^2 }{q^4} {\bf {\ctrsq} T_1^2} + 8 \frac{
 \mB^2
\pDmag^2 }{\mDs^2 q^2} {\bf {\ctrsq} T_2^2} }\nn\\
&&{+\frac{16 \left(\mB^4+\mDs^4-2 \mB^2
\left(\mDs^2+\pDmag^2\right)\right)}{q^4} {\bf {\ctrsq}  T_2^2}}\nn\\
&&{+\frac{32 \mB^4 \pDmag^4 }{\left(-\mB^2
\mDs+\mDs^3\right)^2 q^2} {\bf {\ctrsq} T_3^2}}\nn\\
&&{+\frac{16  \mB^2 \pDmag^2 \left(\mB^2+3
\mDs^2-q^2\right) }{\mDs^2 \left(-\mB^2+\mDs^2\right) q^2}
{\bf {\ctrsq} T_2 T_3}}\bigg]
\end{eqnarray}
\begin{eqnarray}
b_\ell^{D^\ast} (+) &=& \frac{8 M_B \mbmdqthsq |p_{D^\ast}|}{\mbmc M_{D^\ast}}
{\bf \re \left(\cpl \ctlconj\right) A_0 T_2}  \nonumber \\
&& \quad - \frac{32 M_B^3 |p_{D^\ast}|^3}{\mbmc M_{D^\ast} \mbmdsq} {\bf \re
\left(\cpl \ctlconj\right) A_0 T_3} \nonumber \\
&& + m_\ell \left[\frac{ 4 |p_{D^\ast}| M_B \mbmd\mbmdqsq}{M_{D^\ast} \mbmc q^2}
{\bf \re \left( \call \cplconj\right) A_0 A_1} \right.\nn\\
&&\left. - \frac{16}{\mbmc}\frac{|p_{D^\ast}|^3 M_B^3}{\mbmd M_{D^\ast} q^2}
{\bf \re \left(\call \cplconj\right) A_0 A_2} \right. \nn 
\eea
\bea 
&& \left. - \frac{8 M_B \mbmdqthsq |p_{D^\ast}|}{M_{D^\ast} q^2} {\bf \re
\left(\call \ctlconj\right) A_0 T_2} \nonumber \right. \\
&&\left. \quad + \frac{32 M_B^3 |p_{D^\ast}|^3}{q^2 M_{D^\ast} \mbmdsq} {\bf \re
\left(\call \ctlconj\right) A_0 T_3} \right. \nn\\
&& \left. +  \frac{ 32 \mB (-\mB+\mDs)
{\pDmag} }{q^2} {\bf \re\left(\cvr\ctrconj\right) V T_2}\right.\nn\\
&&+ \frac{ 32 \mB (\mB+\mDs) {\pDmag} }{q^2} {\bf \re\left(\car \ctrconj\right)
A_1
T_1}\bigg]\nn\\
&& + m_\ell^2 \left[- \frac{4|p_{D^\ast}| M_B \mbmd}{M_{D^\ast} q^4} \mbmdqsq
{\bf \calsq A_0 A_1} \right.\nn\\
&&\left. + \frac{16 |p_{D^\ast}|^3 M_B^3}{\mbmd M_{D^\ast} q^4}{\bf \calsq A_0
A_2} \right.\nn\\
&&{+ \frac{64  \mB \left(-\mB^2+\mDs^2\right) {\pDmag}
}{q^4} {\bf {\ctrsq} T_1 T_2}}\bigg]
\end{eqnarray}

\begin{eqnarray}
c_\ell^{D^\ast} (+) &=& -\frac{32 M_B^2 \pds}{q^2} {\bf \ctlsq T_1^2} - \frac{2
\left(4 \mds - q^2\right)\lambda}{\mds q^2} {\bf \ctlsq T_2^2} + \frac{32 M_B^4
|p_{D^\ast}|^4 }{\mds \mbmdsq^2 } {\bf \ctlsq T_3^2} \nonumber \\
&& \quad- \frac{16 M_B^2 \pds \mbmdqthsq}{\mds \mbmdsq} {\bf \ctlsq T_2 T_3} \nn
\\
&& - m_\ell \left[\frac{32 M_B^2 \pds}{q^2 \mbmd} {\bf \re\left(\cvl \ctlconj\right)
V T_1}  \nn \right. \\ 
&& \left. -\frac{8 M_B^2 \left(M_B + M_{D^\ast}\right)\pds }{q^2 \mds}{\bf
\re\left(\call \ctlconj\right) A_1 T_2} \right. \nn \\
&& \left.+ \frac{8 M_B^2 \mbmdqsq \pds}{q^2 \mds \left(M_B - M_{D^\ast}\right) }
{\bf \re\left(\call \ctlconj\right) A_1 T_3 } \nn \right. \\
&& \left. + \frac{8 M_B^2 \mbmdqthsq \pds}{q^2 \mds \mbmd }{\bf \re\left(\call
\ctlconj\right) A_2 T_2} \nn \right. \\ 
&& \left. - \frac{32 M_B^4 |p_{D^\ast}|^4}{q^2 \mds \mbmd \mbmdsq} {\bf
\re\left(\call \ctlconj\right) A_2 T_3} \right. \nn \\
&& \left. + \frac{32 \mB^2 {\pDmag}^2}{(\mB+\mDs) q^2}
{\bf \re\left(\cvr\ctrconj\right) V T_1}\right. \nn\\
&& + \frac{ 8 \mB^2 (\mB+\mDs) \pDmag^2 }{\mDs^2
q^2}{\bf \re\left(\car \ctrconj\right) A_1 T_2}\nn\\
&&- \frac{8 \mB^2 \pDmag^2
\left(-\mB^2+\mDs^2+q^2\right) }{(\mB-\mDs) \mDs^2 q^2} {\bf
\re\left(\car \ctrconj\right) A_1 T_3}  \nn\\
\eea
\bea 
&&+\frac{8\mB^2 \pDmag^2 \left(\mB^2+3
\mDs^2-q^2\right) }{\mDs^2 (\mB+\mDs) q^2} {\bf \re\left(\car \ctrconj\right) A_2
T_2}\nn\\
&&- \frac{32  \mB^4 \pDmag^4 }{(\mB-\mDs)
\mDs^2 (\mB+\mDs)^2 q^2} {\bf \re\left(\car \ctrconj\right) A_2 T_3}\bigg]\nn\\
&& + m_\ell^2 \left[ - \frac{8 \pds M_B^2}{\mbmd^2 q^2} {\bf \cvlsq \vsq} +
\frac{\mbmd^2 \lambda}{2 \mds q^4} {\bf \calsq \aosq} \right. \nn\\
&& \left. + \frac{8 |p_{D^\ast}|^4 M_B^4}{\mds \mbmd^2 q^4} {\bf \calsq \atsq}
\right.\nn\\ 
&& \left.- \frac{4 \pds M_B^2}{\mds q^4} \mbmdqsq {\bf \calsq A_1
A_2}\right.\nn\\
&&\left. +\frac{ 32 \mB^2 \pDmag^2 }{q^4} {\bf \ctrsq T_1^2 } +\frac{8 \mB^2 \pDmag^2 \left(4 \mDs^2-q^2\right)
}{\mDs^2 q^4} {\bf  \ctrsq T_2^2 }\right. \nn\\
&&\left.-\frac{32  \mB^4 \pDmag^4 }{\left(-\mB^2 \mDs+\mDs^3\right)^2 q^2} {\bf \ctrsq T_3^2}\right.\nn\\
&&\left.+\frac{16  \mB^2 \pDmag^2 \left(\mB^2+3 \mDs^2-q^2\right) }{\mDs^2 \left(\mB^2-\mDs^2\right) q^2} {\bf \ctrsq T_2 T_3}\right]
\end{eqnarray}
\section{Contribution of the Tensor operator ${\cal O}^{cb\ell}_{\rm TL}$}
\label{tensorL}

\subsection{$\bdtaunu$}
\begin{table}[h!]
\footnotesize
\tabulinesep=1.2mm
\hspace*{-1cm}
\begin{tabu}{|c|c|c|c|c|c|}
\hline
$C_{TL}^\tau$ & $P_\tau (D)$ & \multicolumn{4}{c|}{$R_{D}$ [bin]} \\ 
\cline{3-6}
$\in$ [0.240, 0.796]  & $\in$    [0.125, 0.254]   & $[m_\tau^2-5]$ GeV$^2$ & $[5-7]$ GeV$^2$ & $[7-9]$ GeV$^2$& $[9-(M_B-M_{D})^2]$ GeV$^2$ \\
\cline{2-6}
$C_{TL}^\tau$     &     $\mathcal{A}_{FB}^{D}$  &   \multirow{2}{*}{[0.178, 0.233]} &  
\multirow{2}{*}{[0.673, 0.907]} &  \multirow{2}{*}{[1.135, 1.533]} &  \multirow{2}{*}{[1.989, 2.508]} \\
 $\in$ [-3.500, -3.052]    &  $\in$    [$-0.451$, $-0.404$]          &                            &                            &                             &                             \\
\hline
\end{tabu}
\caption{Predictions for $P_\tau(D)$, $\mathcal{A}_{FB}^{D}$ and binwise values of $R_{D}$
for a range of $C_{TL}^\tau$ for which $R_D$ is experimentally satisfied within $1\sigma$. The  range of the WCs is given in 
the first column. The values in the subsequent columns are only for the range of $C_{TL}^\tau$ closest to the SM value of 0, viz. 
the positive range. \label{b2d-tensor}}
\end{table}

In this section we investigate the effect of the tensor operator ${\cal O}^{cb\ell}_{\rm TL}$ on the $\bdtaunu$ decay. 
In the first column of table \ref{b2d-tensor}, we show the range of $C_{TL}^\tau$ that explains $R_D$ within $1\sigma$. 
In the subsequent columns, we show the predictions of $P_\tau (D)$, $\mathcal{A}_{FB}^{D}$ and binwise $R_D$ for the allowed range of $C_{TL}^\tau$ 
that is closest to zero (i.e., $C_{TL}^\tau \in$ [0.240, 0.796]).  A comparison with the left plot of Fig.~\ref{rds-vs-wc} reveals 
that $P_\tau (D)$ in this case is quite different from the other cases and thus, can completely distinguish the tensor operator 
from the vector or scalar operators. Similarly, $\mathcal{A}_{FB}^{D}$ can also be used to distinguish the tensor from the vector operator, however, 
there exists some degeneracy with the scalar operator.

\begin{figure}[h!]
\includegraphics[scale=0.6]{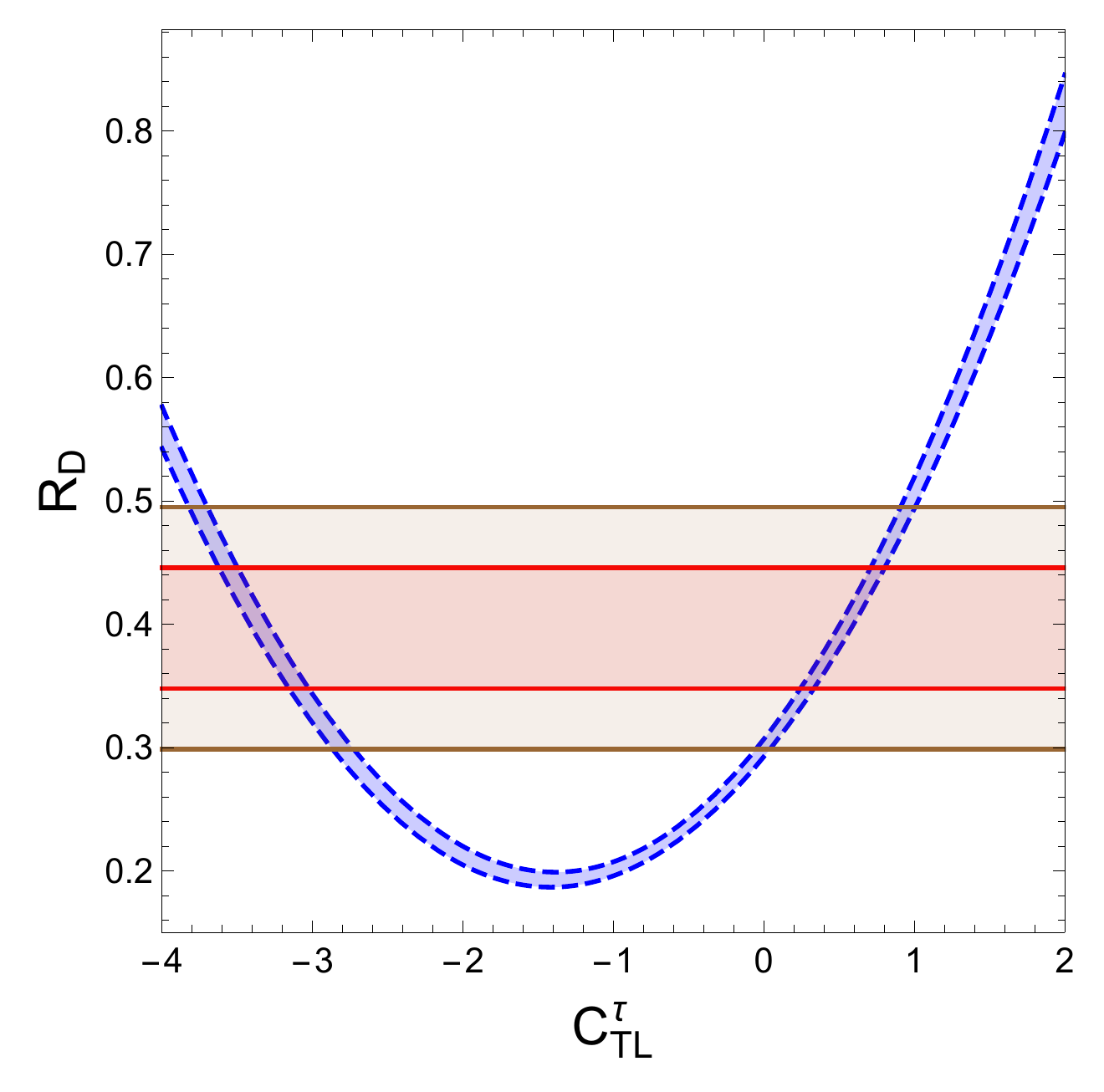}
\includegraphics[scale=0.6]{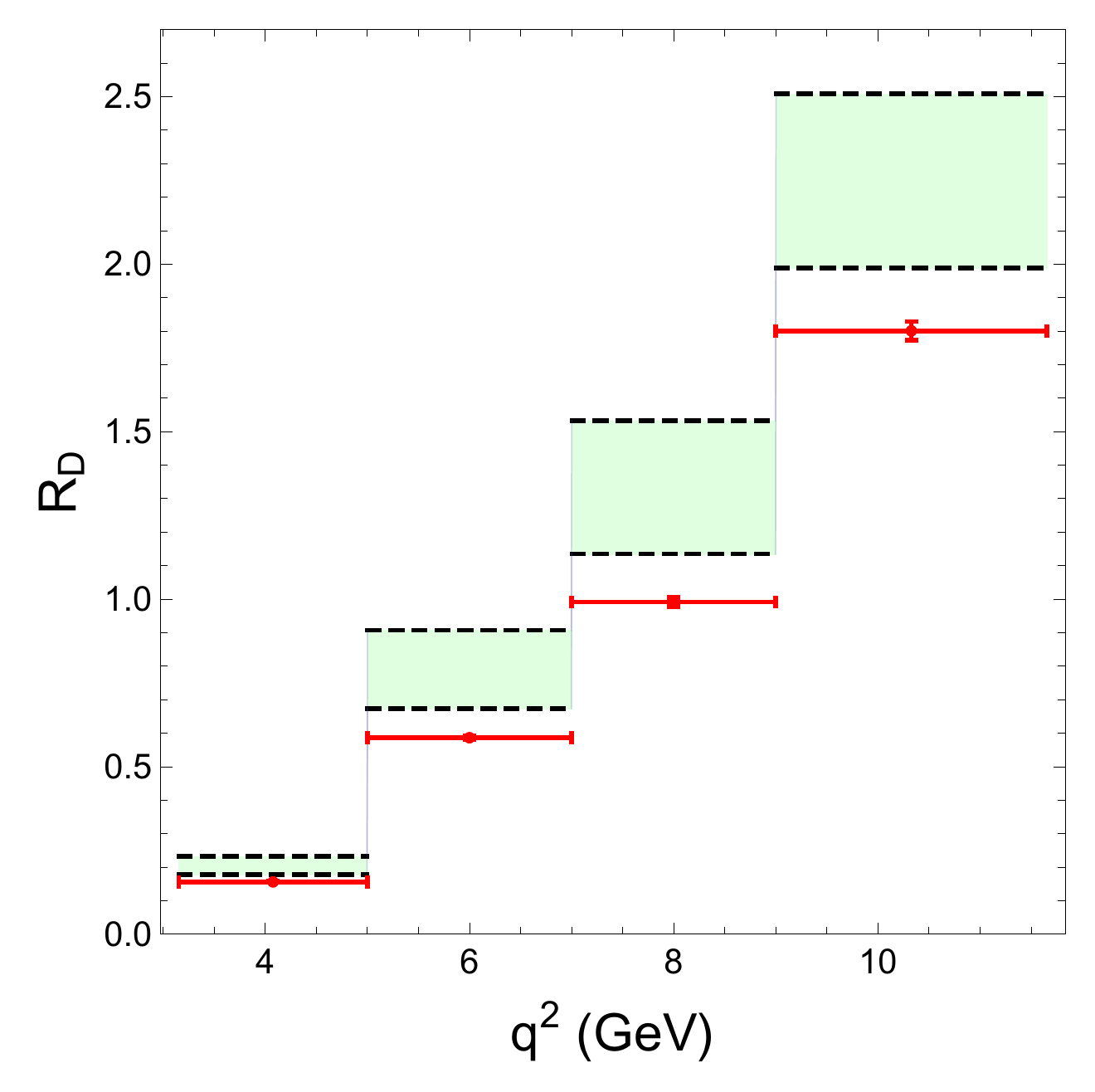}
\caption{ The left panel shows the dependence of $R_D$ with respect to the variation of the WCs $C_{TL}^\tau$ and 
the right panel shows the prediction for $R_D$ in four different bins of $q^2$ from table \ref{b2d-tensor}. \label{Fig:b2d-tensor}}
\end{figure}

The variation of $R_D$ as a function of $C_{TL}^\tau$  is also shown in the left plot of Fig.~\ref{Fig:b2d-tensor}. The predictions 
for binwise $R_D$ for the tensor operators are graphically presented in the right plot of Fig.~\ref{Fig:b2d-tensor}.

\subsection{$\bdstaunu$}
\begin{table}[h!]
\footnotesize
\tabulinesep=1.2mm
\hspace*{-1cm}
\begin{tabu}{|c|c|c|c|c|c|}
\hline
$C_{TL}^\tau$ & $P_\tau (D^\ast)$ & \multicolumn{4}{c|}{$R_{D^\ast}$ [bin]} \\ 
\cline{3-6}
$\in$ [-0.120, -0.058]  & $\in$    [-0.481, -0.441]   & $[m_\tau^2-5]$ GeV$^2$ & $[5-7]$ GeV$^2$ & $[7-9]$ GeV$^2$& $[9-(M_B-M_{D^*})^2]$ GeV$^2$ \\
\cline{2-6}
$C_{TL}^\tau$     &     $\mathcal{A}_{FB}^{D^\ast}$  &  \multirow{2}{*}{[0.113, 0.129]} &  
\multirow{2}{*}{[0.368, 0.423]} &  \multirow{2}{*}{[0.531, 0.610]} &  \multirow{2}{*}{[0.620, 0.715]} \\
 $\in$ [0.709, 0.834]    &  $\in$    [$-0.016$, 0.034]          &                            &                            &                             &                             \\
\hline
\end{tabu}
\caption{Predictions for $P_\tau(D^\ast)$, $\mathcal{A}_{FB}^{D^\ast}$ and binwise values of $R_{D^\ast}$
for a range of $C_{TL}^\tau$ for which $R_{D^\ast}$ is experimentally satisfied within $1\sigma$. 
The corresponding range of the WCs is given in the first column. The values in the subsequent columns are only for the 
range of $C_{TL}^\tau$ closest to the SM value of 0, viz. the negative range. \label{rdsbin-tensor}}
\end{table}

\begin{figure}[h!]
\includegraphics[scale=0.6]{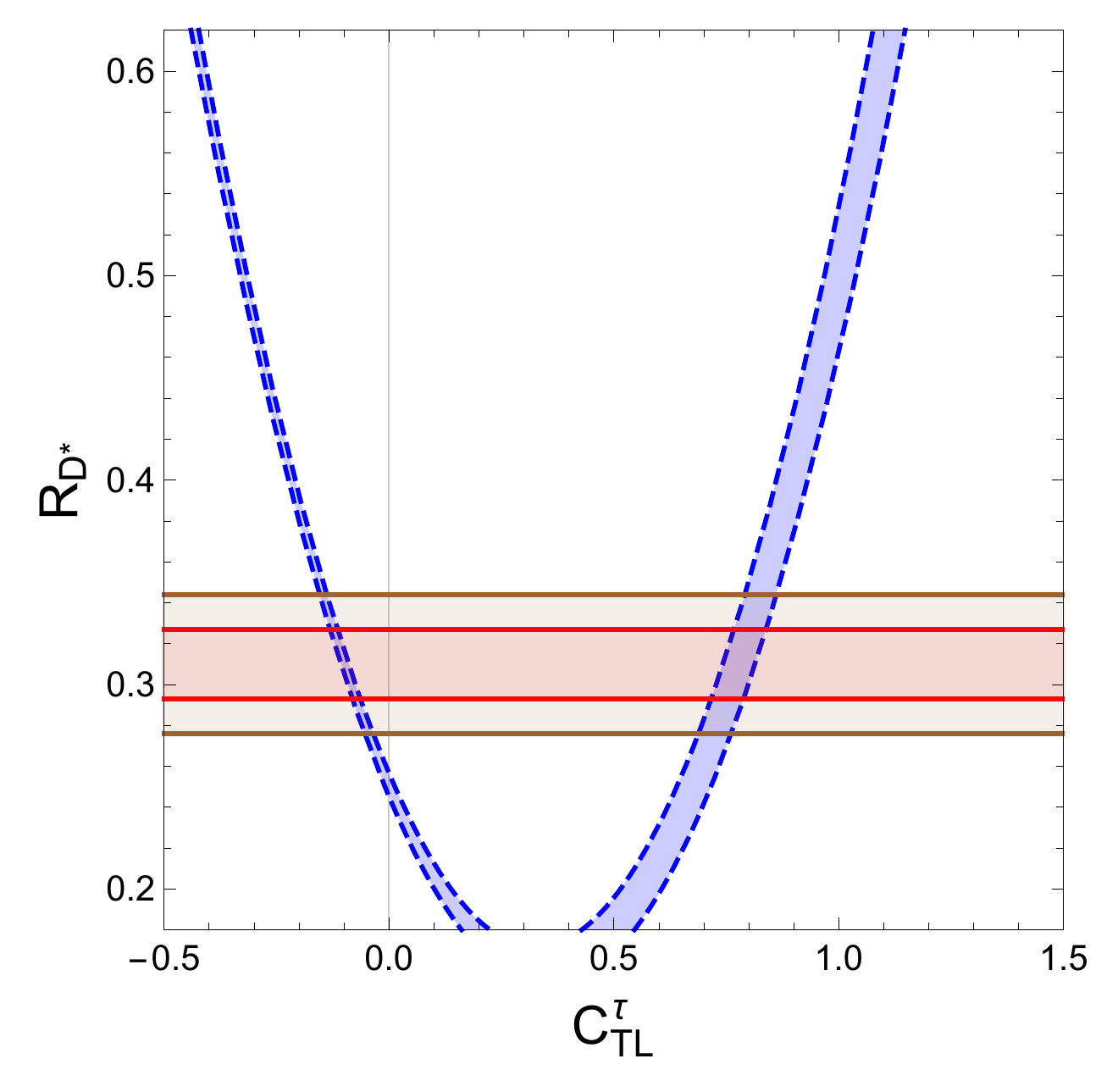}
\includegraphics[scale=0.6]{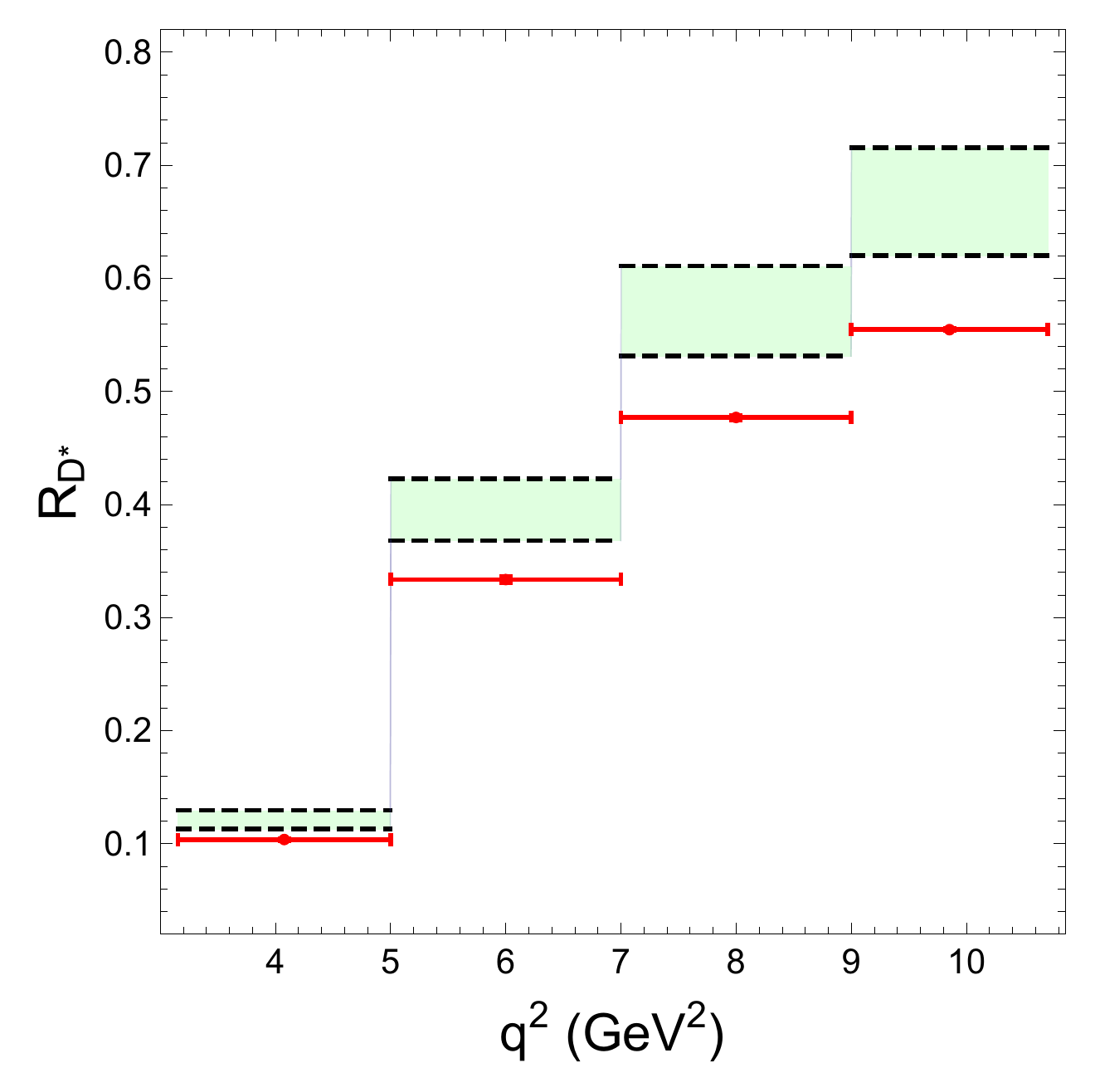}
\caption{ The left panel shows the dependence of $R_{D^*}$ with respect to the variation of the WCs $C_{TL}^\tau$ and 
the right panel shows the prediction for $R_{D^*}$ in four different bins of $q^2$ from table \ref{rdsbin-tensor}. \label{Fig:b2ds-tensor}}
\end{figure}

The  range of $C_{TL}^\tau$ that explains $R_D^\ast$ within $1\sigma$ is shown in the first column of table \ref{rdsbin-tensor}. 
The resulting values for $P_\tau (D^*)$, $\mathcal{A}_{FB}^{D^\ast}$ and binwise $R_D^*$ are shown in the subsequent columns. 
In the left plot of Fig.~\ref{Fig:b2ds-tensor} we also show the dependence of $R_D^\ast$ as a function of $C_{TL}^\tau$. The right 
plot shows the binwise $R_D^*$ graphically. 

A quick look at the allowed ranges for $C_{TL}$ in the $B \to D$ (Table~\ref{b2d-tensor}) and the $B \to D^\ast$ (Table~\ref{rdsbin-tensor}) cases
shows that there is a region of overlap, around 0.7-0.8, which allows one to explain both the anomalies simultaneously.

\section{$\rm SU(3)_C \times SU(2)_L \times U(1)_Y$ gauge invariance }
\label{warsaw}

In table \ref{table-dim6} we show how the WCs of the operators in this paper are related to the WCs of the gauge invariant 
dimension 6 operators of \cite{Grzadkowski:2010es}. We use the following set of notations: 
\begin{itemize}
\item 
Greek letters $\mu, \nu, \cdots$ are used to denote Lorentz indices.
\item
SU(2) fundamental indices are denoted by $a, b, \cdots$ and $I, J \cdots $ will be used to denote adjoint indices. 
\item
To represent quark (lepton) flavors, we use $i, j, k \cdots $ ($m, n \cdots $). 
\item
A tilde (e.g. $\tilde{\mathcal{C}}$) is used to denote high energy Wilson coefficients.

\item
The notation for the operators is as given in \cite{Grzadkowski:2010es}.

\item definition of the quark mixing matrices ($f$ and $m$ denote flavour and mass bases)
\bea
u_L^f =  V_L^u u_L^m \\
u_R^f =  V_R^u u_R^m\\
d_L^f =  V_L^d d_L^m \\
d_R^f =  V_R^d d_R^m
\eea

\end{itemize}

\begin{table}[h!]
\centering
\tabulinesep=1.6mm
\hspace*{-5mm}\begin{tabu}{p{10mm}p{2mm}cccp{2mm}c|c|c}
\hline
\multicolumn{7}{c|}{WCs in this work}          &         WCs in \cite{Grzadkowski:2010es}         &        Operator structure   \\
\hline
\hline
\multirow{7}{*}{$\hspace*{-4mm}\dfrac{2 G_F V_{cb}}{\sqrt{2}} \, \times$}   &  
\multirow{7}{*}{$\begin{cases}  \phantom{1} \\ \phantom{1} \\ \phantom{1} \\  \phantom{1} \\ \phantom{1} \\ \phantom{1} \\ \phantom{1} \\ \phantom{1} 
\\ \phantom{1} \\ \phantom{1} \\ \phantom{1} \end{cases}$ }  & 
\multirow{3}{*}{$\Delta C^{cb\tau *}_9$} & \multirow{3}{*}{=} & \multirow{3}{*}{$-\Delta C^{cb\tau *}_{10} $} & 
\multirow{7}{*}{$\hspace*{-4mm}\begin{rcases*} \phantom{1} \\ \phantom{1} \\ \phantom{1} \\ \phantom{1} \\ \phantom{1} \\ \phantom{1} 
\\ \phantom{1} \\ \phantom{1} \\ \phantom{1} \\ \phantom{1} \\ \phantom{1} \end{rcases*}$ }  & 
 \multirow{3}{*}{=} & \multirow{2}{*}{$\dfrac{1}{2} 
[V_L^{d \, \dagger}]_{3i}\left[ -\dfrac{g^2 v^2}{2 M_W^2} \left( \tilde{C}^{(3) ij, 33 \,\,\dagger}_{\phi q} +\right. \right. $}    & 
\multirow{1}{*}{$\left[\phi^\dagger i \overleftrightarrow{D}_\mu^I \phi\right]\left[\bar{q}_i^2 \, \frac{\sigma^I}{2} \gamma^\mu \, q_j^1\right]$}\\[1mm] 
&&&&&&&  &  $\left[\phi^\dagger i \overleftrightarrow{D}_\mu^I \phi\right]\left[\bar{\ell}_i^1 \, \frac{\sigma^I}{2} \gamma^\mu \, \ell_j^2\right]$\\
&&&&&&& $\left. \left. \tilde{C}^{(3)33,{ij}}_{\phi \ell} \right)  + 2 \tilde{C}_{\ell q}^{(3) ij \,33} \right]  [V_L^{u}]_{j2}$  &  
$\left[\bar{q}_i^2 \gamma^\mu q_j^1\right]\left[\bar{\ell}_3^1 \gamma_\mu \ell_3^2 \right]$\\
\cline{3-5}
\cline{7-7}
\cline{8-9}
&   &  $C^{cb\tau \, ' *}_9$  &   =   &   $- C^{cb\tau \, ' *}_{10} $   &    &   =    &   $-\dfrac{1}{2} 
[V_R^{u \, \dagger}]_{2i} \frac{g^2 v^2}{2M_W^2} \tilde{C}^{ij \, 33 \, \dagger}_{\phi u d}  [V_R^{d}]_{j3} $   &  
\multirow{1}{*}{$[i\tilde{\phi}^\dagger D_\mu \phi]\left[\bar{u}_p \gamma^\mu d_r\right]$}
\\
\cline{3-5}
\cline{7-7}
\cline{8-9}
&  &  ${C^{cb\tau}_s}^\ast$     &  =     &  $- {C^{cb\tau}_{p}}^\ast$    &     &    =  & $\dfrac{1}{2}
[V_L^{d \, \dagger}]_{3i} \, {\tilde C}_{\ell e q u}^{(1) ij, 33} [V_R^u]_{j2}$ &  $\left(\bar{\ell}_3^1 e_3 \right)
\left(\bar{q}_i^2 u_j\right)$  \\
\cline{3-5}
\cline{7-7}
\cline{8-9}
&   &${C^{cb\tau \, '}_s}^\ast$ &   = & $-{C^{cb\tau \, '}_{p}}^\ast $ &     & = & $\dfrac{1}{2} [V_R^{d \, \dagger}]_{3i} \,  {\tilde C}_{\ell e d q}^{i j, 33} [V_L^u]_{j2}$    &  $\left(\bar{\ell}_3^1 e_3 \right) \left(\bar{d}_i q_j^1\right) $   \\
\cline{3-5}
\cline{7-7}
\cline{8-9}
&   &${C^{cb\tau}_T}^\ast$  &  = & $-{C^{cb\tau}_{T5}}^\ast $ &     & = &   $\dfrac{1}{2}[V_L^{d \, \dagger}]_{3i} \, {\tilde C}_{\ell equ}^{(3)ij, 33} \, [V_R^{u}]_{j2}$  &   
$\left(\overline{\ell}_3^1 \sigma_{\mu\nu}  e_3 \right) \epsilon_{12} \left(\overline{q}_i^2 \sigma^{\mu\nu} u_j \right)$  \\
\hline
\end{tabu}
\caption{Correspondence of our operators with those in reference \cite{Grzadkowski:2010es}. The mixing of different lepton 
flavours are ignored.\label{table-dim6}}
\end{table}

\section{RG Running of Wilson Coefficients}
\label{RG}

In this section, we note the renormalisation group (RG) running of the 
couplings and the Wilson coefficients. The QCD coupling above
the $m_b$ scale is given by $\alpha_s^{(5)}$ and that above the $m_t$ scale is given by $\alpha_s^{(6)}$. These are given by
\begin{equation}
\alpha_s^{(5)}(\mu) = \frac{\alpha_s (m_b)}{1 - \beta^{(5)}_0 
\frac{\alpha_s(m_b)}{2\pi} {\rm ln} \left(\frac{\mu}{m_b}\right)} \quad \quad 
\alpha_s^{(6)}(\mu) = \frac{\alpha_s (m_t)}{1 - \beta^{(6)}_0 
\frac{\alpha_s(m_t)}{2\pi} {\rm ln} \left(\frac{\mu}{m_t}\right)}
\label{eqn:qcd_running}
\end{equation}
where 
$\beta_0^{(n_f)} = 11 - \frac{2 n_f}{3} $. 

In order to calculate the running of the Wilson Coefficients to a high scale $M$, we need 
to calculate the beta functions for the different operators - the scalar, vector and tensor 
operators. 
The calculation is sketched below (for a good review on the subject, see \cite{Skiba:2010xn}) \\
\begin{figure}[h]
\begin{center}
\includegraphics[scale=0.45]{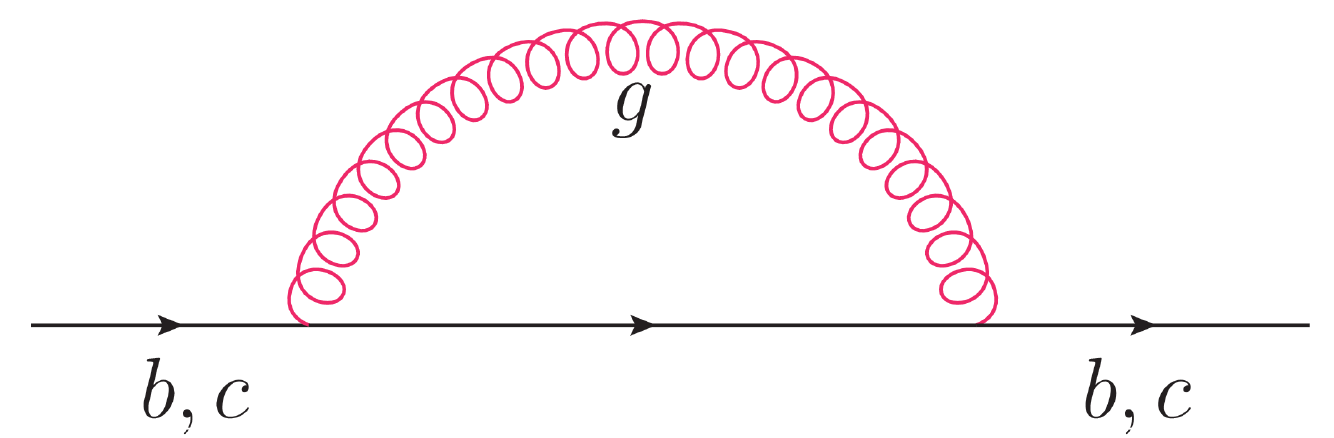}
\includegraphics[scale=0.7]{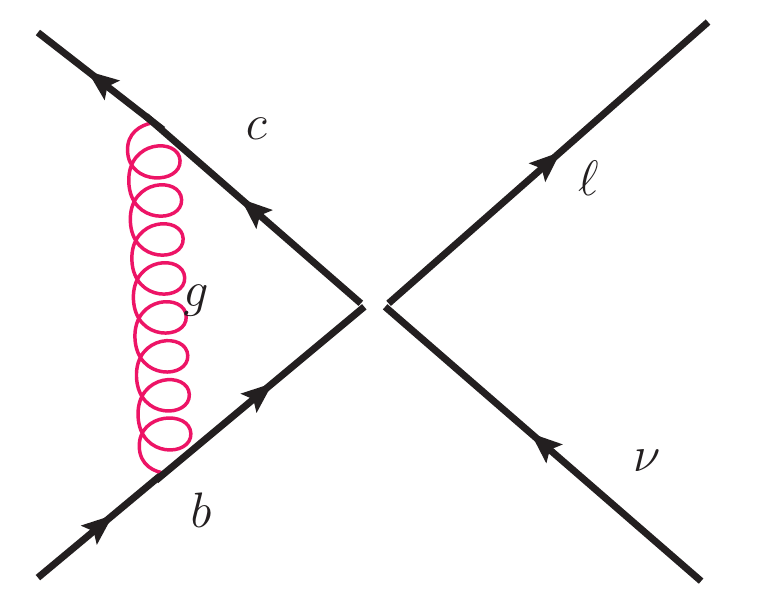}
\end{center}
\caption{Vertex Correction and self energy diagrams.}
\label{fig:RG_running}
\end{figure}
Firstly, we need to consider the self-energy correction for the $b$ or $c$ quarks (left diagram in Fig.~\ref{fig:RG_running}). 
This is given by 
\begin{eqnarray}
 \Sigma (p) &=& i\int \frac{d^4k}{(2\pi)^4} \left(i g_s \gamma^\mu T^a\right) \frac{i (\cancel{p}+\cancel{k}+m_{b/c})}{(p+k)^2 - m_{b/c}^2} \left(i g_s \gamma^\nu T^b\right) \frac{(-i g_{\mu\nu}\delta_{ab})}{k^2} \nonumber\\
 &=& \frac43 \left(-\frac{\alpha_s}{4\pi}\cancel{p} + 
\underbrace{\frac{\alpha_s\, m_{b/c}}{\pi}}_{dropped} \right) \frac1\epsilon + 
{\rm finite} \label{self-corr}
\end{eqnarray}
where $p$ is the momentum of the incoming (or outgoing) quark. \\
From Feynman diagram on the right of Fig.~\ref{fig:RG_running}, we find that the vertex correction in $d$ dimensions ($d = 4 - 2\epsilon$) is given by
\begin{eqnarray}
\Gamma_{\rm Had}(p,p') &=& i \int \frac{d^dk}{(2\pi)^d} \left(i g_s \gamma^\lambda 
T^a\right) \frac{i}{\cancel{p} + \cancel{k} - m_{b}} i {\cal F} \frac{i}{\cancel{p}' + 
\cancel{k} - m_{c}} \left(i g_s \gamma^\sigma T^b\right) (-i\delta_{ab} 
g_{\lambda\sigma}) \frac{1}{k^2} \nonumber\\*
&=& ig_s^2 C_2(3) \int \frac{d^dk}{(2\pi)^d} \frac{ \gamma_\lambda   \left( 
\cancel{p} + \cancel{k} + m_b \right) {\cal F} \left( \cancel{p}' + \cancel{k} +m_c 
\right) \gamma^\lambda }{k^2\left((p +k)^2 + m_b^2 \right) \left((p' +k)^2 + m_c^2 \right)}
\end{eqnarray}

where $C_2(3)=\frac43$ and ${\cal F} = 1, \gamma_\mu, \sigma_{\mu\nu}$ for scalar, 
vector and tensor operators and $p$ ($p'$) is the on-shell momentum of the $b$ ($c$) quark. A few things are noteworthy and enlisted below:
\begin{itemize}
 \item 
As the denominator has mass dimension 6, divergence will appear only when the
numerator is a function of loop momentum with mass dimension greater than
and equals to two.
\item 
The general form of the numerator is 
\begin{eqnarray}
N &=& \gamma_\lambda \left(\cancel{p'}+\cancel{k} + m_b   \right) {\cal F}
\left(\cancel{p}+\cancel{k} + m_c  \right) \gamma^\lambda \nonumber\\
&=& \gamma_\lambda \cancel{k} {\cal F} \cancel{k}\gamma^\lambda + {\rm
finite}
\end{eqnarray}
\begin{itemize}
\item
For scalar
\begin{equation} 
N= 4 k^2 
\end{equation}
\item
For vector
\begin{equation*}
N= \gamma_\lambda \cancel{k} \gamma_\mu \cancel{k}\gamma^\lambda = - k^2
\gamma_\lambda \gamma_\mu \gamma^\lambda + 2 k_\mu \gamma_\lambda
\cancel{k}\gamma^\lambda = 2 k^2 \gamma_\mu - 4 k_\mu \cancel{k} \nonumber
\end{equation*}
Using 
\begin{equation*}
\int d^4k k^\mu k^\nu f(k^2) =  \frac14 g^{\mu\nu} \int d^4k k^2 f(k^2)
\end{equation*}
we get
\begin{equation}
N= k^2 \gamma_\mu
\end{equation} 
\item
For tensor
\begin{equation}
N= \gamma_\lambda \cancel{k} \sigma_{\mu\nu} \cancel{k}\gamma^\lambda {\bf 
\rightarrowtail} k^2
\frac14 \gamma_\lambda \gamma_\rho \sigma_{\mu\nu} \gamma^\rho \gamma^\lambda = 
0 
\end{equation}
\end{itemize}
where we used the previous integral formula in the second step.
\end{itemize}

Putting this back and using Feynman parameterisation and neglecting quark masses, we have 
the following formula
\begin{eqnarray}
\Gamma_{\rm Had} &=& ig_s^2 C_2(3) {\cal N} {\cal F} \int_0^1 d\zeta  \int 
\frac{d^dk}{(2\pi)^d} \frac{ 1 }{\left[\zeta\left(p +k \right)^2 + (1- \zeta ) 
\left(p' +k \right)^2 \right]^2}\nonumber\\
&=& i\frac{16 \pi}{3}\alpha_s {\cal N} {\cal F} \int_0^1 d\zeta  \int 
\frac{d^d\ell}{(2\pi)^d} \frac{1}{\left(\ell^2 +\Delta\right)^2} \nonumber\\
&&\text{where $\ell = k + p + (1-\zeta)(p'-p)$ and $\Delta =  \zeta (1-\zeta)(p'-p)^2$} \nonumber \\
&=& i\frac{16 \pi}{3}\alpha_s {\cal N} {\cal F} \int_0^1 d\zeta 
\frac{i}{(4\pi)^2} \left( \frac{2}{\epsilon} + {\rm finite} \right) \nonumber\\
&=& - \frac{\alpha_s}{4\pi} \frac{8 {\cal N}}{3} {\cal F} \frac{1}{\epsilon} + {\rm finite} \label{vertex-corr}
\end{eqnarray}
where ${\cal N} = 4,1,0$ for ${\cal F} = 1, \gamma^\mu, \sigma^{\mu \nu}$ respectively.
The bare effective Lagrangian to the lowest power in derivatives is 
\begin{equation}
\mathcal{L}_{\rm eff}^{\rm bare} = i \bar{\psi}_0 \cancel{\partial} \psi_0 + \mathcal{C} \bar{c}_0 \mathcal{F} b_0 \bar{\ell}_0 \mathcal{F}' \nu_{\ell 0}
\label{eff-Lag}
\end{equation}
where $\psi_0$ is any bare quark or lepton field, $\mathcal{C}$ is the Wilson coefficient to the six-dimensional operator and $\mathcal{F}$, $\mathcal{F}'$ are Dirac operators. \\
We redefine the quantities in the bare Lagrangian as 
\begin{equation}
\psi_0 = \sqrt{Z_\psi} \psi; \quad \quad \mathcal{C}_0 = \mu^{2\epsilon} Z_{\mathcal{C}} \mathcal{C}
\end{equation}
where $\psi$ represents any quark field. The QCD contributions to the 
different quark fields will be equal to each other.
Then Eqn. \ref{eff-Lag} can then be written as 
\begin{eqnarray}
\mathcal{L}_{\rm eff}^{\rm ren} &=& i Z_\psi \bar{\psi}\cancel{\partial}\psi + {\mathcal{C}} \ Z_{\mathcal{C}} Z_\psi^2 \mu^{2\epsilon} \ \bar{c} \mathcal{F} b \  \bar{\ell} \mathcal{F}' \nu_\ell \nonumber \\
&=& i \bar{\psi}\cancel{\partial}\psi + i (Z_\psi - 1) \bar{\psi}\cancel{\partial}\psi + {\mathcal{C}} \mu^{2\epsilon}\  \bar{c} \mathcal{F} b \  \bar{\ell} \mathcal{F}' \nu_\ell + {\mathcal{C}} \  (Z_{\mathcal{C}} Z_\psi^2 - 1)\  \mu^{2\epsilon}\  \bar{c} \mathcal{F} b \  \bar{\ell} \mathcal{F}' \nu_\ell \nonumber 
\end{eqnarray}
Absorbing the divergences in Eqn. \ref{self-corr} and Eqn. \ref{vertex-corr} in the counter terms, we find that
\begin{eqnarray}
Z_\psi = 1 - \frac43 \frac{\alpha_s}{4\pi}\frac1\epsilon \mbox{ and } Z_{\mathcal{C}} = 1- \frac83 \frac{\alpha_s}{4\pi} ({\cal N}-1) \frac1\epsilon
\end{eqnarray}
Using the RG equations, the $\beta$-function turns out to be
\begin{eqnarray}
\beta_{\mathcal{C}} &=& -2 \epsilon {\mathcal{C}} -  \frac{\mu}{Z_{\mathcal{C}}} {\mathcal{C}} \frac{d\, Z_{\mathcal{C}}}{d\mu} \nonumber\\
&=&  \frac83 \frac{1}{4\pi} ({\cal N}-1) {\mathcal{C}}  \frac{\mu}{Z_{\mathcal{C}}}
\frac{d\, \alpha_s}{d\mu} \frac1\epsilon  \nonumber\\
&=& - \frac83 \frac{\alpha_s}{4\pi} ({\cal N}-1){\mathcal{C}}
\end{eqnarray}
Thus, 
\begin{equation}
\beta_{\mathcal{C}}^S = {\bf -8}   \frac{\alpha_s}{4\pi} {\mathcal{C}} ,\,\, \beta_{\mathcal{C}}^V = {\bf 0}, 
\mbox{ and } \beta_{\mathcal{C}}^T =  {\bf\frac83} \, \frac{\alpha_s}{4\pi} {\mathcal{C}}
\label{eqn:beta-function}
\end{equation}
where the superscripts $S$, $V$ and $T$ on the $\beta$ denote scalar, vector and tensor couplings. 
\begin{figure}[t]
\begin{center}
\hspace{-9mm}\includegraphics[scale=0.48]{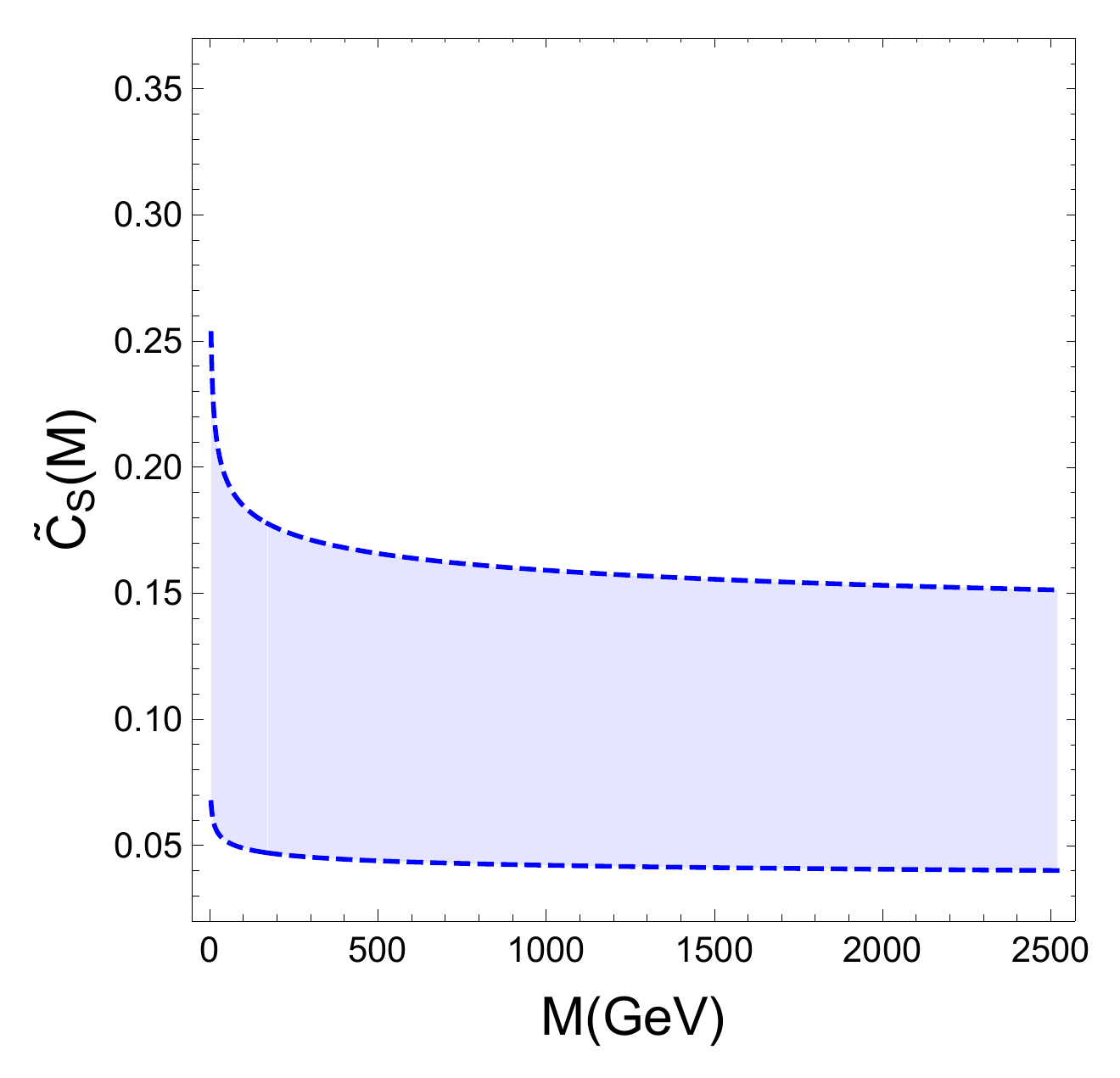}
\includegraphics[scale=0.48]{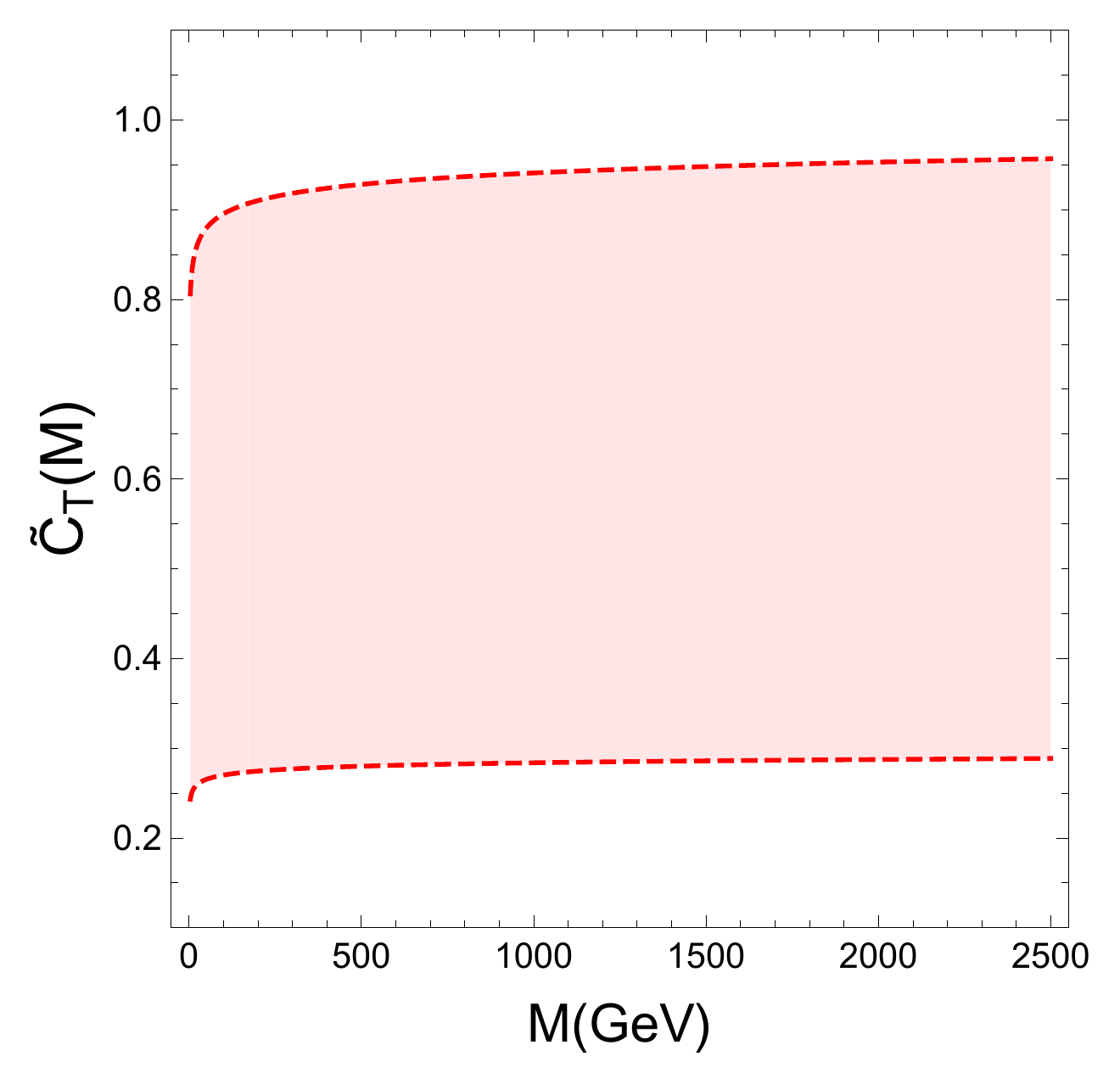}
\end{center}
\caption{Plot of the running of the Scalar (left) and Tensor (right) Wilson Coefficients. The range of the running is from $m_b$ to 2.5 TeV. As a demonstration, the range of the initial values used are the ones mentioned in the text for $B \to D$ decay.}
\label{fig:wilson-run}
\end{figure}
The running of the Wilson Coefficients can be found by solving the $\beta$-function equation given in Eqn. \ref{eqn:beta-function}. Solving, we get, 
\begin{equation}
\tilde{C}(m_b)  = \left[\frac{\alpha_s(m_t)}{\alpha_s(m_b)}\right]^{\frac{\gamma}{2\beta_0^{(5)}}} \left[\frac{\alpha_s(M)}{\alpha_s(m_t)}\right]^{\frac{\gamma}{2\beta_0^{(6)}}} \tilde{C}(M)
\end{equation}

Thus, the scalar and tensor WCs are given by:
\begin{eqnarray}
\tilde{C}_S(M)  &=& \left[\left[\frac{\alpha_s(m_t)}{\alpha_s(m_b)}\right]^{\frac{\gamma_S}{2\beta_0^{(5)}}} \left[\frac{\alpha_s(M)}{\alpha_s(m_t)}\right]^{\frac{\gamma_S}{2\beta_0^{(6)}}}\right]^{-1} \tilde{C}_S(m_b) \\
\tilde{C}_T(M)  &=& \left[\left[\frac{\alpha_s(m_t)}{\alpha_s(m_b)}\right]^{\frac{\gamma_T}{2\beta_0^{(5)}}} \left[\frac{\alpha_s(M)}{\alpha_s(m_t)}\right]^{\frac{\gamma_T}{2\beta_0^{(6)}}}\right]^{-1} \tilde{C}_T(m_b)
\end{eqnarray}
where 
\begin{equation}
\gamma_S = -8 \quad \quad \quad \gamma_T = \frac{8}{3}
\end{equation}
which are simply the boldfaced coefficients in Eqn. \ref{eqn:beta-function}. This is 
plotted in Fig. \ref{fig:wilson-run}.

%
%
\providecommand{\href}[2]{#2}\begingroup\raggedright\endgroup
\end{document}